\RequirePackage{lineno}
\documentclass[aps,prd,twocolumn,floatfix,superscriptaddress,balancelastpage,nofootinbib,amsmath,preprintnumbers]{revtex4-1}[11pt]
\pdfoutput=1
\usepackage{graphicx}
\usepackage{amsmath,amssymb,mathrsfs}
\usepackage{bbm}
\usepackage{color}
\usepackage{xcolor}
\usepackage{dsfont}
\usepackage{cancel}
\usepackage{dsfont}
\usepackage{epstopdf}
\usepackage{epsfig}
\usepackage{bm}
\usepackage{dcolumn}
\usepackage{hyperref}
\usepackage{booktabs}
 \usepackage{array,multirow}
 \usepackage{comment}

\usepackage[utf8]{inputenc}

\newcommand{\ben}{\begin{enumerate}}
\newcommand{\een}{\end{enumerate}}
\newcommand{\bit}{\begin{itemize}}
\newcommand{\eit}{\end{itemize}}

\newcommand{\beqa}{\begin{eqnarray}}
\newcommand{\eeqa}{\end{eqnarray}}
\newcommand{\beq}{\begin{equation}}
\newcommand{\eeq}{\end{equation}}
\newcommand{\bay}{\begin{array}}
\newcommand{\eay}{\end{array}}

\def\ifmath#1{\relax\ifmmode #1\else $#1$\fi}

\def\gsim{\ \rlap{\raise 3pt \hbox{$>$}}{\lower 3pt \hbox{$\sim$}}\ }
\def\lsim{\ \rlap{\raise 3pt \hbox{$<$}}{\lower 3pt \hbox{$\sim$}}\ }

\def\ls#1{\ifmath{_{\lower1.5pt\hbox{$\scriptstyle #1$}}}}
\def\lsup#1{^{\lower 6pt\hbox{$\scriptstyle#1$}}}

\def\bracket#1#2 {\mathinner{\langle{#1}|{#2}\rangle}}

\def\bracket#1#2 {\mathinner{\langle{#1}|{#2}\rangle}}

\newcommand{\bea}{\begin{eqnarray}}
\newcommand{\eea}{\end{eqnarray}}

\graphicspath{{figs/}}
\begin{document}
\title{Solar Neutrinos as a Signal and Background in Direct-Detection Experiments Searching for Sub-GeV Dark Matter With Electron Recoils}

\author{Rouven Essig}
\email{rouven.essig@stonybrook.edu}
\affiliation{C.N. Yang Institute for Theoretical Physics, Stony Brook University, Stony Brook, NY 11794}

\author{Mukul Sholapurkar}
\email{mukul.sholapurkar@stonybrook.edu}
\affiliation{C.N. Yang Institute for Theoretical Physics, Stony Brook University, Stony Brook, NY 11794}

\author{Tien-Tien Yu}
\email{tien-tien.yu@cern.ch}
\affiliation{Theoretical Physics Department, CERN, CH-1211 Geneva 23, Switzerland}
\affiliation{Department of Physics and Institute of Theoretical Science, University of Oregon, Eugene, Oregon 97403}

\preprint{YITP-SB-17-36, CERN-TH-2017-194}

\begin{abstract}
Direct-detection experiments sensitive to low-energy electron recoils from sub-GeV dark matter interactions 
will also be sensitive to solar neutrinos via coherent 
neutrino-nucleus scattering (CNS), since the recoiling nucleus can produce a small ionization signal. 
Solar neutrinos constitute both an interesting signal in their own right and a potential background to a dark matter search that cannot be controlled or reduced by 
improved shielding, material purification and handling, or improved detector design.  
We explore these two possibilities in detail for semiconductor (silicon and germanium) and xenon targets, 
considering several possibilities for the unmeasured ionization efficiency at low energies. 
For dark-matter-electron-scattering searches, neutrinos start being an important background for exposures 
larger than $\sim$1--10~kg-years in silicon and germanium, and for exposures larger than $\sim$0.1--1~kg-year in xenon.  
For the absorption of bosonic dark matter (dark photons and axion-like particles) by electrons, neutrinos 
are most relevant for masses below $\sim$1~keV and again slightly more important in xenon. 
Treating the neutrinos as a signal, we find that the CNS of $^8$B neutrinos can be observed with $\sim$2$\sigma$ significance with exposures 
of $\sim$2, 7, and 20 kg-years in xenon, germanium, and silicon, respectively, assuming there are no other backgrounds. 
We give an example for how this would constrain non-standard neutrino interactions. 
Neutrino components at lower energy can only be detected if the ionization efficiency is sufficiently large.  
In this case, observing pep neutrinos via CNS requires exposures $\gtrsim$10--100~kg-years in silicon or germanium 
($\sim$1000~kg-years in xenon), and observing CNO neutrinos would require an order of magnitude more exposure.  
Only silicon could potentially detect $^7$Be neutrinos. 
These measurements would allow for a direct measurement of the electron-neutrino survival probability over a wide energy range.  
\end{abstract}

\maketitle

\tableofcontents

\section{Introduction}  

Dark matter (DM) direct-detection experiments typically search for recoiling nuclei from DM-nucleus scattering events.  
Upcoming experiments will soon have sufficiently low thresholds and large enough exposures to be sensitive to solar neutrinos, which 
can scatter coherently off nuclei~\cite{Freedman:1973yd}.  
Moreover, solar neutrinos will eventually be a dominant background when probing sufficiently small DM-nucleon cross sections 
(this is sometimes called the ``neutrino floor'').  
Solar neutrinos were first mentioned as a background to direct-detection experiments more than 30 years ago~\cite{Cabrera:1984rr}, 
and have been explored in detail since then, see e.g.~\cite{Drukier:1986tm,Monroe:2007xp,Vergados:2008jp,Strigari:2009bq,Gutlein:2010tq,Billard:2013qya,Baudis:2013qla,Ruppin:2014bra}.  

Of increasing interest in the last few years is to expand DM searches to masses well below the GeV-scale, 
for which the energy of a recoiling nucleus typically falls below current detector thresholds.  
A particularly promising strategy is to search for 
DM interactions with electrons, using various materials~\cite{Essig:2011nj,Essig:2012yx,Graham:2012su,Essig:2015cda,Hochberg:2015pha,Hochberg:2015fth,Derenzo:2016fse,Hochberg:2016ntt,Essig:2017kqs,Tiffenberg:2017aac,Hochberg:2017wce,Aprile:2014eoa,An:2014twa,Hochberg:2016ajh,Hochberg:2016sqx,Bloch:2016sjj,Aguilar-Arevalo:2016zop,Knapen:2017ekk,Cavoto:2017otc}.  
The resulting small ionization signals can be detected with new, low-threshold detectors~\cite{Tiffenberg:2017aac,Romani:2017iwi}, and the 
first generation of new experiments with exposures $\sim100$-gram-years will be operating soon.  
For more details and reference, see~\cite{Battaglieri:2017aum,Alexander:2016aln}.  

Direct-detection experiments searching for small ionization signals will also be sensitive to solar neutrinos via coherent 
neutrino-nucleus scattering (CNS)~\cite{Essig:2011nj}, since 
the recoiling nucleus can produce a small ionization signal.\footnote{Solar neutrinos can also scatter directly off 
electrons, but the resulting electron recoils are typically at much higher energies than the electron recoil energies of interest from 
DM~\cite{Essig:2011nj}.  
We will include them in our analysis below, but they are subdominant.}  
While many challenges will need to be overcome to control both radioactive and detector-specific backgrounds as exposures 
approach $\mathcal{O}$(kg-year) and increase beyond that, solar neutrinos present both an interesting signal in their own right, as well as 
a background (to a DM search) that cannot be controlled or reduced by 
improved shielding, material purification and handling, or improved detector design.  
We thus look ahead and analyze the prospects for detecting and understanding the properties of solar neutrinos as well how they 
would eventually limit the sensitivity of direct-detection experiments 
sensitive only to electron recoils.\footnote{Experiments that are able to distinguish low-energy electron recoils from nuclear recoils 
will be able to distinguish coherent solar-neutrino scattering from DM-electron scattering; 
we do not consider this possibility here.  Also, see~\cite{Strigari:2016ztv} for a discussion of the neutrino backgrounds for 
low-threshold nuclear recoil searches.}

In this paper, we have {\bf two specific aims}. {\bf First}, we will calculate the {\it neutrino backgrounds} for two semiconductor targets, 
silicon and germanium, as well as for xenon
(in an appendix, we discuss briefly the scintillating targets NaI, CsI, and GaAs~\cite{Derenzo:2016fse}).  
We will discuss two distinct classes of DM models, 
which lead to very different electron-recoil spectra: (i) MeV-to-GeV mass DM that \textit{scatters} off electrons, for both momentum-dependent and 
momentum-independent DM interactions, and (ii) eV-to-keV bosonic 
DM, including dark photons ($A'$) and axion-like particles (ALPs), that are \emph{absorbed} by electrons.  
For both classes of DM models, we will present the exposure-dependent discovery limits assuming that the only background is 
from solar neutrinos.  
Since a significant uncertainty in estimating the solar neutrino background is how much ionization is generated by low-energy nuclear recoils, 
$E_{\rm NR}\lesssim 1$~keV, we present our results under different assumptions for the low-energy ionization efficiency. 

\begin{figure}[t!]
\includegraphics[width=0.48\textwidth]{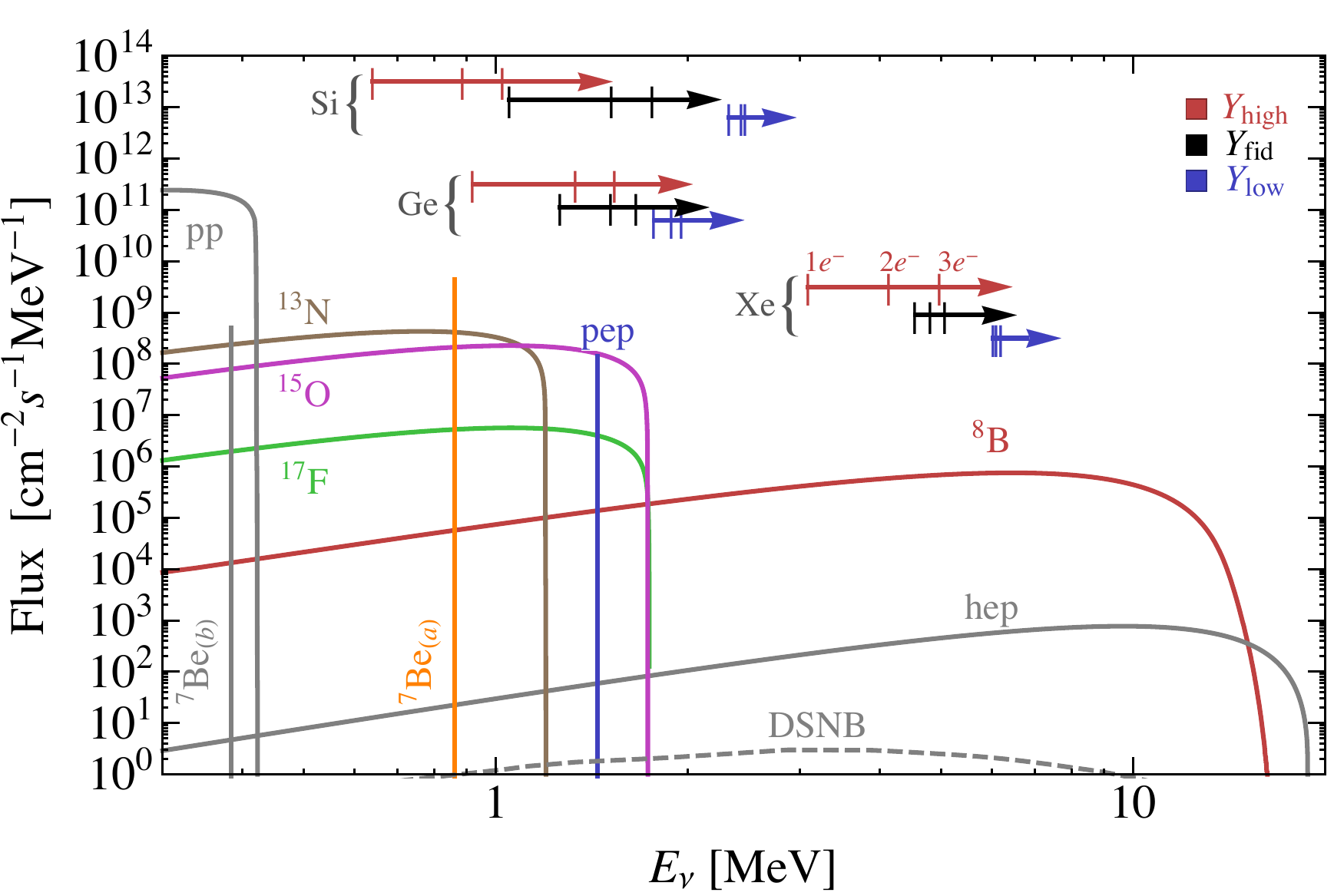}
\caption{Various components of the neutrino flux on Earth as a function of neutrino energy. The fluxes that contribute 
to the background of the direct detection of sub-GeV DM are shown in color and are dominated by the solar neutrinos.  
We use the solar neutrino model BS05(OP)~\cite{Bahcall:2004pz,Bahcall:2004mq}. 
The gray dashed line is the contribution from diffuse supernova neutrinos (DSNB). 
The atmospheric neutrinos are not shown as they are subdominant over the plot's energy range. The horizontal colored lines 
show the neutrino-energy thresholds for seeing at least 1, 2, or 3 electrons (indicated with vertical bars) in silicon, germanium, and xenon, 
under different assumptions of the ionization efficiency (high, fiducial, and low), which are given in 
Sec.~\ref{subsubsec:Y-semi} and Fig.~\ref{fig:conv} for silicon and germanium, 
and in Sec.~\ref{subsubsec:Y-xenon} and Fig.~\ref{fig:convxe} for xenon. 
}
\label{fig:flux}
\end{figure}

Our {\bf second} specific aim is to treat the {\it neutrinos as the signal} and analyze how well future direct-detection experiments 
can measure some of the solar-neutrino components via CNS. 
(Only the CNS of laboratory-produced neutrinos have been detected recently, by the COHERENT collaboration~\cite{Akimov:2017ade}.) 
There are two main process chains that produce neutrinos, the proton-proton and the Carbon-Nitrogen-Oxygen (CNO) cycles.  
The former produces the pp, pep, hep, $^8$B, and $^7$Be neutrino components, which have all been measured through (non-coherent) 
$\nu$-electron scattering~\cite{Agostini:2017cav,Aharmim:2011vm,Abe:2016nxk,Bellini:2014uqa,Collaboration:2011nga,Bellini:2013lnn,Bergstrom:2016cbh,Aharmim:2006wq}, while the latter produces the $^{13}\rm{N}$, $^{15}\rm{O}$, and $^{17}\rm{F}$ components, which have not yet been measured; see Fig.~\ref{fig:flux} and Table~\ref{tab:nuflux}.  
We will investigate how well future direct-detection experiments sensitive to electron recoils could measure the CNS scattering 
of $^8$B, pep, and $^7$Be components, which dominate in different parts of the neutrino energy spectrum, as well as the 
(subdominant) CNO-cycle components.    
Fig.~\ref{fig:flux} shows the neutrino-energy thresholds for seeing at least 1, 2, or 3 electrons in silicon, germanium, and xenon, 
under different assumptions of the ionization efficiency.  

Measurements of these neutrino components are interesting for several reasons.  
First, direct-detection experiments could measure the $^8$B spectrum 
and (depending on the ionization efficiencies in silicon, germanium, and xenon) 
also probe lower energies than existing SNO measurements 
(SNO detects $^8$B neutrinos via CNS that break apart a deuteron via neutral-current, inelastic scattering; there was no spectral information~\cite{Aharmim:2011vm}).  
Combining this with existing Borexino measurements of $^8$B neutrinos scattering elastically off electrons, 
which only probes the electron-neutrino component of the solar flux, the electron-neutrino survival probability can be directly measured as 
a function of energy.  
In particular, this could yield a first measurement of the survival probability in the transition region between where 
vacuum oscillations dominate at low energies to where the matter (MSW~\cite{Wolfenstein1,Wolfenstein2,Mikheyev-Smirnov}) 
effect dominates at high energies. 
In addition, detecting $^7$Be and pep neutrinos at lower energies 
would directly measure the survival probability in the vacuum-oscillation-dominant region.  
Besides being a welcome test of our current understanding of solar neutrinos, 
this would also strongly constrain any non-standard neutrino interactions (NSI), 
see e.g.~\cite{Harnik:2012ni,Dutta:2015vwa,Dutta:2017nht,AristizabalSierra:2017joc}, 
and we will provide one specific example in this paper.   
Moreover, these measurements would probe for other new physics beyond 
the Standard Model (SM), including a neutrino magnetic moment~\cite{Marciano:1977wx,Lee:1977tib} and sterile neutrinos.  

Second, measuring the solar neutrino fluxes produced by the CNO cycle would inform us of the Sun's metallicity 
(i.e., the abundance of elements heavier than helium), 
and a precise measurement could help solve the solar metallicity problem (also called the ``solar abundance'' 
problem)~\cite{Bahcall:2004pz,Grevesse1998,Asplund:2004eu,Asplund:2009fu,Bergstrom:2016cbh,Song:2017kvf,Cerdeno:2017xxl}.  
This problem arose about a decade ago when new measurements of the solar surface revealed the elements C, N, and O to be less abundant 
than predicted previously by standard solar models.  Standard solar models can account for these lower abundances, but only at the 
cost of becoming incompatible with helioseismic data~\cite{Basu:2009,Serenelli:2016dgz}.   
We study therefore how well the CNO fluxes could be measured in future. 

The outline of the paper is as follows. Sec.~\ref{sec:nu-background} calculates the neutrino signal, assuming various ionization efficiencies. 
In Sec.~\ref{sec:DMsignal}, we briefly review DM-electron scattering and absorption.  
In Sec.~\ref{sec:likelihood}, we describe the log-likelihood analysis with which we compare the electron recoil 
spectra from DM absorption or scattering with the electron recoil spectra from CNS.  
We also describe our analysis procedure for detecting the CNS of $^8$B, pep, $^7$Be$_{(a)}$, and CNO 
neutrinos.  
We present our results for neutrinos as a background and signal in Sec.~\ref{sec:results-DM} and 
Sec.~\ref{sec:results-coherent}, respectively, and in Sec.~\ref{nsi}, we show constraints on some NSI parameters assuming a detection of 
$^8$B in xenon. We conclude in Sec.~\ref{sec:conclusions}.  
Appendix~\ref{app:scintillators} presents the CNS rates for several 
scintillating targets.\footnote{Scintillating targets, such as GaAs, are potentially excellent target materials for sub-GeV DM-electron scattering experiments. However, the lack of low-energy data on the ionization efficiency results in large uncertainties in the conversion from nuclear to electron recoil energy.  A more detailed analysis of the solar neutrino signal in these targets is thus beyond the scope of this work.}, while  Appendix~\ref{app:2e-threshold} shows results assuming an experimental energy threshold of 2 electrons for 
DM-electron scattering. In Appendix~\ref{app:magnetic_moment}, we briefly discuss searches for a neutrino magnetic moment. 

\setlength{\tabcolsep}{0.5em} 
{\renewcommand{\arraystretch}{1.3}
\begin{table}[t]
\caption{Solar neutrino fluxes and their respective uncertainties (in parentheses) 
from the BS05(OP) solar neutrino model~\cite{Bahcall:2004pz,Bahcall:2004mq}.}
\begin{center}
\begin{center}
\begin{tabular}{r l c}
\hline
\multicolumn{2}{c}{Solar neutrino component}&Flux [cm$^{-2}$s$^{-1}$] \\ \hline
$pp:$&$p+p\to {^2\rm{H}}+e^++\nu_e$&5.99$\times$10$^{10}~$(0.7\%)\\
pep:&$ p+e^-\to {^2\rm{H}}+\nu_e$&1.42$\times$10$^{8}~$(1.3\%)\\
${^7\rm{Be}_{(a)}}:$&${^7\rm{Be}}+e^-\to {^7\rm{Li}}+\nu_e$&4.34$\times$10$^{9}~$(5.3\%)\\
${^7\rm{Be}_{(b)}}:$&${^7\rm{Be}}+e^-\to {^7\rm{Li}}+\nu_e$&4.99$\times$10$^{8}~$(5.3\%)\\
${^8\rm{B}}:$&${^8}{\rm B}\to {^8\rm{Be}}+e^++\nu_e$&5.69$\times$10$^{6}~$(11.6\%)\\
hep:&${^3\rm{He}}+p\to ^4\rm{He}+e^++\nu_e$&7.93$\times$10$^{3}~$(2.0\%)\\
${^{13}\rm{N}}:$&${^{13}\rm{N}}\to {^{13}\rm{C}}+e^++\nu_e$&3.07$\times$10$^{8}~$(26.2\%)\\
${^{15}\rm{O}}:$&${^{15}}\rm{O}\to {^{15}\rm{N}}+e^++\nu_e$&2.33$\times$10$^{8}~$(26.2\%)\\
${^{17}\rm{F}}:$&${^{17}}\rm{F}\to {^{17}\rm{O}}+e^++\nu_e$&5.84$\times$10$^{6}~$(48.3\%)\\
\hline
\end{tabular}
\end{center}
\end{center}
\label{tab:nuflux}
\end{table}%

\begin{figure*}[t!]
\includegraphics[width=0.32\textwidth]{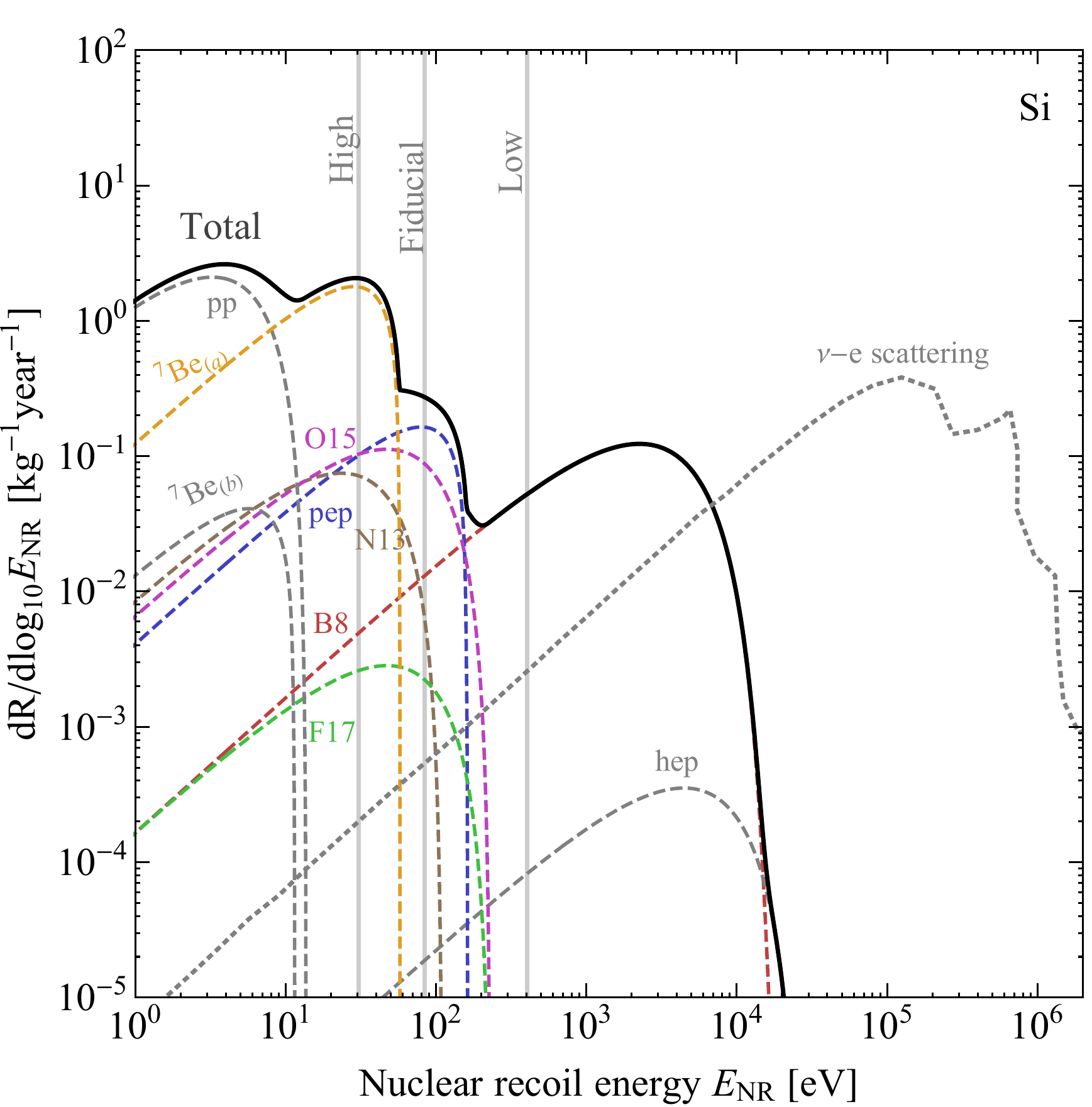}
\includegraphics[width=0.32\textwidth]{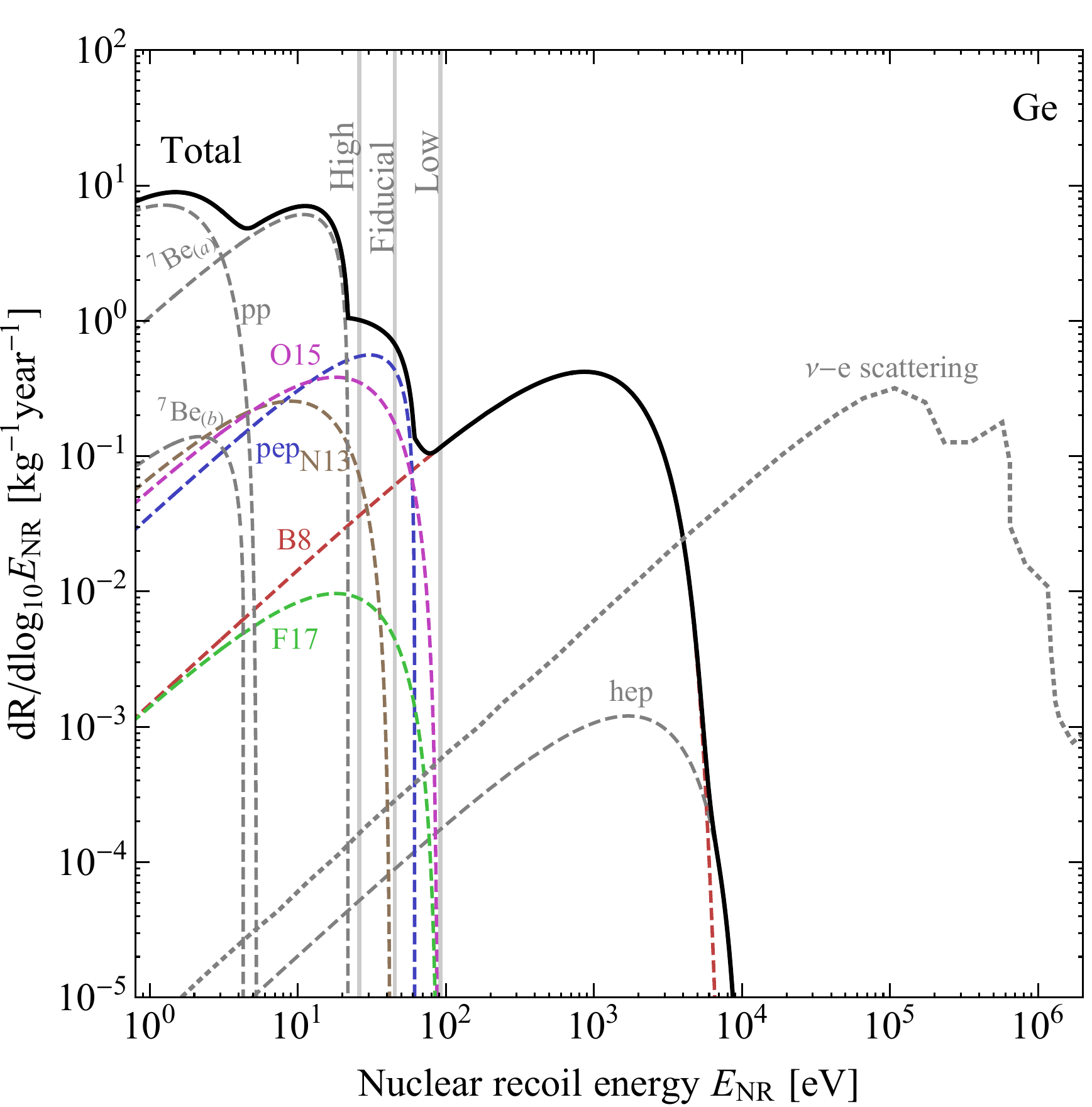}
\includegraphics[width=0.32\textwidth]{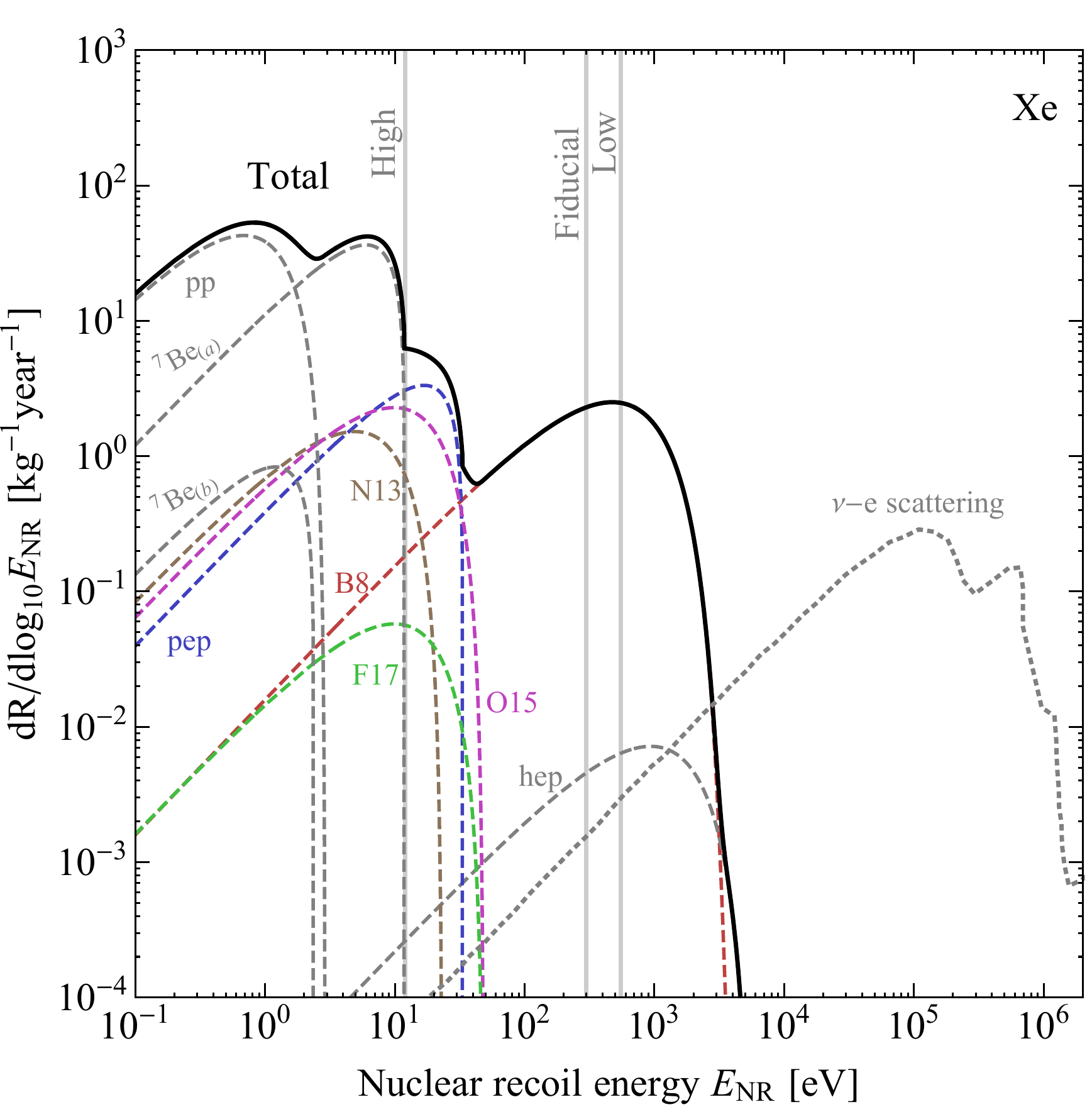}
\caption{Coherent scattering rates for the individual and total solar neutrino components off silicon ({\bf left}), germanium ({\bf center}), and xenon ({\bf right}) nuclei. 
The vertical lines denote the minimum nuclear recoil energy needed to generate a non-zero ionization signal for 
three different ionization efficiencies (for details 
see Sec.~\ref{subsubsec:Y-semi} and Fig.~\ref{fig:conv} for silicon and germanium, 
and Sec.~\ref{subsubsec:Y-xenon} and Fig.~\ref{fig:convxe} for xenon). 
Gray dotted lines show the neutrino-electron scattering rates for the three elements. Note that for the neutrino-electron scattering rates, the x-axis corresponds to electron recoil energy.}
\label{fig:events_si_ge}
\end{figure*}

\section{The Ionization Signal from Solar Neutrino-Nucleus Scattering} \label{sec:nu-background}

\subsection{Neutrino flux}\label{sec:nuflux}

The neutrino flux observed on Earth is composed primarily of solar, atmospheric, and diffuse supernova neutrinos. For low-mass DM, 
we are interested in neutrino energies  $\lesssim$10 MeV, where the solar neutrino flux dominates over the atmospheric and diffuse 
supernova neutrino fluxes, see Fig.~\ref{fig:flux}. 
Hence, in this work we will only consider the contribution of solar neutrinos as a background. 
We use the fluxes given by the high-metallicity solar neutrino model BS05(OP),
together with their respective uncertainties~\cite{Bahcall:2004pz,Bahcall:2004mq}, 
see Table~\ref{tab:nuflux}.\footnote{Other high-metallicity models, such as the GS98-SFII model~\cite{Grevesse1998}, 
have similar fluxes and would thus yield similar results.  
Low-metallicity models mainly predict lower CNO fluxes, and would affect some of our results in Sec.~\ref{sec:results-coherent}. 
}
The neutrinos must have sufficient energy to produce an ionization signal consisting of at least 1 electron.  
In Fig.~\ref{fig:flux}, the flux components that contribute non-negligibly to the DM background are shown in color, while the others are 
shown in gray only for completeness.

\subsection{Coherent neutrino scattering}
The differential cross section for coherent neutrino-nucleus scattering for a nucleus of mass $m_N$ is given by
\begin{eqnarray}\label{neutrino1}
\frac{d\sigma}{dE_{\rm{NR}}}&=& \frac{G_{F}^{2}}{4\pi}Q_{w}^{2}m_{N}\left(1-\frac{m_{N}E_{\rm{NR}} }{2E_{\nu}^{2}}\right)F^{2}(E_{\rm{NR}}), 
\end{eqnarray}  
where $E_{\rm{NR}}$ is the nuclear recoil energy, $E_{\nu}$ is the neutrino energy, $G_{F}$ is the Fermi constant, $Q_{w}=N-Z(1-4\sin^2\theta_w)$ is the weak nuclear hypercharge for $N$ neutrons and $Z$ protons, $\theta_w$ is the weak mixing angle, and $F(E_{\rm{NR}})$ is the standard Helm form factor~\cite{LewinSmith}. In the coherent elastic scattering case, the recoil energies $E_{\rm{NR}}$ are low and 
$F(E_{\rm{NR}}) \simeq 1$. The minimum neutrino energy $E_{\nu}^{\rm min}$ that produces a recoil energy $E_{\rm{NR}}$ is given by
\begin{eqnarray}\label{neutrino2}
E_{\nu}^{\rm min}&=& \sqrt{\frac{m_{N}E_{\rm{NR}}}{2}}\,.
\end{eqnarray}
The differential scattering rate for a detector of mass $M$ and exposure time $T$ is then given by
\begin{eqnarray}\label{neutrino3}
\frac{dR}{dE_{\rm{NR}}}&=& N_{T}MT \int_{E_{\nu}^{\rm min}} \frac{d\sigma}{dE_{\rm{NR}}}  \frac{dN_{\nu}}{dE_\nu} dE_{\nu}\,, 
\end{eqnarray}  
where $N_{T}$ is the number of target nuclei per unit mass and $\frac{dN_{\nu}}{dE_\nu}$ is the neutrino flux.  Fig.~\ref{fig:events_si_ge} shows the rate of neutrino-nucleus scattering events expected per kg-year as a function of nuclear recoil energy in silicon, germanium, and xenon, 
respectively. 
The three vertical lines labelled ``High'', ``Fiducial'', and ``Low'' indicate the minimum nuclear recoil energy that will lead to an ionization signal 
under three different assumptions for the ionization efficiency (discussed next in Sec.~\ref{subsec:yield}).  
The solar neutrino-electron scattering rates are shown in dotted lines and are not an important background 
for sub-GeV DM searches (moreover, our calculation of the neutrino-electron scattering rates do not include atomic binding effects, which yield a sizable suppression at low recoil energies~\cite{Chen:2016eab}).  

\begin{figure*}[t!]
\includegraphics[width=0.48\textwidth]{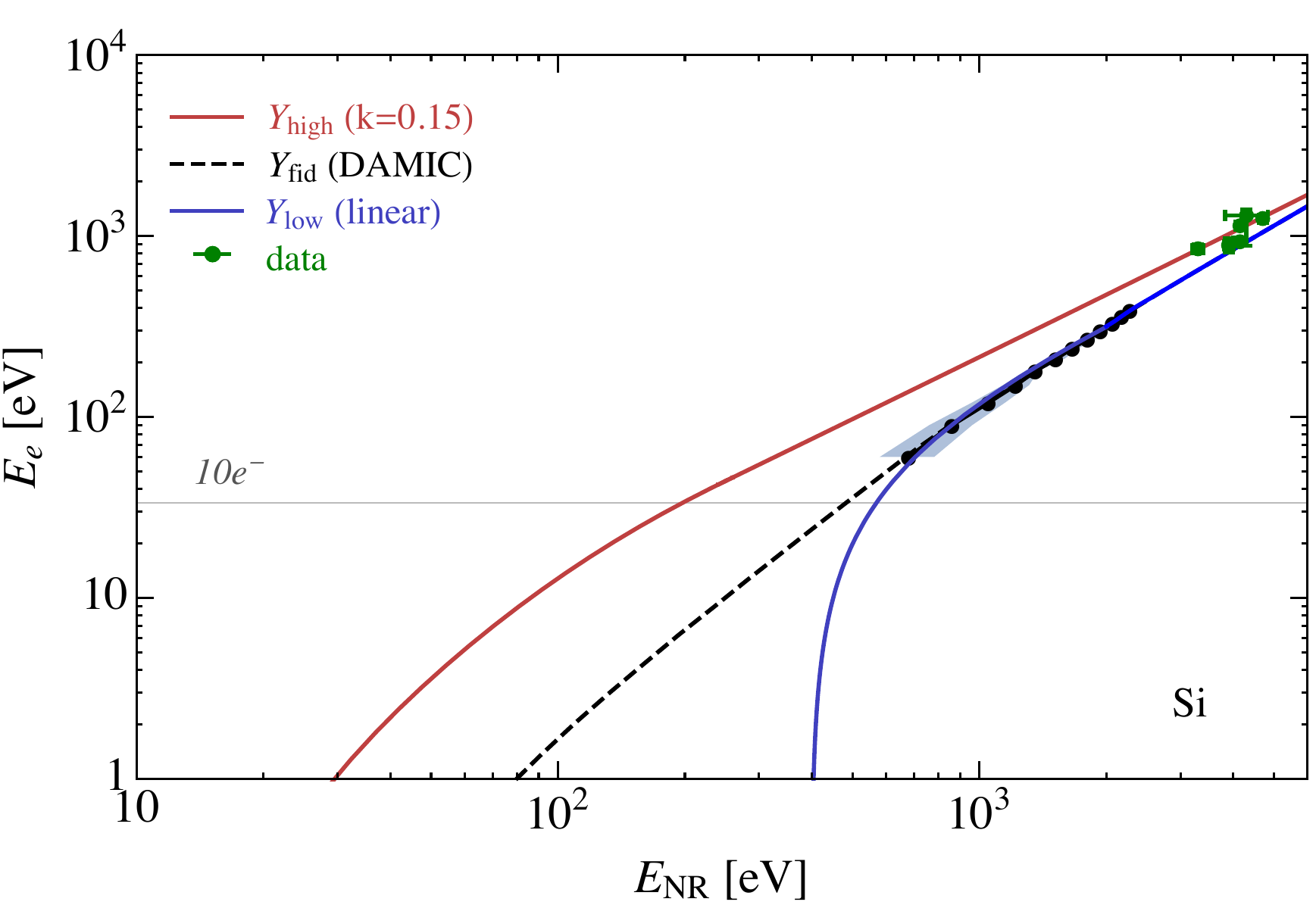} ~~
\includegraphics[width=0.48\textwidth]{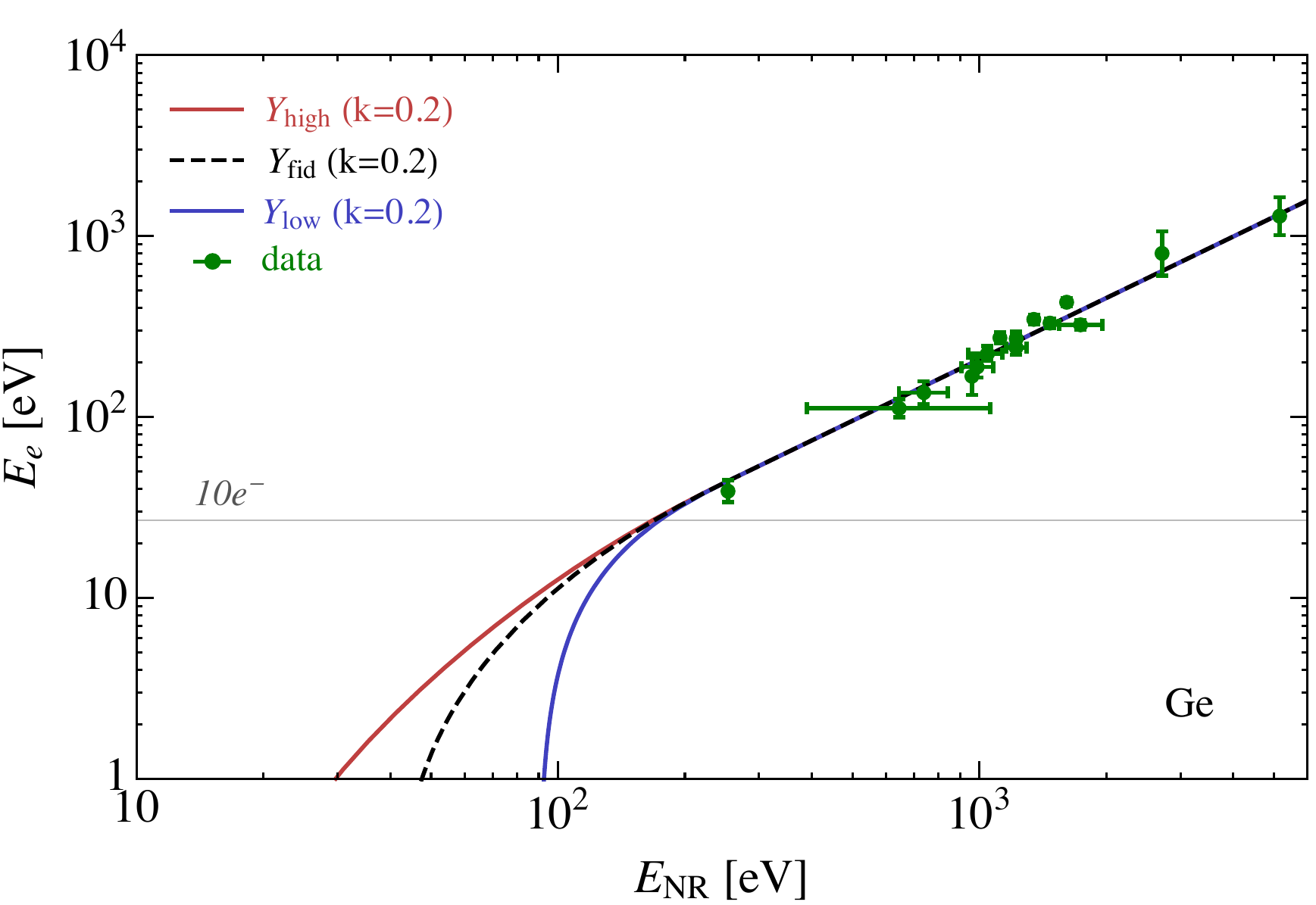}
\caption{Models of the ionization efficiency to convert nuclear recoil energy $E_{\rm{NR}}$ to ionization energy $E_e$ for silicon ({\bf left}) and germanium ({\bf right}) as defined in Table~\ref{tab:conversionsige}. The red solid (black dashed, blue solid) lines 
represent our modeling of a high (fiducial, low) ionization efficiency. 
The horizontal gray line denotes the ionization energy that corresponds to 10 electrons. The data (green points) are from~\cite{PhysRevA.41.4058,PhysRevD.42.3211,PhysRevA.45.2104} for silicon and from~\cite{PhysRevLett.15.245,1968PhRvL..21.1430C,Jones:1975zze,Messous:1995dn,COGENT} for germanium. 
In addition, for silicon, the black dots and shaded blue region show data from the DAMIC collaboration~\cite{Chavarria:2016xsi}.
}
\label{fig:conv}
\end{figure*}

\subsection{Ionization Efficiency} \label{subsec:yield}
A nucleus that recoils after being struck by a solar neutrino can convert some of its energy to an ionization signal and thereby 
produce a background to searches for DM that scatters off, or is absorbed by, electrons.  
At the low energies of interest for sub-GeV DM searches, there are significant uncertainties in the ionization efficiency from a 
nuclear recoil.  
We thus show our results for three different ionization efficienies, which we expect to span a reasonable range that likely includes the true 
ionization efficiency.  
We refer to the three efficiencies as ``high'', ``fiducial'', and ``low'', depending on whether a given nuclear recoil yields a large, medium, or low 
amount of charge, and denote them with $Y_{\rm high}$, $Y_{\rm fid}$, and $Y_{\rm low}$, respectively.  
In the following two subsections, we discuss our treatment of the semiconductors (silicon and germanium) and xenon targets, respectively.  

\subsubsection{Ionization efficiencies for semiconductors}\label{subsubsec:Y-semi}
The ionization energy $E_e$ produced when a nucleus recoils with energy $E_{\rm NR}$ 
is given by 
\beq
E_e = Y E_{\rm{NR}}\,, 
\eeq
where $Y$ is the quenching factor, which depends on $E_{\rm{NR}}$ as well as the detector material. 
At high energies, the quenching factor can be theoretically estimated by the Lindhard model~\cite{Lindhard}, 
\begin{eqnarray}\label{yield1}
Y_{\rm{Lindhard}}(E_{\rm{NR}})&=& \frac{kg(\epsilon)}{1+kg(\epsilon)},\\
 g(\epsilon)&=&3 \epsilon^{0.15}+0.7 \epsilon^{0.6}+\epsilon, \\
 \epsilon&=&11.5Z^{-7/3}E_{\rm{NR}},
 \end{eqnarray}  
where $Z$ is the atomic number of the recoiling nucleus and $E_{\rm{NR}}$ is given in keV. The original description by Lindhard sets $k=0.133Z^{2/3}A^{-1/2}$, where $A$ is the mass number of the nucleus. However, experimental data give a 
range of values for $k$, which is therefore usually treated as a free parameter.

\setlength{\tabcolsep}{0.5em} 
{\renewcommand{\arraystretch}{1.3}
\begin{table}[t]
\caption{
Analytic expressions for the quenching factor for the high, fiducial, and low ionization efficiency models to convert between nuclear recoil 
energy $E_{\rm{NR}}$ and ionization energy $E_e$, as seen in Fig.~\ref{fig:conv}, in silicon and germanium.}
\begin{center}
\begin{tabular}{l c c c}
\hline
& $E_{\rm{NR}}$ [eV] & quenching $Y(E_{\rm{NR}})$\\ \hline
\multirow{8}{*}{\rotatebox[origin=c]{90}{\large{silicon}}}
\multirow{3}{*}{~~~~~~high}& 0-15  & 0 \\
 & 15-250  & $0.18\left[1-e^{-(E_{\rm{NR}} - 15)/71.3}\right]$\\
& $>250$  & $Y_{\rm{Lindhard}}(E_{\rm{NR}})$\\ \cline{2-3}
&0-40& 0 \\
~~~~~~~~~~{fiducial}& 40-675 & $(1.49\times10^{-3} E_{\rm{NR}}^{0.65}-0.01)$ \\ 
& $>675$  & empirical fit of DAMIC data\\ \cline{2-3}

\multirow{2}{*}{~~~~~~~~~~low} & 0-300  & 0\\
& $>300$  & $E_{\rm{NR}}^{-1}(0.20 E_{\rm{NR}}-78.37)$ \\ 
\hline
\multirow{9}{*}{\rotatebox[origin=c]{90}{\large{germanium}}}
\multirow{3}{*}{~~~~~~high}&0-15&0\\
&$15-254$  &$0.18\left[1-e^{-(E_{\rm{NR}} - 15)/71.03}\right]$ \\
 & $>254$  &$Y_{\rm{Lindhard}}(E_{\rm{NR}})$\\ \cline{2-3}
\multirow{3}{*}{~~~~~~~~~~fiducial}&0-40&0\\
&$40-254$  & $0.18\left[1-e^{-(E_{\rm{NR}} - 40)/60.9}\right]$ \\
& $>254$& $Y_{\rm{Lindhard}}(E_{\rm{NR}})$ \\ \cline{2-3}

\multirow{3}{*}{~~~~~~~~~~low}& 0-90& 0\\
&$90-254$  &$0.18\left[1-e^{-(E_{\rm{NR}} - 90)/42.42}\right]$ \\
& $>254$  &$Y_{\rm{Lindhard}}(E_{\rm{NR}})$ \\ \hline

\end{tabular}
\end{center}
\label{tab:conversionsige}
\end{table}%

\begin{figure*}[t!]
\includegraphics[width=0.32\textwidth]{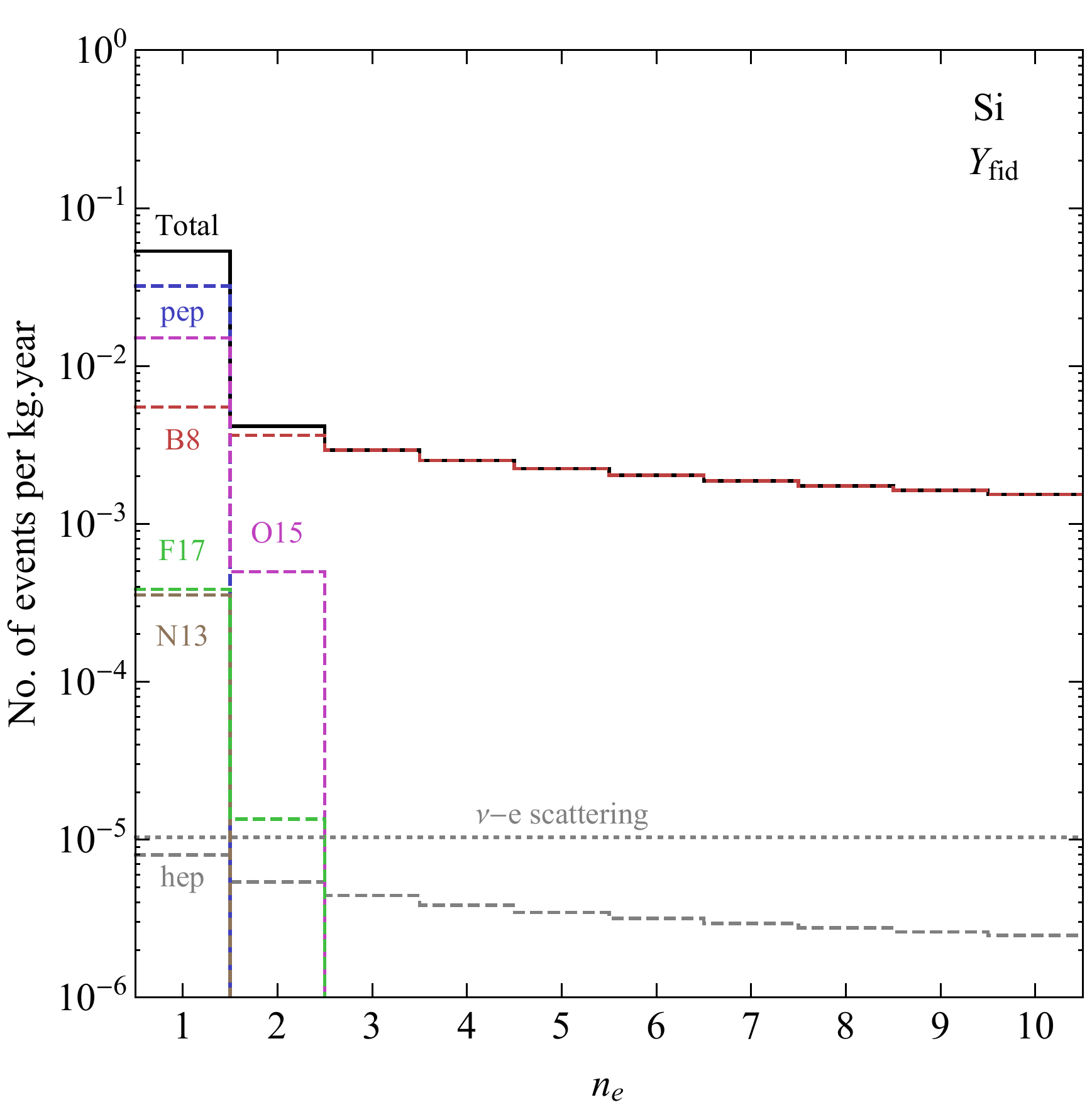}
\includegraphics[width=0.32\textwidth]{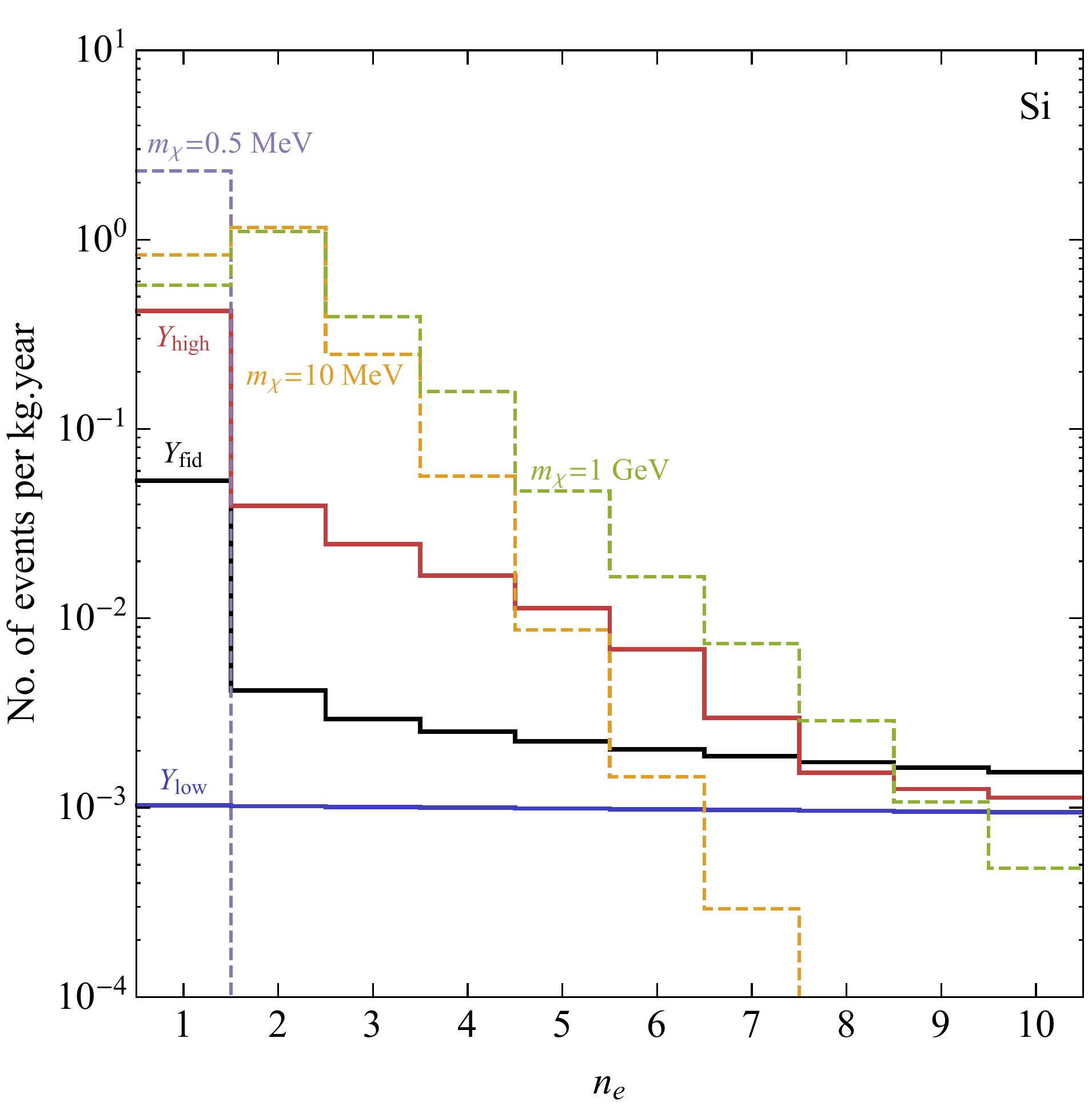}
\includegraphics[width=0.32\textwidth]{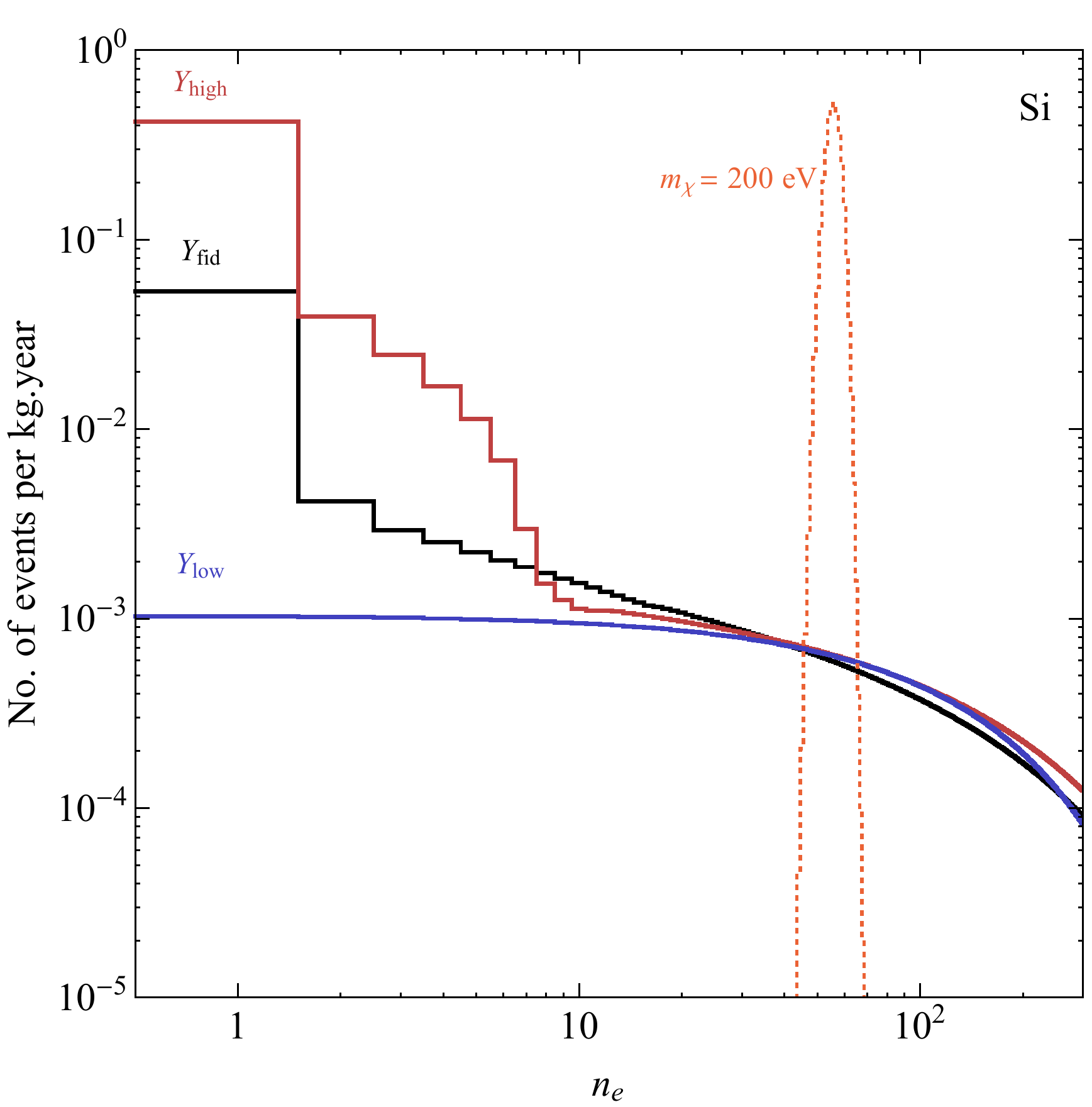}\\
\includegraphics[width=0.32\textwidth]{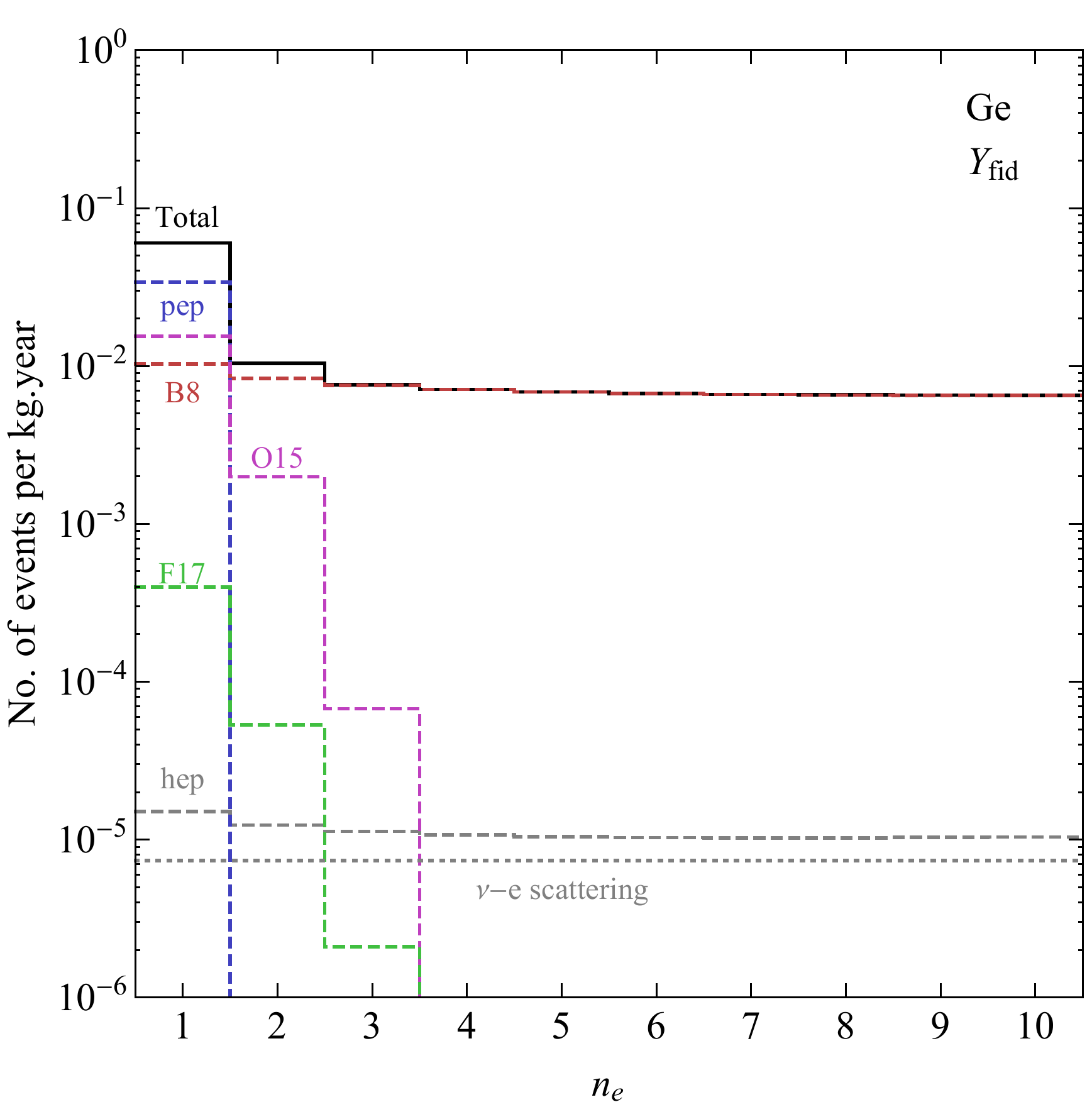}
\includegraphics[width=0.32\textwidth]{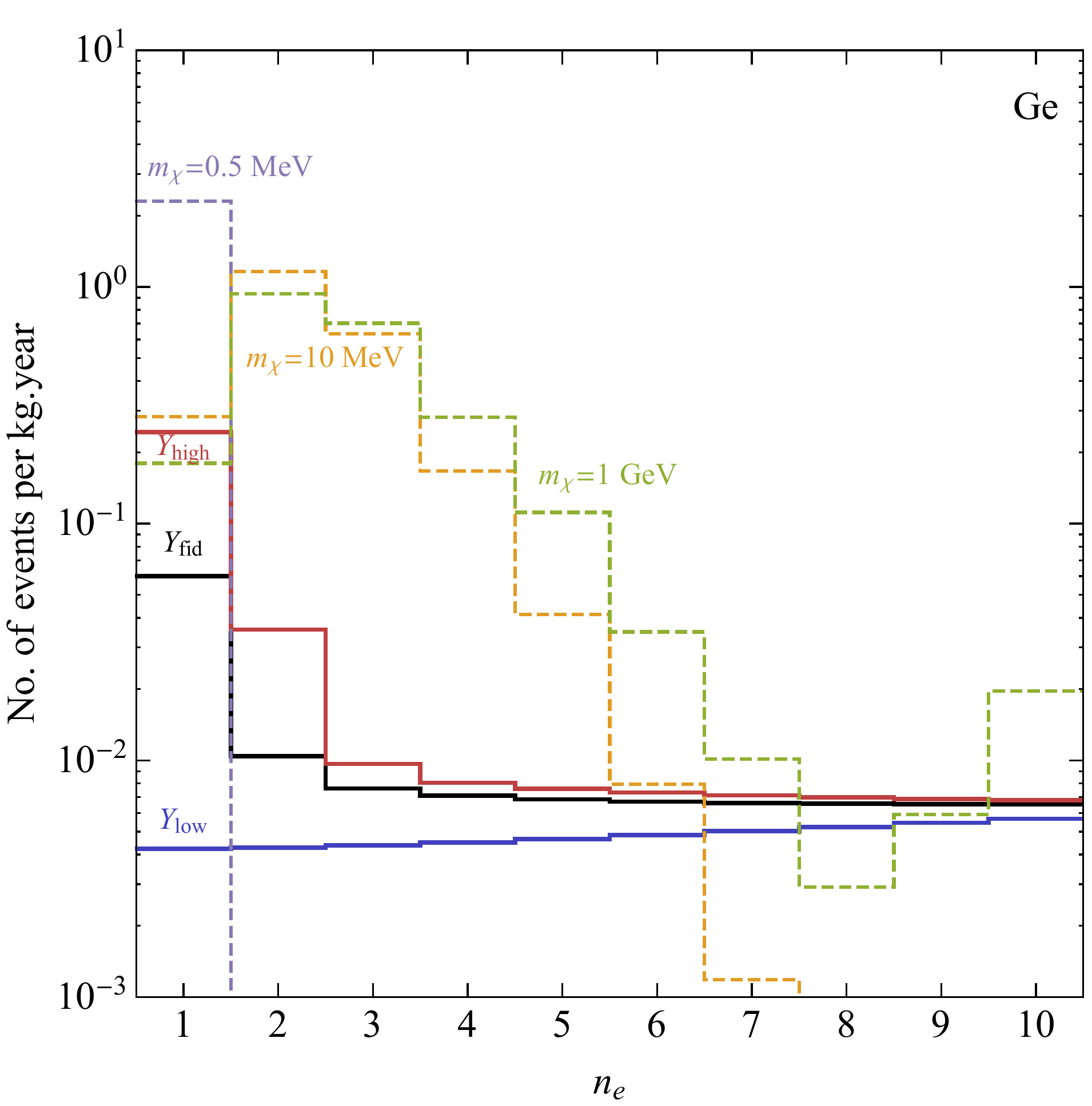}
\includegraphics[width=0.32\textwidth]{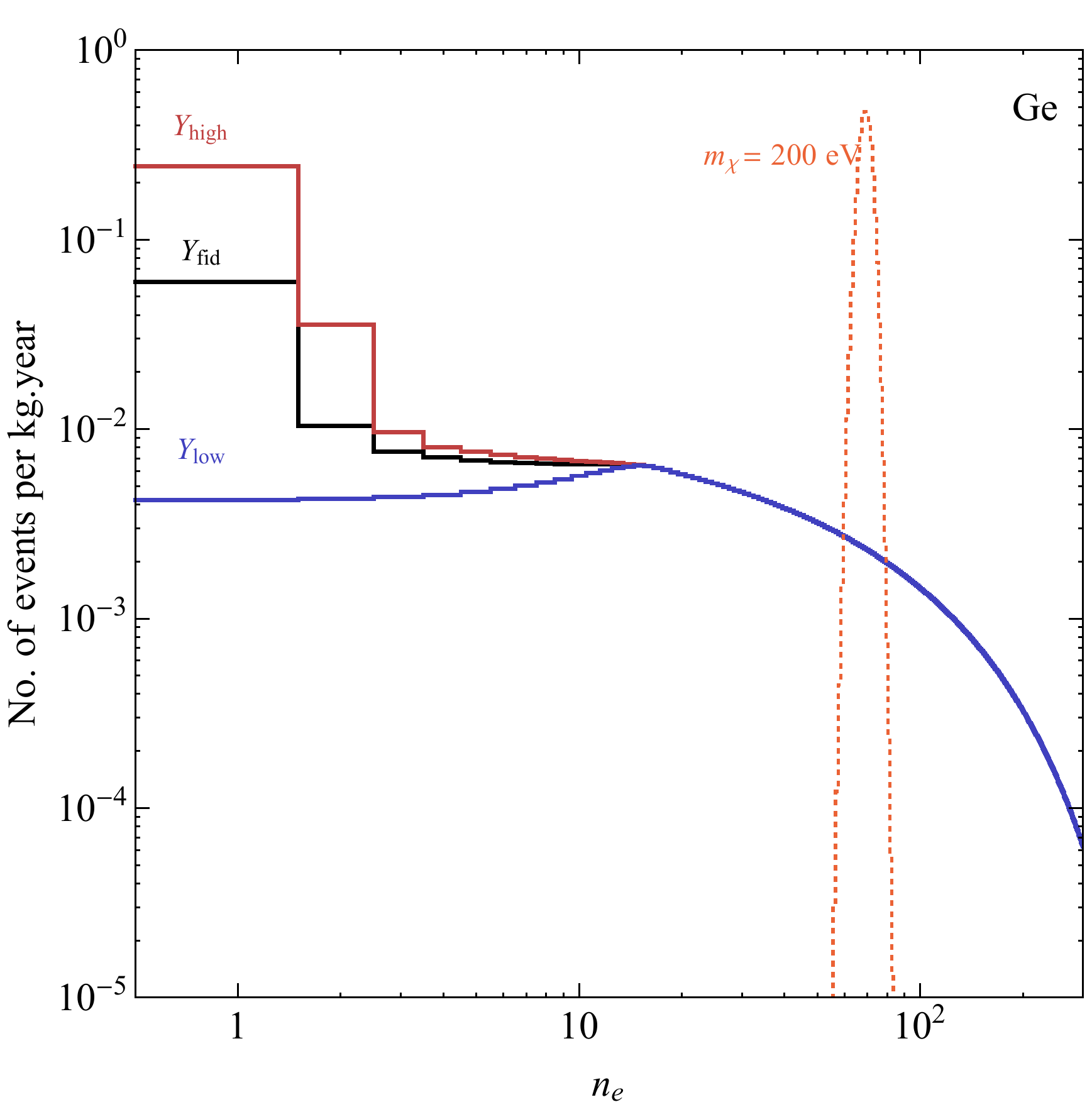}
\caption{Ionization spectra (number of events versus number of electrons) produced by solar neutrinos  
scattering coherently off {\it nuclei} normalized to 1~kg-year, in silicon ({\bf top}) and germanium ({\bf bottom}).  The {\bf left} plots assume the fiducial ionization-efficiency model, with the black line showing the total number of events and the colored lines showing 
the various components. 
The {\bf middle} and {\bf right} plots show the total neutrino flux assuming the low (blue), fiducial (black), and high (red) ionization-efficiency  models. For illustration, we include the DM-scattering spectra for $m_\chi=0.5$~MeV (purple, dashed), 10~MeV (orange, dashed), and 1~GeV (green, dashed) in the {\bf middle} panels, as well as the DM-absorption spectrum for $m_\chi=200$~eV (red, dotted) in the {\bf right} panels.  
To normalize the DM spectra, we set the DM-electron scattering cross section ($\overline\sigma_e$) to the 90\% confidence-level limit assuming no background and an exposure of 1 kg-year.  
For DM-electron scattering, only the first few bins are relevant, while for DM absorption, the signal shape is modeled by Gaussian centered around the mass ({\emph i.e.} 200~eV in the right panel) minus the binding energy. See text for more details.}
\label{fig:neutrinorates}
\end{figure*}

For germanium, the Lindhard model with $k$ between 0.1 and 0.2 is consistent with the data~\cite{BENOIT2007558,Scholz:2016qos}. Ref.~\cite{Barker:2012ek} showed that $k=0.2$ provides a good fit to the quenching data for $E_{\rm NR}\sim 1-10$~keV~\cite{Ahmed:2012vq},
 while $k=0.1$ provides a good fit for $E_{\rm NR}\gtrsim 500$~keV. 
Since we are interested in low energies, $E_{\rm NR}\sim$~eV--keV, we set $k=0.2$. 
In Fig.~\ref{fig:conv} (right), we show the experimental data, which is only available for $E_{\rm NR} \gtrsim 250$~eV~\cite{PhysRevLett.15.245,1968PhRvL..21.1430C,Jones:1975zze,Messous:1995dn,COGENT}, corresponding to $E_e\gtrsim 50$ eV. 
However, sub-GeV DM scattering peaks at $E_e\sim\ $few eV~\cite{Essig:2015cda}, and so we must extrapolate the Lindhard model to lower energies. For our ``fiducial'' model, we smoothly extrapolate the Lindhard model with $k=0.2$ to a cutoff of $E_{\rm{NR}}=40$~eV, which is approximately 2--3 times the minimum energy required to dislocate the germanium atom from the lattice site~\cite{PhysRevB.57.7556}. 
In order to estimate the systematic uncertainty in the neutrino backgrounds, we also define a ``high'' ionization efficiency model that has  
a cutoff of $E_{\rm{NR}}=15$~eV and a ``low'' ionization efficiency model that has a cutoff of $E_{\rm{NR}}=90$~eV.  In the latter case, only 
neutrinos from B$^8$ and hep contribute, see Fig.~\ref{fig:events_si_ge}.
The three ionization efficiencies are shown in Fig.~\ref{fig:conv} (right), while analytical expressions are given in Table~\ref{tab:conversionsige}.  

For silicon, previous data above $E_{\rm{NR}}=3$~keV is fit well with the Lindhard model with $k=0.15$.  However, recent data from the DAMIC collaboration, which spanned the energy range $0.68-2.28$~keV (see Fig.~\ref{fig:conv} (left)), was not consistent with Lindhard~\cite{Chavarria:2016xsi}. Since we are interested in lower energies than the data, we have to extrapolate.  
For our ``fiducial'' model, we extrapolate the DAMIC data as was done by the SuperCDMS collaboration in~\cite{Agnese:2016cpb}. 
This leads to a 40~eV nuclear recoil energy cut-off, which is approximately 2--3 times the minimum energy required to dislocate the silicon atom from the lattice site~\cite{PhysRevB.57.7556}. For our ``low'' ionization efficiency model, we follow 
the DAMIC collaboration~\cite{Aguilar-Arevalo:2016ndq} 
in extrapolating their data linearly in $E_e$ vs $E_{\rm NR}$, which gives a cut-off below which $Y=0$ of $E_{\rm NR} = 300$~eV.  
For our ``high'' ionization efficiency model, we extrapolate Lindhard with $k=0.15$ to lower energies, with a 15~eV 
nuclear recoil energy cut-off.  
Although this model lies above the DAMIC data points, the energy range of interest for sub-GeV DM scattering is below a few hundred eV. Since there is no experimental data below 0.68 keV, the ``high" ionization-efficiency model is a possible model at these lower energies and offers a reasonable upper bound to the neutrino background.  
The three ionization efficiencies are shown in Fig.~\ref{fig:conv} (left), while analytical expressions are given in Table~\ref{tab:conversionsige}.  

We next use these efficiencies to calculate the ionization event rates in silicon and germanium as follows. 
For a given model  defined by a quenching function $Y(E_{\rm{NR}})$, the differential electron ionization energy is 
\begin{eqnarray}\label{conv2}
dE_e&=& Y(E_{\rm{NR}}) \, dE_{\rm{NR}} + E_{\rm{NR}} \, \frac{dY(E_{\rm{NR}})}{dE_{\rm{NR}}} \, dE_{\rm{NR}}\, .
\end{eqnarray}
Let $R_{N}$ denote the rate of scattering on nuclei and $R_{e}$ the rate at which we observe ionized electrons. The differential rate of ionization events is given by,
\begin{eqnarray}\label{conv3}
dR_{e}=\frac{dR_{N}}{dE_{\rm{NR}}} \times dE_{\rm{NR}}\,.
\end{eqnarray}
 Now, dividing Eq.~(\ref{conv3}) by Eq.~(\ref{conv2}), we find  
\begin{eqnarray}\label{conv4}
\frac{dR_{e}}{dE_e}&=&\frac{dR_{N}}{dE_{\rm{NR}}}\times \frac{1}{ (Y(E_{\rm{NR}}) + E_{\rm{NR}}\frac{dY(E_{\rm{NR}})}{dE_{\rm{NR}}})} .
\end{eqnarray}

Starting from the band-gap energy as the minimum energy needed to produce at least one electron-hole pair (0.67~eV in germanium and 1.1~eV in 
silicon~\cite{ExptGaps,Klein:1968}), we can now integrate the differential rate in intervals of the average energy required to produce an electron-hole pair 
(2.9~eV in germanium and 3.6~eV in silicon~\cite{ExptGaps,Klein:1968}).  
Fig.~\ref{fig:neutrinorates} shows the resulting solar-neutrino rate binned into the observed number of electrons, $n_e$, 
assuming an exposure of 1~kg-year for each ionization efficiency.  

\begin{figure}[t!]
\includegraphics[width=0.48\textwidth]{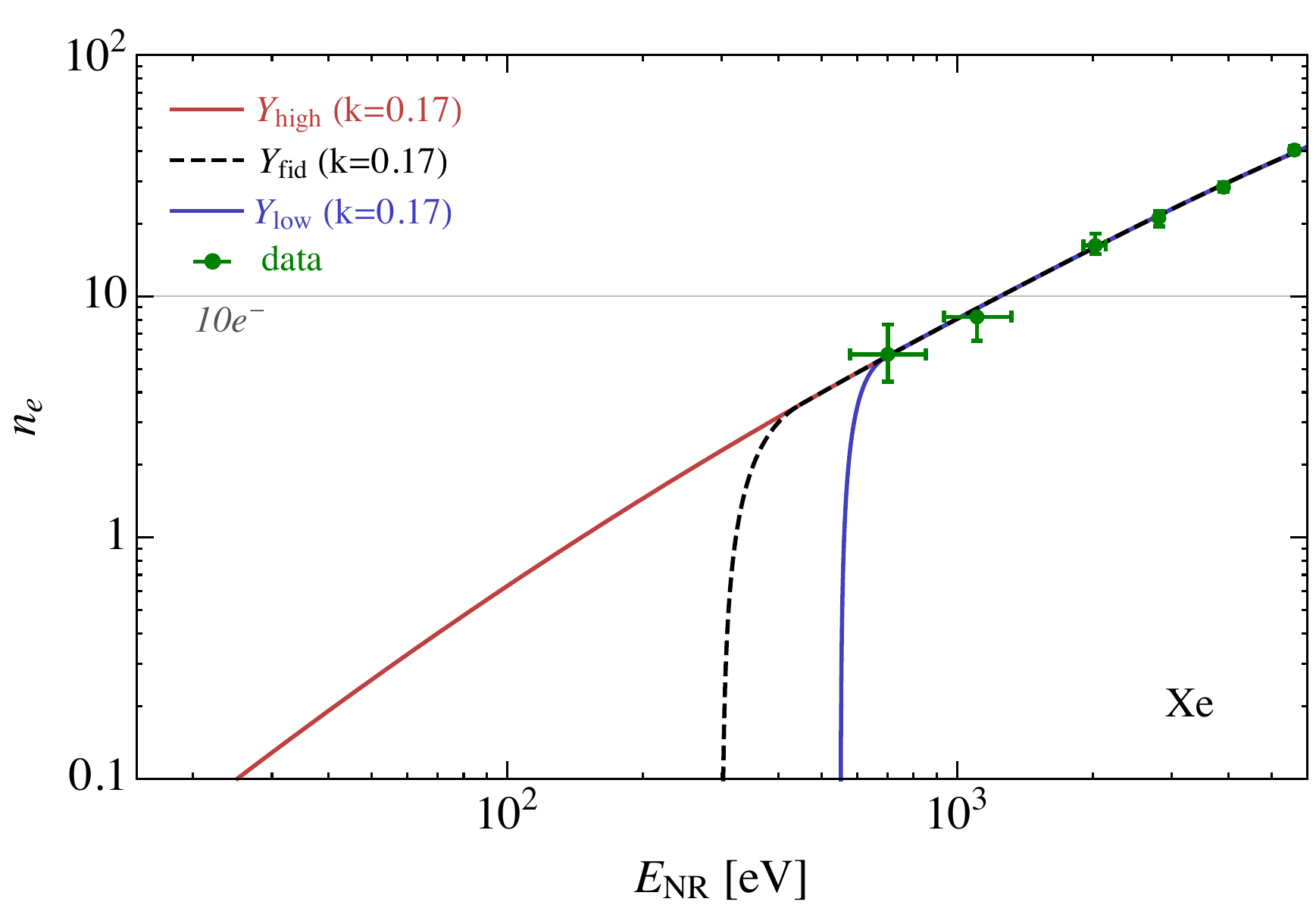} 
\caption{Models of the ionization efficiency to convert the nuclear recoil energy $E_{\rm{NR}}$ to the number of electrons $n_e$ for xenon, as defined in Table~\ref{tab:conversionxe}. 
The red solid (black dashed, blue solid) lines correspond to our high (fiducial, low) ionization efficiency models. 
The horizontal gray line denotes 10 electrons. 
The green points are LUX D-D neutron data~\cite{Akerib:2015rjg}.
}
\label{fig:convxe}
\end{figure}

\subsubsection{Ionization efficiencies for xenon}\label{subsubsec:Y-xenon}
For the case of xenon, we use the model in~\cite{Wang:2016obw}, which gives an average number of electrons
produced as a function of $E_{\rm NR}$. This model fits the charge-yield data obtained by LUX~\cite{Akerib:2015rjg} at an electric field of 
181~V/cm, see Fig.~\ref{fig:convxe}. 
The charge yield, $Q_{y}$, which is defined as the number of electrons ionized per eV of nuclear recoil energy in keV, is given by 
\begin{multline}\label{xeion}
Q_y = \frac{Y}{W_{i}} \times \\ \exp{\left[-\frac{\ln 2}{t_{c}}\left(t_{pa}+   \alpha (\ln E_{\rm{NR}})+\beta (\ln E_{\rm{NR}})^2\right)\right]}\,,
\end{multline} 
where $Y$ is the Lindhard quenching factor described in Eq.~(\ref{yield1}), the average recombination time is $t_{c}=15$~ns, the parent recombination time is $t_{pa}=1.5$~ns, $\alpha =3.617$~ns, and $\beta = 1.313$~ns~\cite{Wang:2016obw}. The average energy expended per electron-hole pair, $W_i$, is given by~\cite{Wang:2016obw}, 
\begin{eqnarray}\label{w}
W_{i} &=& 14.94 +8.35 \times \frac{N_{ex}}{N_{i}}\,,
\end{eqnarray}
where $N_{ex}/N_{i}$ is the ratio of excited to ionized atoms, 
\begin{eqnarray}\label{exctoionratio}
\frac{N_{ex}}{N_{i}}&=& \frac{1- \exp{(-I/E_e)}}{3+\exp{(-I/E_e)}}\,.
\end{eqnarray}
The mean ionization potential for xenon is $I = 555.57$~eV and $E_e=Y \times E_{\rm{NR}}$ is the electron-equivalent recoil energy given by the Lindhard quenching of the nuclear recoil energy.

The observed charge yield is 6 electrons for $E_{\rm NR} \approx 700$~eV, which is the lowest available data~\cite{Akerib:2015rjg}. 
We again define three extrapolations to lower energies. For our ``fiducial'' model, we assume a cut-off of 300~eV and consider a smooth exponential extrapolation of the model in the energy region $300-450$~eV. 
To model a ``high'' ionization efficiency, we assume a cut-off of 12~eV (this is close to the ionization energy for xenon 
of 12.1~eV~\cite{Aprile:2009dv}), and for a ``low'' 
ionization efficiency, we consider a cut-off of 550~eV. 
For each model, $Q_y \times E_{\rm{NR}}$ gives the average number of electrons $n_{e}$ as a function of $E_{\rm{NR}}$. The functional forms for the three conversion schemes are shown in Fig.~\ref{fig:convxe} and their analytic form is given 
in Table~\ref{tab:conversionxe}.
\setlength{\tabcolsep}{0.5em} 
{\renewcommand{\arraystretch}{1.3}
\begin{table}[t!]
\caption{Analytic expressions for the number of ionized electrons $n_{e}$ as a function of nuclear recoil energy $E_{\rm{NR}}$ 
for the high, fiducial, and low ionization efficiency models, shown in Fig.~\ref{fig:convxe}, for xenon. $Q_{y}$ here refers to the charge yield model described in Eq.~\ref{xeion}.}
\begin{center}
\begin{tabular}{l c c c}
\hline
& $E_{\rm{NR}}$ [eV] & $n_{e}$\\ \hline
\multirow{8}{*}{\rotatebox[origin=c]{90}{\large{xenon}}}
\multirow{3}{*}{~~~~~~high}& 0-12  & 0 \\
 & 12-20  & $0.11\left[1-e^{-(E_{\rm{NR}} - 12)/7.9}\right]$\\
& $>20$  & $Q_{y} \times E_{\rm{NR}}$\\ \cline{2-3}
&0-300& 0 \\
~~~~~~~~~{fiducial}& 300-450 &$4.27\left[1-e^{-(E_{\rm{NR}} - 300)/82.88}\right]$ \\ 
& $>450$  & $Q_{y} \times E_{\rm{NR}}$\\ \cline{2-3}
&0-550& 0 \\
~~~~~~~~~~{low}& 550-700 &$6.13\left[1-e^{-(E_{\rm{NR}} - 550)/59.02}\right]$ \\ 
& $>700$  & $Q_{y} \times E_{\rm{NR}}$\\
\hline
\end{tabular}
\end{center}
\label{tab:conversionxe}
\end{table}%

\begin{figure*}[t!]
\includegraphics[width=0.32\textwidth]{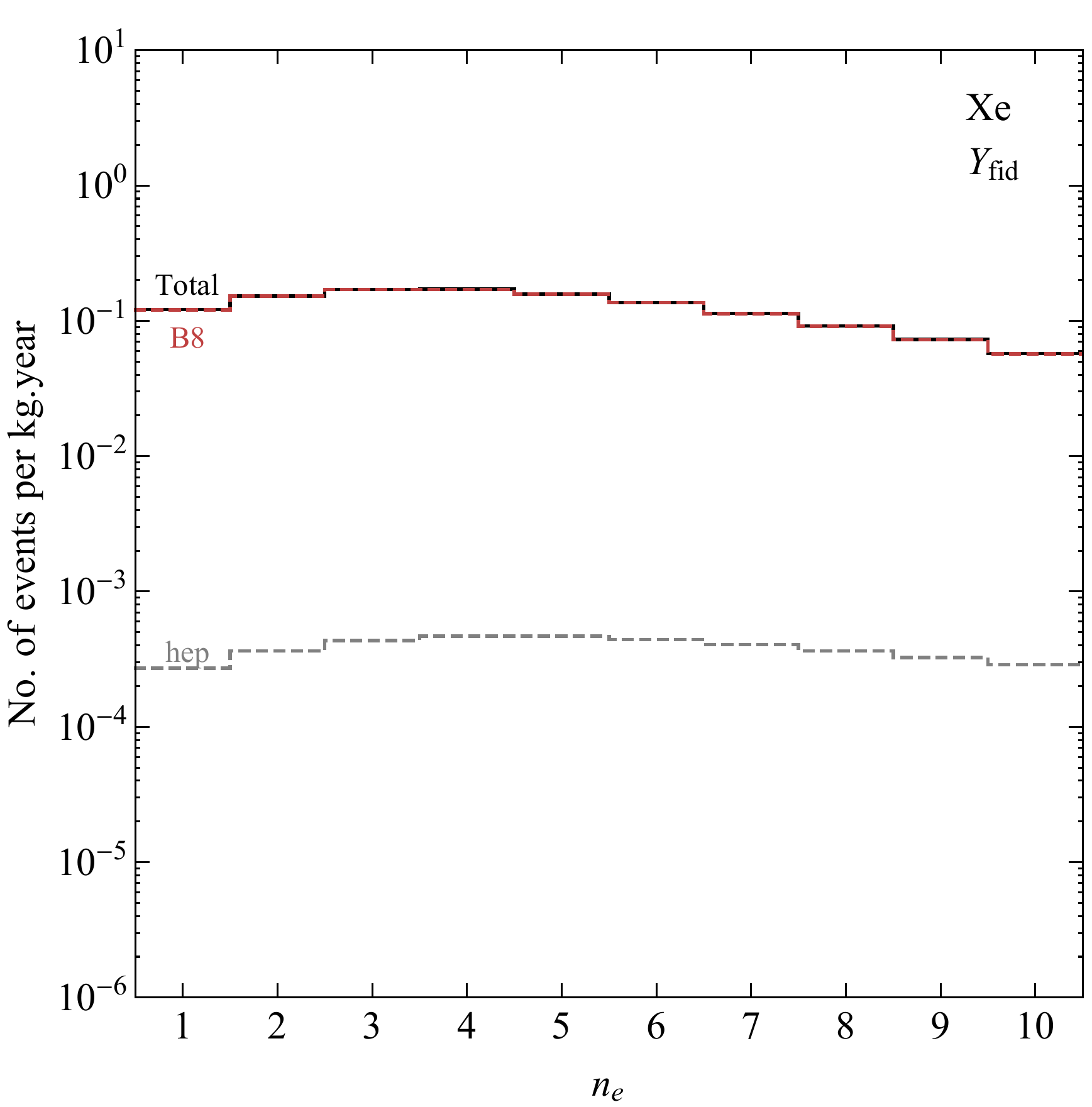}~~
\includegraphics[width=0.32\textwidth]{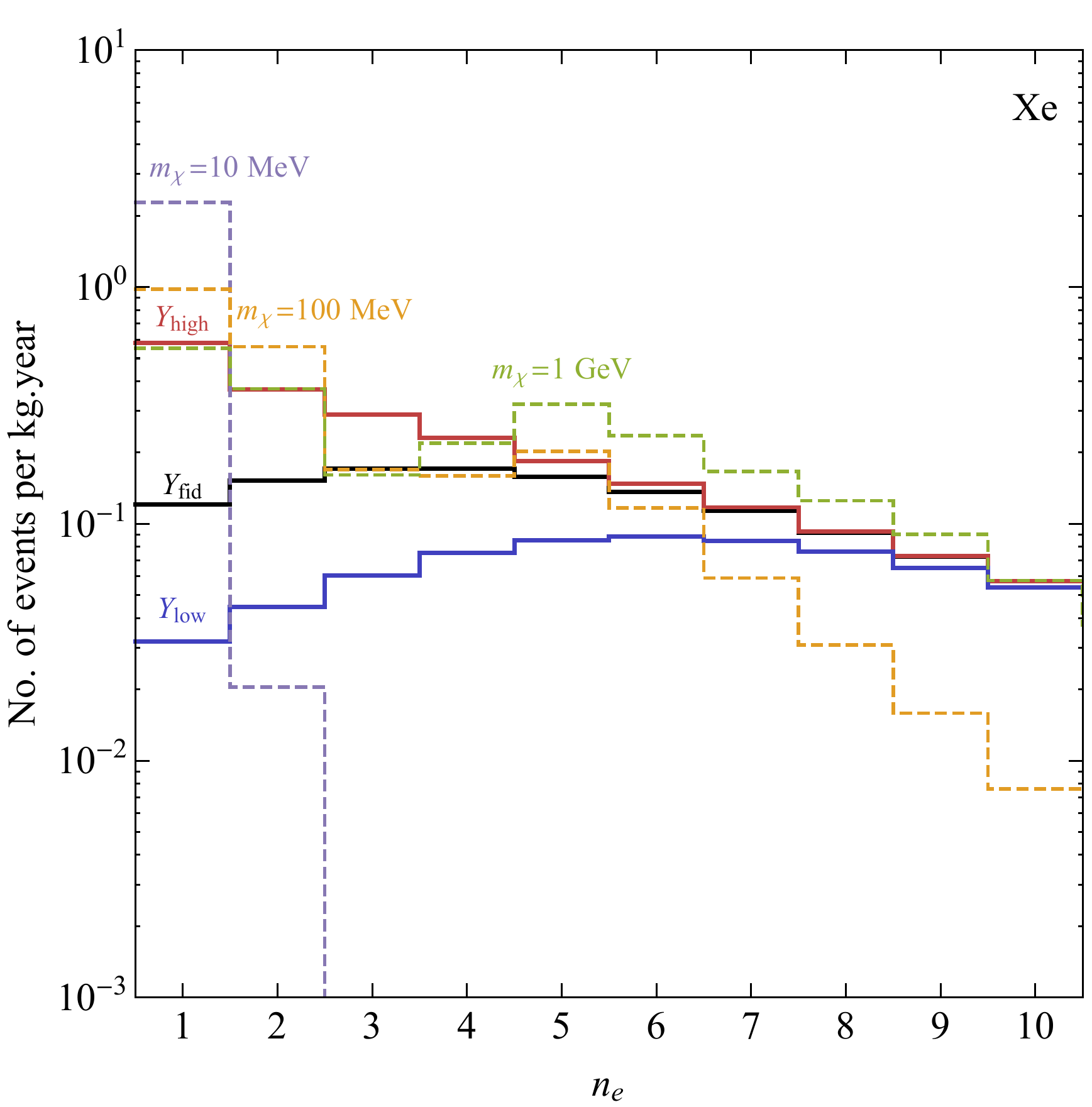}~~
\includegraphics[width=0.32\textwidth]{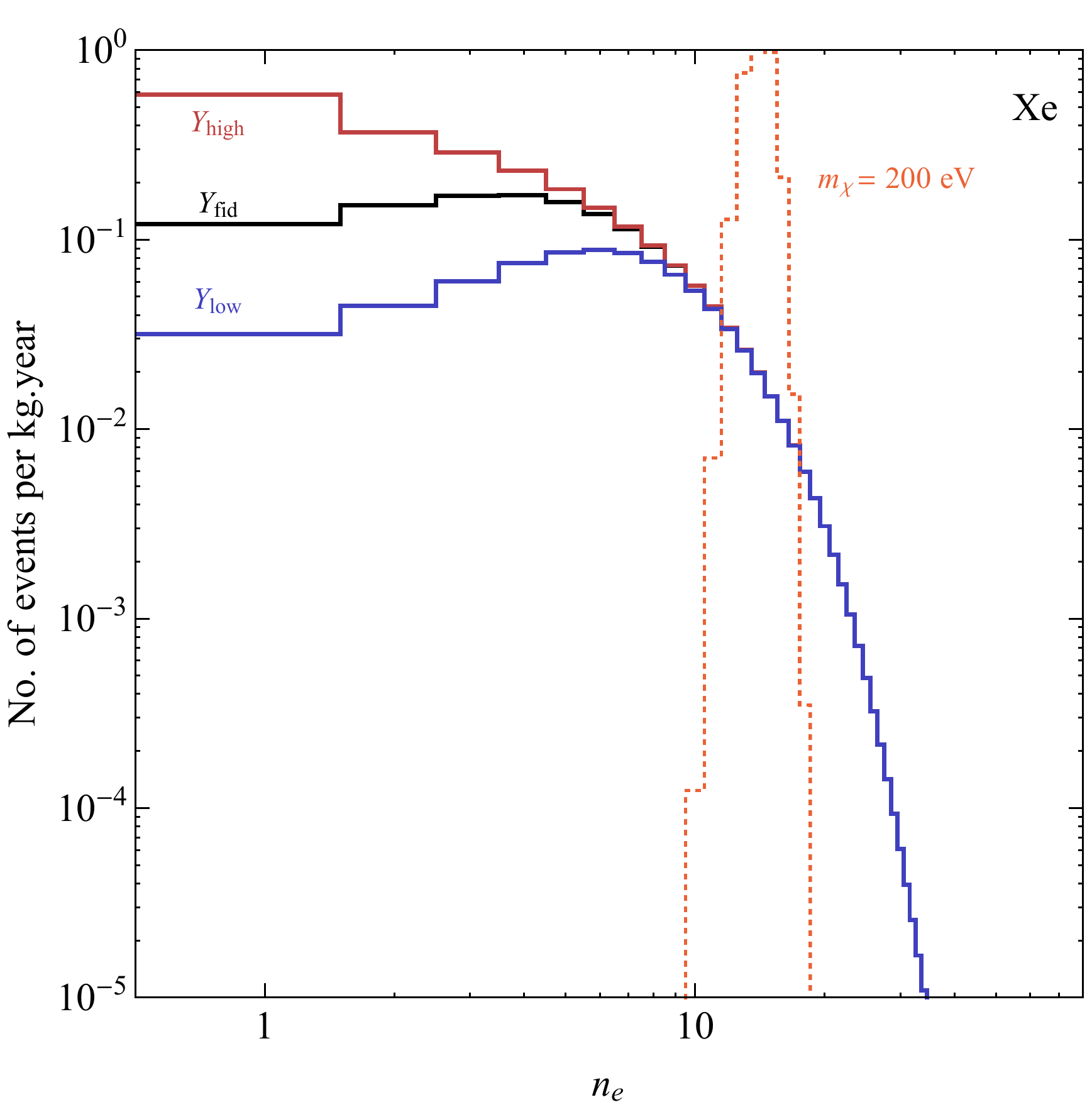}
\caption{
As in Fig.~\ref{fig:events_si_ge}, but for a xenon target: 
Ionization spectrum (number of events versus number of electrons) produced by solar neutrinos  
scattering coherently off xenon {\it nuclei} normalized to 1~kg-year.  
The {\bf left} plots assume the fiducial ionization-efficiency model, with the black line showing the total number of events and the 
colored lines showing the various components. 
The {\bf middle} and {\bf right} plots show the total neutrino flux assuming the low ({\bf{blue}}), fiducial ({\bf{black}}), and high ({\bf{red}}) ionization-efficiency  models. For illustration, we include the DM-scattering spectra for $m_\chi=0.5$~MeV (purple, dashed), 10~MeV (orange, dashed), and 1~GeV (green, dashed) in the {\bf middle} panels, as well as the DM-absorption spectrum for $m_\chi=200$~eV (red, dotted) in the {\bf right} panels.  
To normalize the DM spectra, we set the DM-electron scattering cross section ($\overline\sigma_e$) to the 90\% confidence-level limit assuming no background and an exposure of 1 kg-year.  
}
\label{fig:events_xe}
\end{figure*}

Given a conversion model defined by $n_{e}(E_{\rm{NR}})$, the CNS events between $E_{\rm{NR}}^{\rm min}$ and 
$E_{\rm{NR}}^{\rm max}$ can be translated into the number of events 
observed $n_{i}$ in the $i$-th electron-bin using Poisson statistics as  
\begin{eqnarray}
n_{i} &=& \int_{E_{\rm{NR}}^{\rm min}}^{E_{\rm{NR}}^{max}} \frac{dR}{dE_{\rm{NR}}}  \\ &\times&\exp{(-n_{e}(E_{\rm{NR}}))} \frac{n_{e}^{i}(E_{\rm{NR}})}{i!} dE_{\rm{NR}}\,.
\end{eqnarray}
Fig.~\ref{fig:events_xe} shows the resulting CNS background, together with several DM signal shapes.
\section{Dark Matter Signal}\label{sec:DMsignal}

\subsection{Dark Matter-Electron Scattering}\label{sec:scattering}
For DM-electron scattering, the minimum DM mass that can be probed is found by requiring the DM kinetic energy to be larger than the 
band gap or binding energy. 
For semiconductors, we use the DM spectra and rates from~\cite{Essig:2015cda} (publicly available 
at \cite{ddldm-website}), while for xenon 
we use~\cite{Essig:2017kqs}.  
Moreover, we follow~\cite{Essig:2011nj,Essig:2015cda} in parameterizing the DM-electron scattering cross section, which we review now briefly. 
First, we define the matrix element for the elastic scattering of a DM particle off a free electron as 
\begin{equation}
\overline{|\mathcal M_{\rm free}(\vec q\,)|^2} \equiv \overline{|\mathcal M_{\rm free}(\alpha m_e)|^2} \times |F_{\rm DM}(q)|^2\,,
\label{eq:DM-form-factor}
\end{equation}
where $m_\chi$ is the DM mass, 
$\overline{|\mathcal M|^2}$ is the absolute square of $\mathcal M$, averaged over initial and summed over final particle spins, 
and the DM form factor, $F_{\rm DM}(q)$, gives the momentum-transfer dependence of the interaction. 
Second, we define a reference cross-section at a fixed momentum-transfer of $q=\alpha m_e$ as 
\beq
\overline \sigma_e \equiv \frac{\mu_{\chi e}^2 \overline{|\mathcal M_{\rm free}(\alpha m_e)|^2}}{16 \pi m_\chi^2 m_e^2} \,, 
\label{eq:sigma-bar-e}
\eeq
which parameterizes the strength of the interaction.  For the case of $F_{\rm DM}(q)=1$, $\overline\sigma_e$ is equal to the cross section for free elastic scattering. 
We present our results in the $\overline\sigma_e$ versus $m_\chi$ parameter space, and will consider two DM form factors, $F_{\rm DM}=1$ and 
$F_{\rm DM}=(\alpha m_e/q)^2$.  
Figs.~\ref{fig:neutrinorates} and \ref{fig:events_xe} show a few examples of the DM signal shapes.  

\subsection{Dark Matter Absorption by Electrons }\label{sec:absorption}

For the absorption of bosonic DM by electrons, we consider two DM candidates, ALPs and $A'$s.  
Since the entire rest mass energy of the DM is absorbed and the DM has a negligible kinetic energy, the minimum DM mass that can be probed 
is given by the band gap or binding energy.  We consider DM masses 
up to 1.5~keV for the semiconductors and 0.5~keV for xenon.  
We take the DM absorption rates, DM spectra, and notation from~\cite{Bloch:2016sjj} (see also~\cite{An:2014twa,Hochberg:2016sqx}).  
The right plots in Figs.~\ref{fig:neutrinorates} and \ref{fig:events_xe} show examples of the DM signal shape.  
The signal is a gaussian centered at the mass of the DM particle with width~\cite{Bloch:2016sjj},
\begin{eqnarray}\label{sigma}
\sigma &=& \epsilon_{e}\sqrt{F \langle Q(E_{e}) \rangle},
\end{eqnarray}  
where $\epsilon_{e}$ is the average energy to ionize an electron in the semiconductor (i.e., 2.9~eV for germanium, 3.6~eV for silicon~\cite{ExptGaps,Klein:1968}, and 13.8~eV for xenon~\cite{Angle:2011th}), $F$ is the Fano factor (about 0.13 for both silicon and germanium~\cite{Lepy2000,Lowe1997} and 0.059 for xenon~\cite{Aprile:2009dv}), and $\langle Q(E_{e}) \rangle$ is the mean expected number of ionized electrons.  For our sensitivity estimates, we add the electron bins that are encompassed by the central $2 \sigma$ of the gaussian signal 
(i.e., $\pm1\sigma$, or 68\% of the total).  

\noindent  
{\bf Electron ionization from ALPs.}
ALPs are pseudoscalars whose interactions with electrons are given by the following effective Lagrangian,
\begin{eqnarray}\label{alps}
\mathcal{L}_{a}&=&\frac{1}{2}\partial_{\mu} a\partial^{\mu}a - \frac{1}{2}m_{a}^2 a^2 +i g_{aee}a\overline{e}\gamma_{5}e\,,
\end{eqnarray}
where $m_a$ is the ALP mass and $g_{aee}$ parameterizes the ALP-electron interaction strength.  
Since $g_{aee}$ thus also determines the absorption rates~\cite{Bloch:2016sjj}, 
we present our results in the $g_{aee}$ versus $m_a$ parameter space. 

\noindent  
{\bf Electron ionization from $A'$.}
The $A'$ is a massive gauge vector boson corresponding to a broken dark gauge group $U(1)_{D}$ that kinetically mixes 
with the SM $U(1)$ hypercharge. The relevant part of the low-energy Lagrangian after electroweak symmetry breaking is 
 \begin{eqnarray}\label{aprime}
\mathcal{L}_{A'}&=&-\frac{1}{4}F'_{\mu \nu}F'^{\mu \nu} - \frac{\epsilon}{2}F'_{\mu \nu}F^{\mu \nu} + \frac{1}{2}m_{A'}^2 A'_{\mu}A'^{\mu},
\end{eqnarray}
where $F'_{\mu \nu}$ is the field strength of the dark photon, $F^{\mu \nu}$ is the field strength of SM photon, $m_{A'}$ is the mass of the dark photon, and $\epsilon$ is the kinetic mixing parameter.  
Here $\epsilon$ determines the absorption rates~\cite{Bloch:2016sjj}, and we thus present our results in the $\epsilon$ versus $m_{A'}$ parameter space.  

\section{Analysis}\label{sec:likelihood}

In this section, we describe our calculations to determine the DM scattering rates (parameterized in terms of $\overline\sigma_e$) 
or absorption rates (parameterized by the couplings $g_{aee}$ or $\epsilon$) for which the DM signal becomes statistically 
indistinguishable from the solar neutrino background.  
For a given DM mass, this depends on the detector threshold (measured in terms of number of electrons) and exposure.   
We also describe our calculations to determine the sensitivity to solar neutrinos, treating them as the signal of interest. 

For DM-electron scattering, we perform a hypothesis test using a binned log-likelihood analysis to compute the discovery potential following the method described in~\cite{PhysRevD.85.035006}. For a given detector exposure, threshold, and DM mass, the likelihood function is 
\beqa\label{eq:lh1}
\mathcal{L}(\sigma_{\chi e},\vec{\phi}) & = & \frac{e^{-(\mu_\chi + \sum_{j=1}^{n_{\nu}}\mu_{\nu}^{j})}}{N!}\times \prod_{j=1}^{n_{\nu}} \mathcal{L}(\phi_{j})  \times  \\
& & ~~~~~ \prod_{i=1}^{N} \Big[ \mu_\chi f_{\chi}(n_{i})+ \sum_{j=1}^{n_{\nu}}\mu_{\nu}^{j}f_{\nu}^{j}(n_{i})\Big]\,,  \nonumber
\eeqa
where $\sigma_{\chi e}$ and $\vec{\phi}$ are nuisance parameters corresponding to the DM-electron scattering cross section and neutrino fluxes, respectively, $\mu_\chi=\mu_\chi(\sigma_{\chi e})$ is the expected number of DM events, $\mu_\nu^j=\mu_\nu^j(\vec{\phi})$ 
is the expected number of neutrino events for the $j$-th 
solar-neutrino component, $n_i$ is the bin number (number of electrons) for the $i$-th event, $N$ is the total number of events, and $f_{\chi}$ and $f_{\nu}^j$ are the distribution functions for the DM and neutrino spectra (normalized to one total event), respectively.  We take the individual likelihood functions of solar neutrino source $j$, $\mathcal{L}(\phi_j)$, to be Gaussian distributions of the flux $\phi_j$ around its mean value with relative uncertainty of the flux normalizations as listed in Table~\ref{tab:nuflux}. 

To calculate the DM discovery potential in the presence of the neutrino backgrounds, 
we use the profile likelihood ratio 
\begin{equation}\label{lh2}
\lambda=\frac{\mathcal{L}(\sigma_{\chi e}=0, \hat{\hat{\vec{\phi}}})}{\mathcal{L}(\hat{\sigma}_{\chi e},\hat{\vec{\phi}})}\,.
\end{equation}
The numerator corresponds to the background-only hypothesis ($\sigma_{\chi e} = 0$) and is maximized for 
$\hat{\hat{\vec{\phi}}}$, while the denominator is maximized for $\hat{\sigma}_{\chi e}$ and $\hat{\vec{\phi}}$. 
We define the test statistic as, 
\[   
t = 
     \begin{cases}
       -2 \ln{\lambda} &\quad\hat{\sigma}_{\chi e}>0\\
       0&\quad\hat{\sigma}_{\chi e}<0\,. \\  
     \end{cases}
\]
Using Wilks theorem~\cite{wilks1938}, the distribution of $t$ follows a $\chi^2$ distribution with one degree of freedom and the significance of 
rejecting the background-only hypothesis is given by $\sqrt{t}$-sigma.

For each DM mass, detector threshold, exposure, and model for the ionization efficiency, we generate 200 samples of pseudodata by Poisson fluctuating the expected number of DM plus neutrino events assuming some fixed value for $\overline\sigma_e$.  
For each sample, we perform a log-likelihood analysis to calculate the significance at which the background-only hypothesis can be rejected. This creates a distribution of the significance values, from which we find the significance value that is exceeded by 90\% of the samples. We then 
vary $\overline\sigma_e$ until the latter significance value equals $2\sigma$. 
The final result is a ``{\bf $2\sigma$-discovery limit}'' for $\bar\sigma_e$, which depends on the DM mass, threshold, and exposure.  
We then vary each of these.  
We find that a 1-electron threshold sets the best limit for almost all DM masses; however, we also present the results for a 2-electron threshold in 
Appendix~\ref{app:2e-threshold}, which may be easier to achieve in future experiments. 

For DM absorption, the analysis is similar, except instead of $\overline\sigma_e$ we now have the couplings $g_{aee}$ for 
ALPs or $\epsilon$ for the $A'$.  We treat the signal shape as discussed in Sec.~\ref{sec:absorption}. 

We next consider the solar neutrinos as the signal of interest (assuming no DM signal).  
First, coherent scatters of $^8$B neutrinos off nuclei can be detected for all three elements (silicon, germanium, xenon) for all three conversion models.  This signal is free from contamination by other neutrino components for a sufficiently large threshold (one or a few electrons), 
depending on the ionization efficiency, and we calculate the number of events as a function of exposure for various ionization efficiencies.  

Second, we consider the possibility of detecting the CNO fluxes, which can only be detected if the ionization efficiency is sufficiently high. 
We perform a likelihood analysis, taking the CNO fluxes as the signal, and all other solar neutrino fluxes as background. 
More precisely, we take the sum of $^{13}\rm{N}$ and $^{15}\rm{O}$ as our ``CNO'' signal, since the contribution of $^{17}\rm{F}$ is negligible.  
We perform a likelihood analysis similar to the DM case above, 
and calculate the mean significance and fractional uncertainty of detecting the CNO flux, 
as a function of exposure and thresholds, in silicon, germanium, and xenon.  
We perform a similar calculation also for detecting the $^7{\rm Be}_{\rm (a)}$ and pep CNS signals.

\section{Results: Solar Neutrinos as a Background to Dark Matter Searches with Electron Recoils} \label{sec:results-DM}

\begin{figure*}[t!]
\vspace{-7mm}
\includegraphics[width=0.43\textwidth]{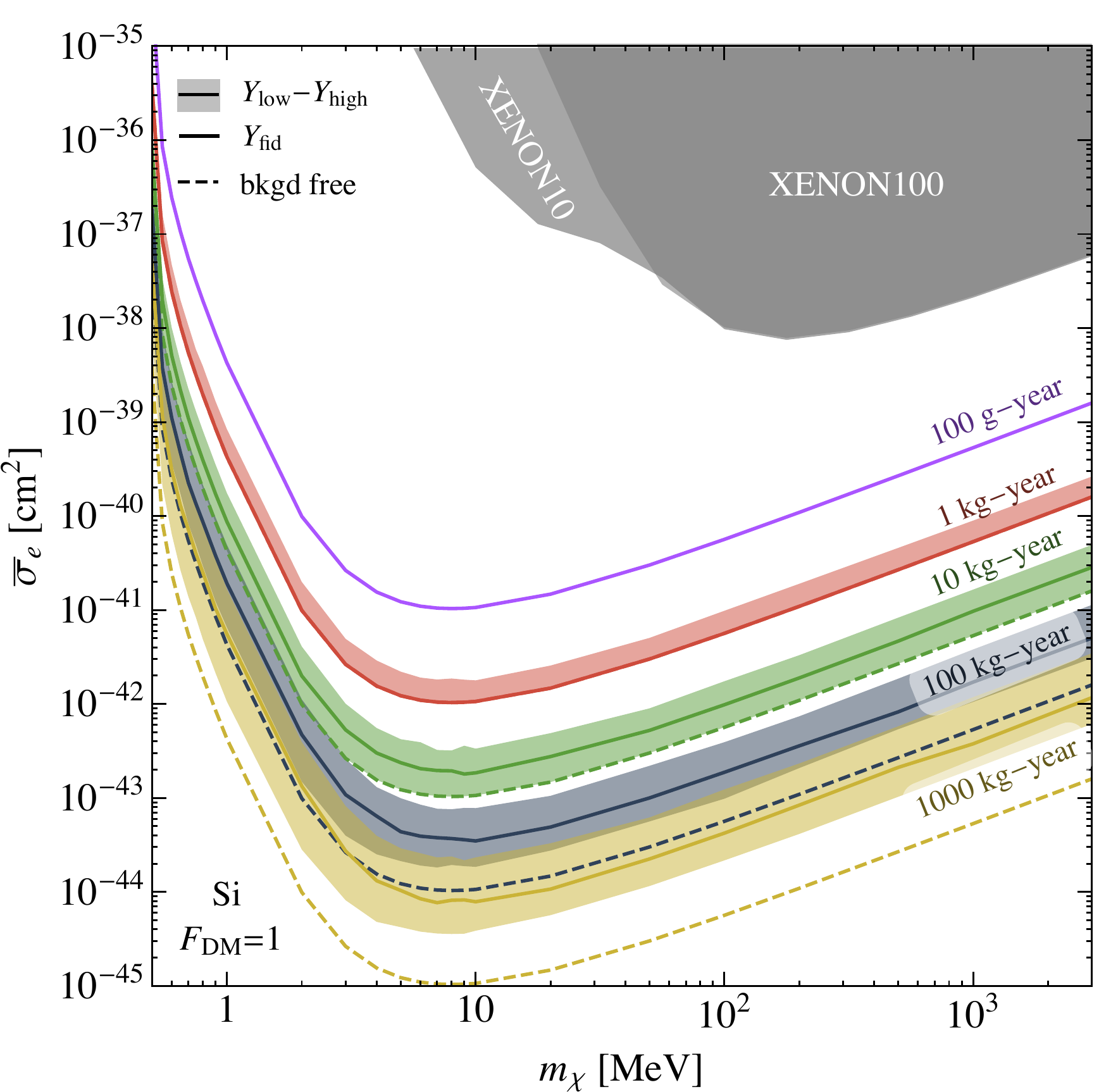} ~~~~~~~
\includegraphics[width=0.43\textwidth]{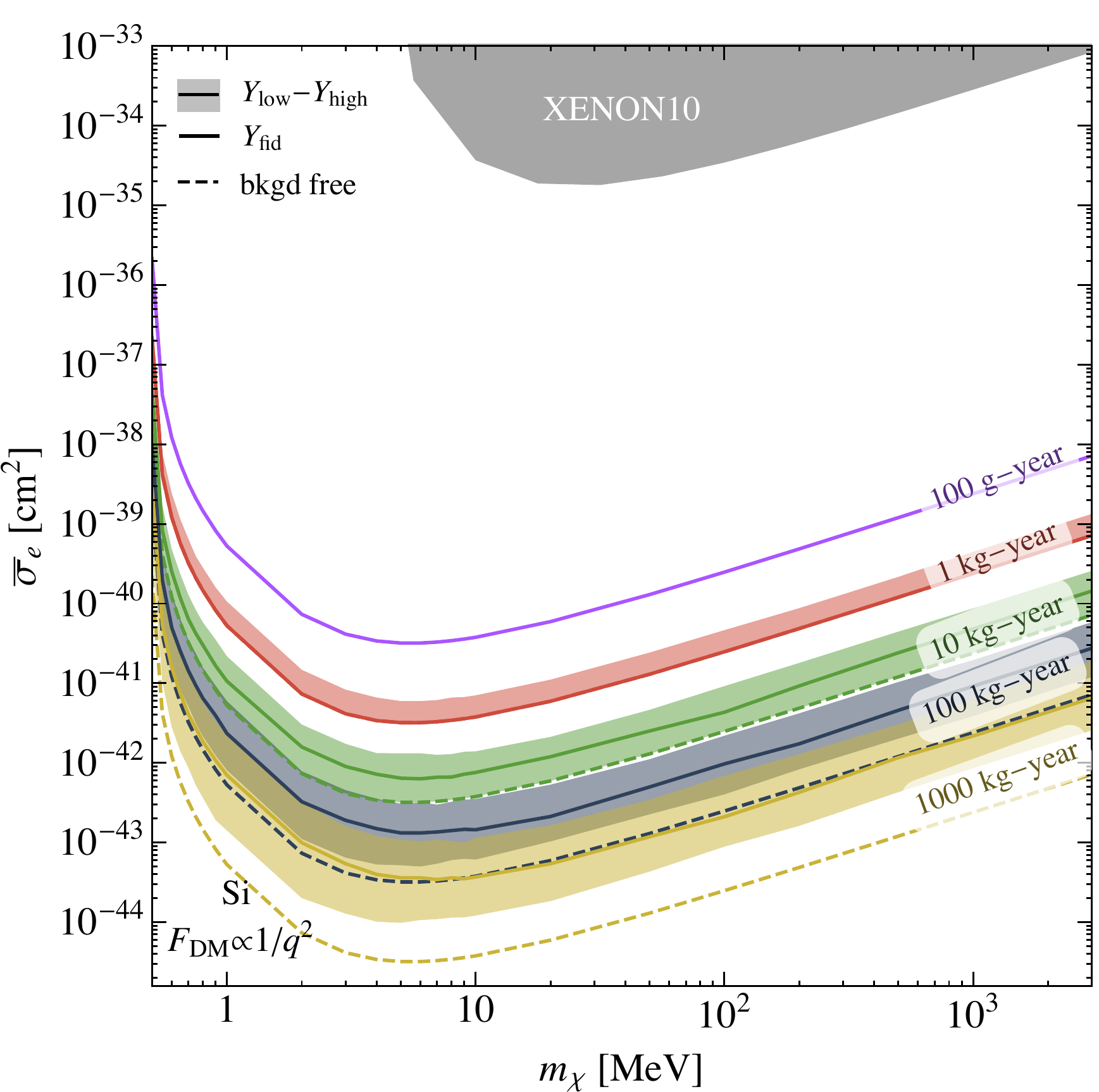}\\
\vspace{-2mm}
\includegraphics[width=0.43\textwidth]{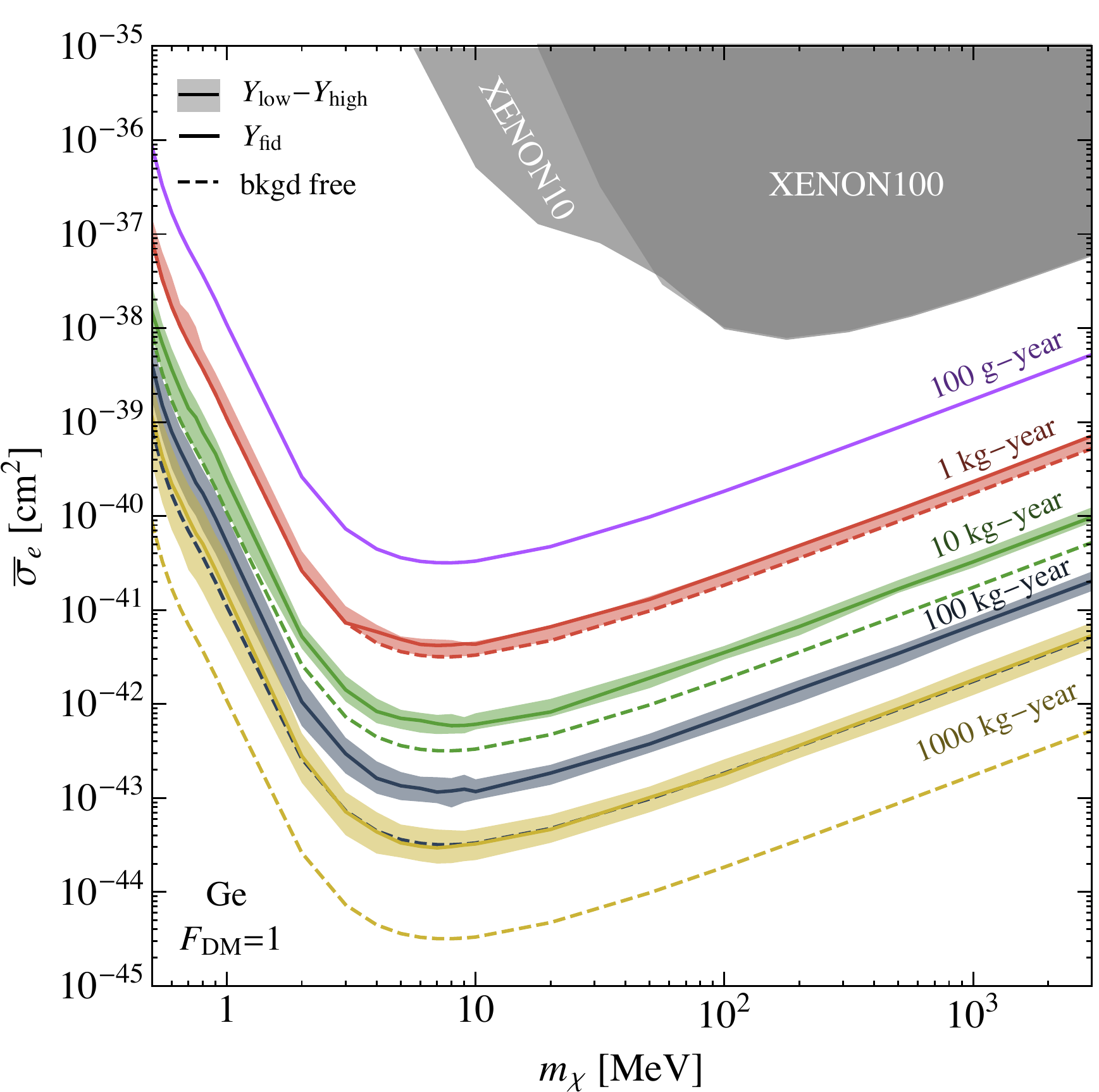} ~~~~~~~
\includegraphics[width=0.43\textwidth]{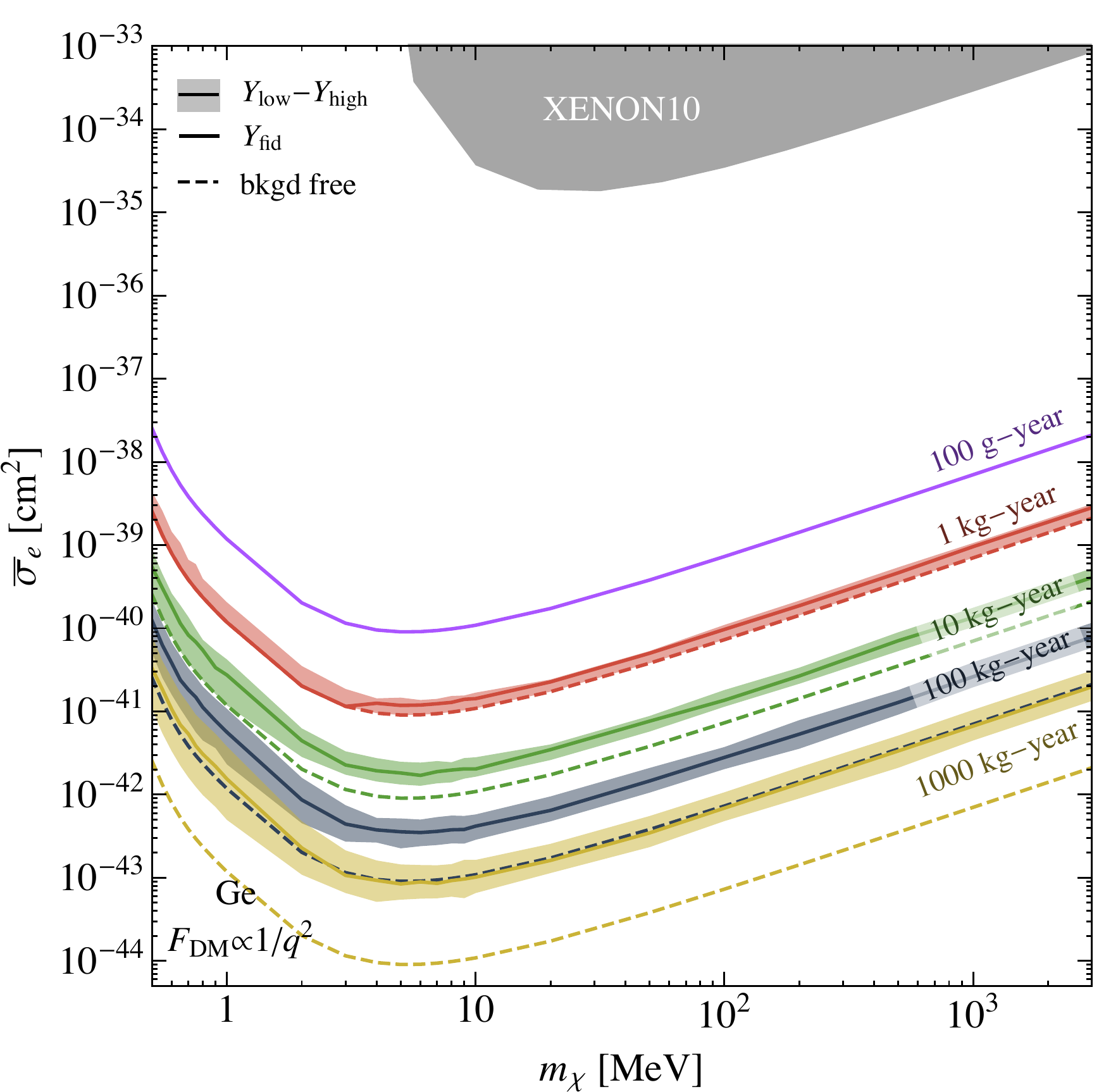} \\
\vspace{-2mm}
\includegraphics[width=0.43\textwidth]{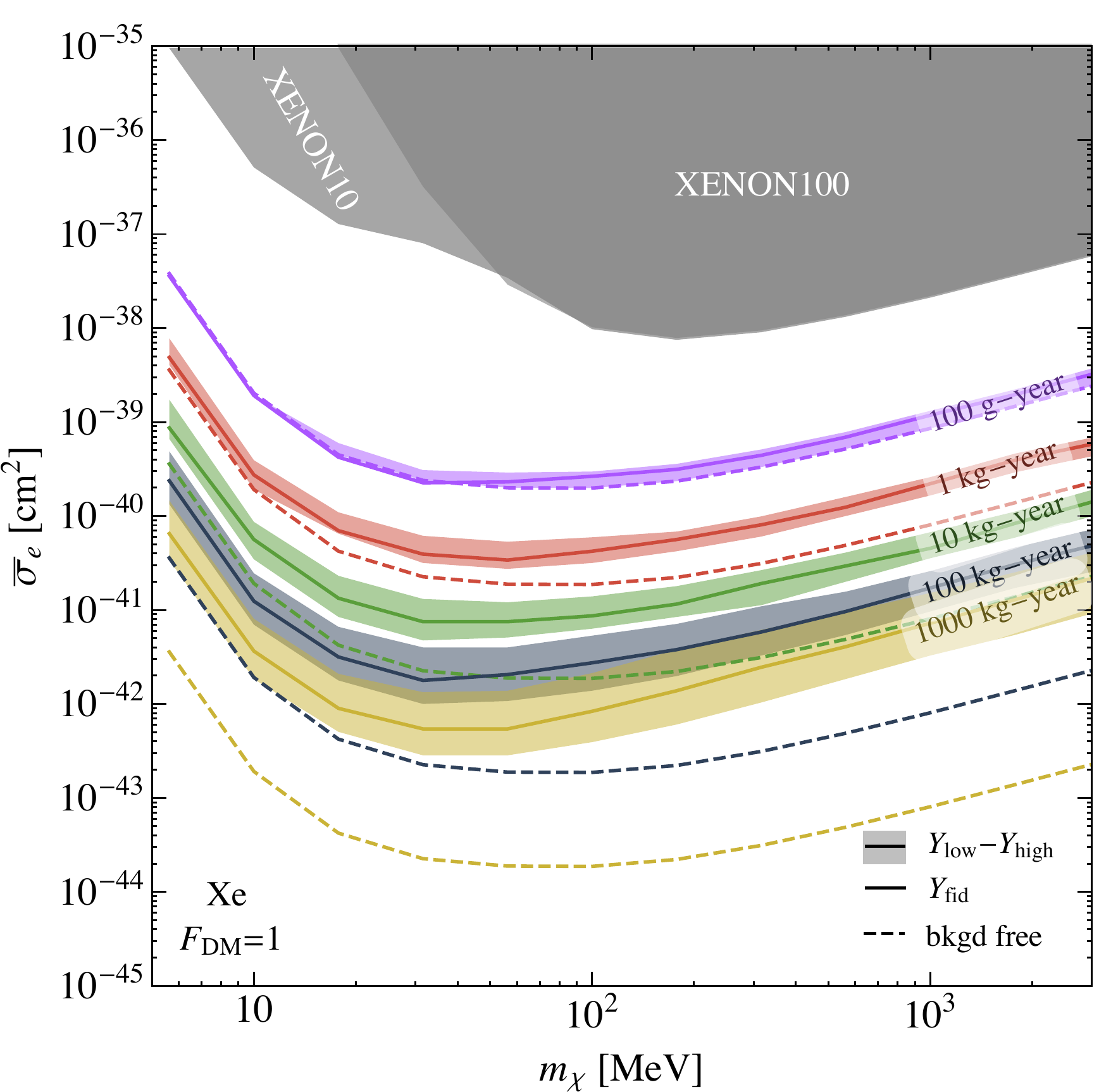} ~~~~~~~
\includegraphics[width=0.43\textwidth]{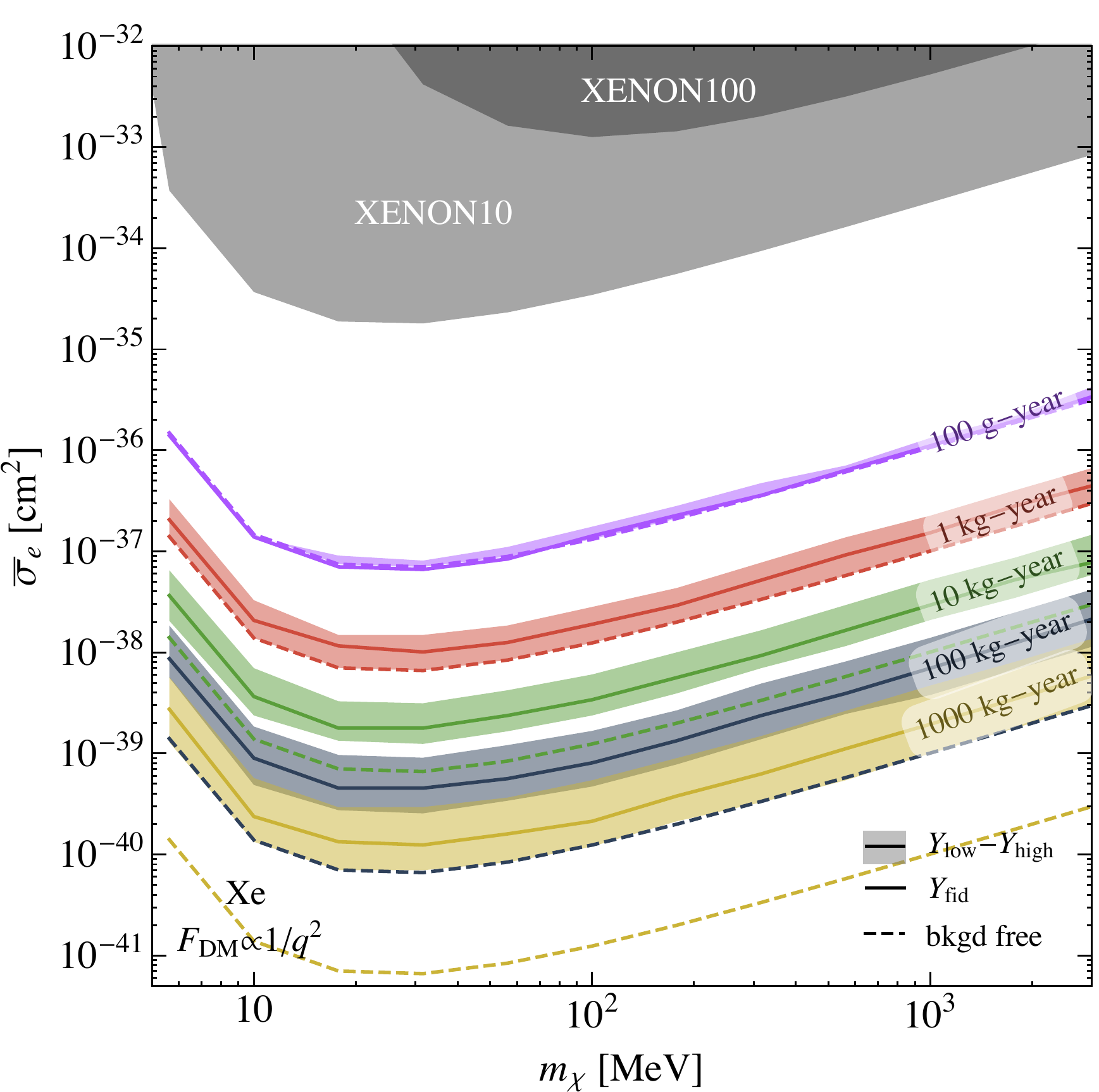}
\caption{Discovery limits for DM-electron scattering in silicon ({\bf top}), germanium ({\bf middle}), and xenon ({\bf bottom}). 
The panels on the {\bf left} ({\bf right}) 
assume the scattering is mediated by a heavy (light) particle, i.e.~$F_{\rm DM}=1$ ($F_{\rm DM} = \alpha^2m_e^2/q^2$). 
Exposures of 0.1, 1, 10, 100, and 1000~kg-years are shown in {\bf purple}, {\bf red}, {\bf green}, {\bf blue}, and {\bf yellow}, respectively. 
The solid line shows the results assuming the fiducial ionization efficiency, while the shaded bands denote the range between the high and low 
ionization efficiencies.  The dashed lines show the background-free 90\%~C.L.~sensitivities. Note that when the background assuming the fiducial ionization efficiency is negligible, the solid line and the dashed line are indistinguishable, making the dashed line disappear.  
The gray shaded region shows the current direct-detection limits on DM-electron scattering from~\cite{Essig:2017kqs}. 
}
\label{fig:limits}
\end{figure*}

\begin{figure*}[t!]
\includegraphics[width=0.48\textwidth]{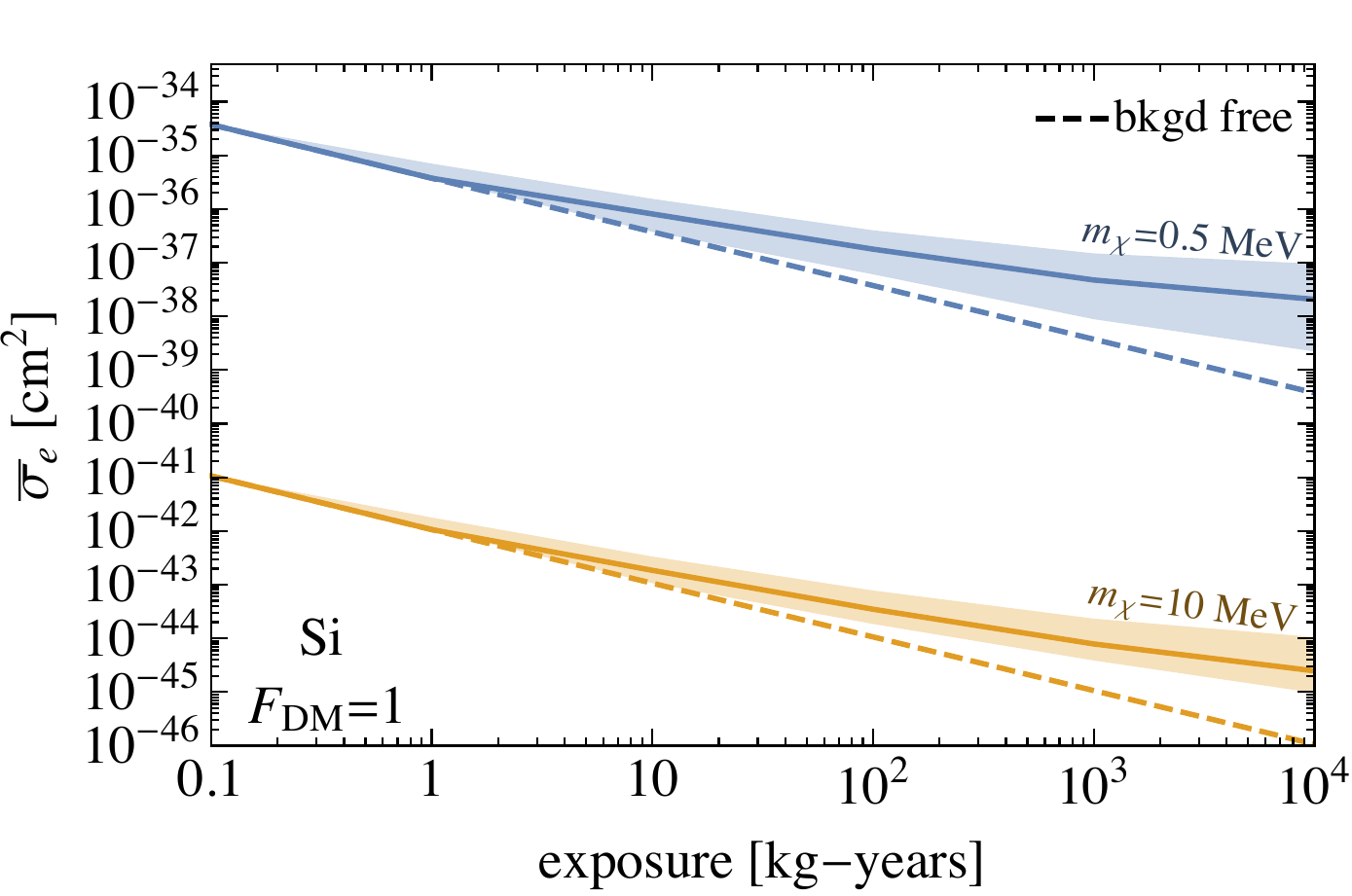}
\includegraphics[width=0.48\textwidth]{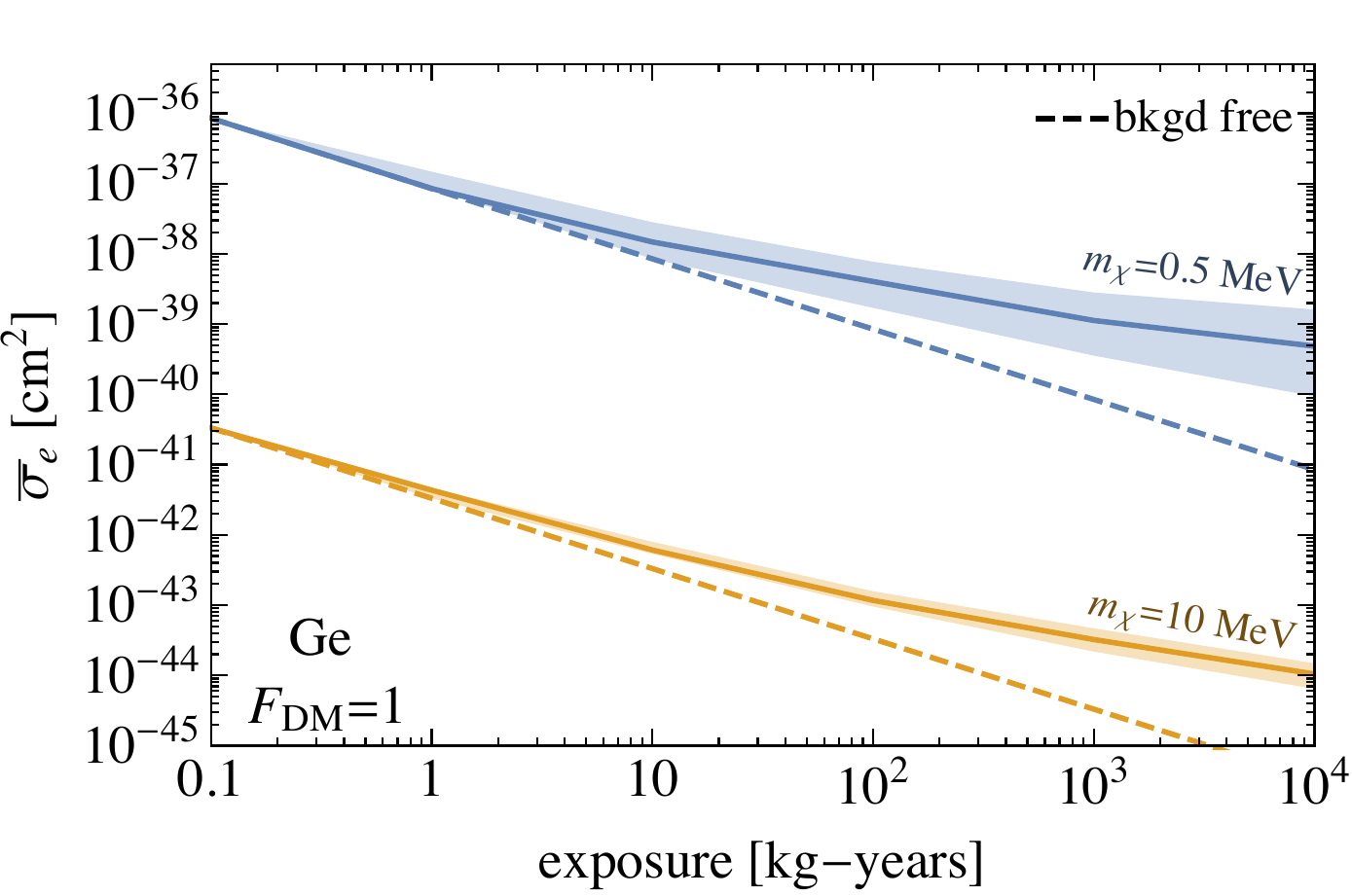}
\caption{Discovery limits for DM-electron scattering as a function of exposure, for a DM mass of 0.5~MeV ({\bf blue}) and 10~MeV ({\bf orange}), for $F_{\rm DM}=1$, in silicon ({\bf left}) and germanium ({\bf right}).  The solid line assumes the fiducial ionization efficiency, while the 
shaded bands denote the range between the high and low ionization efficiencies. The dashed lines show the background-free 90\%~C.L.~sensitivities, which simply scale as 1/exposure. 
}
\label{fig:saturation}
\end{figure*}

\begin{figure*}[t!]
\vspace{-5mm}
\includegraphics[width=0.43\textwidth]{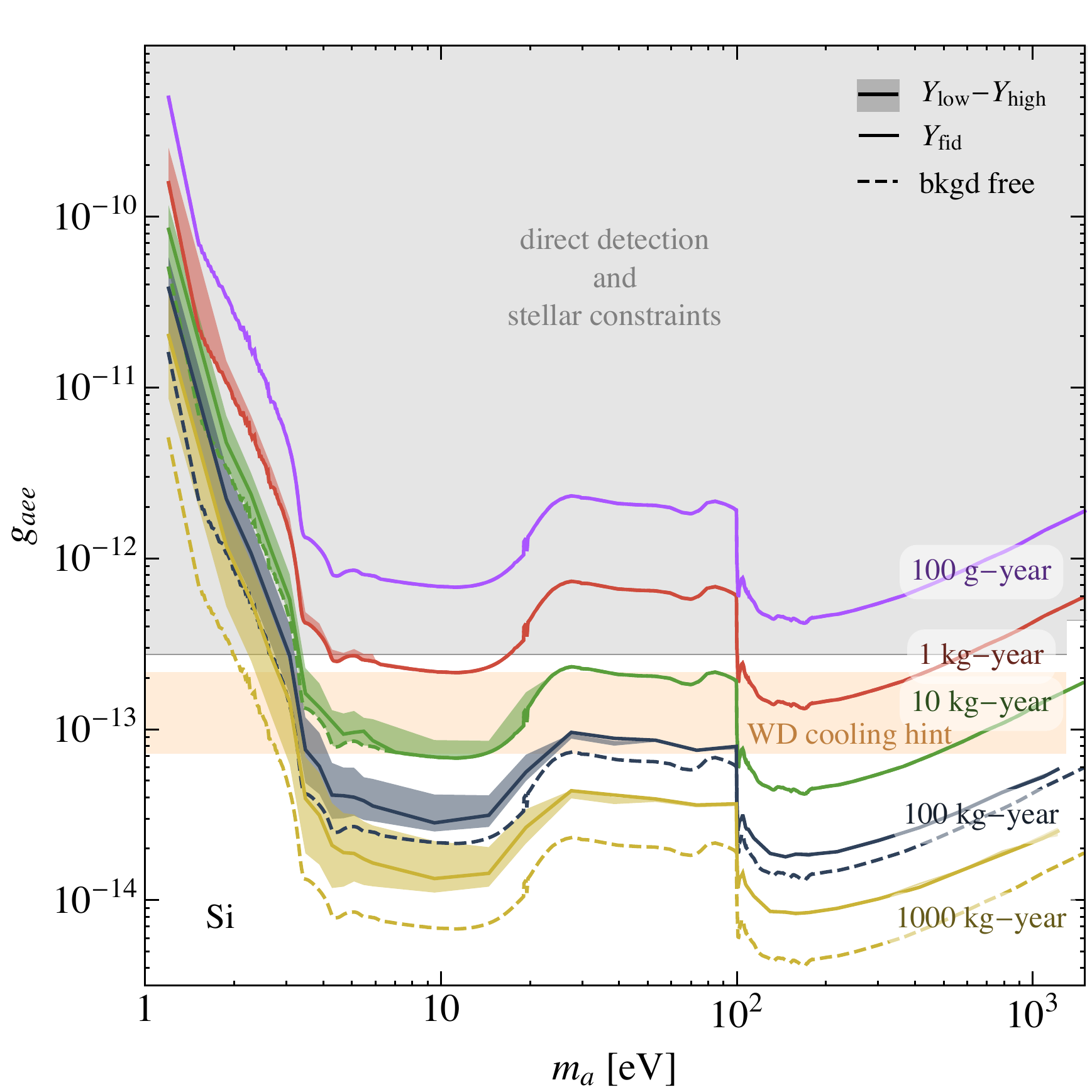} ~~~~~~~
\includegraphics[width=0.43\textwidth]{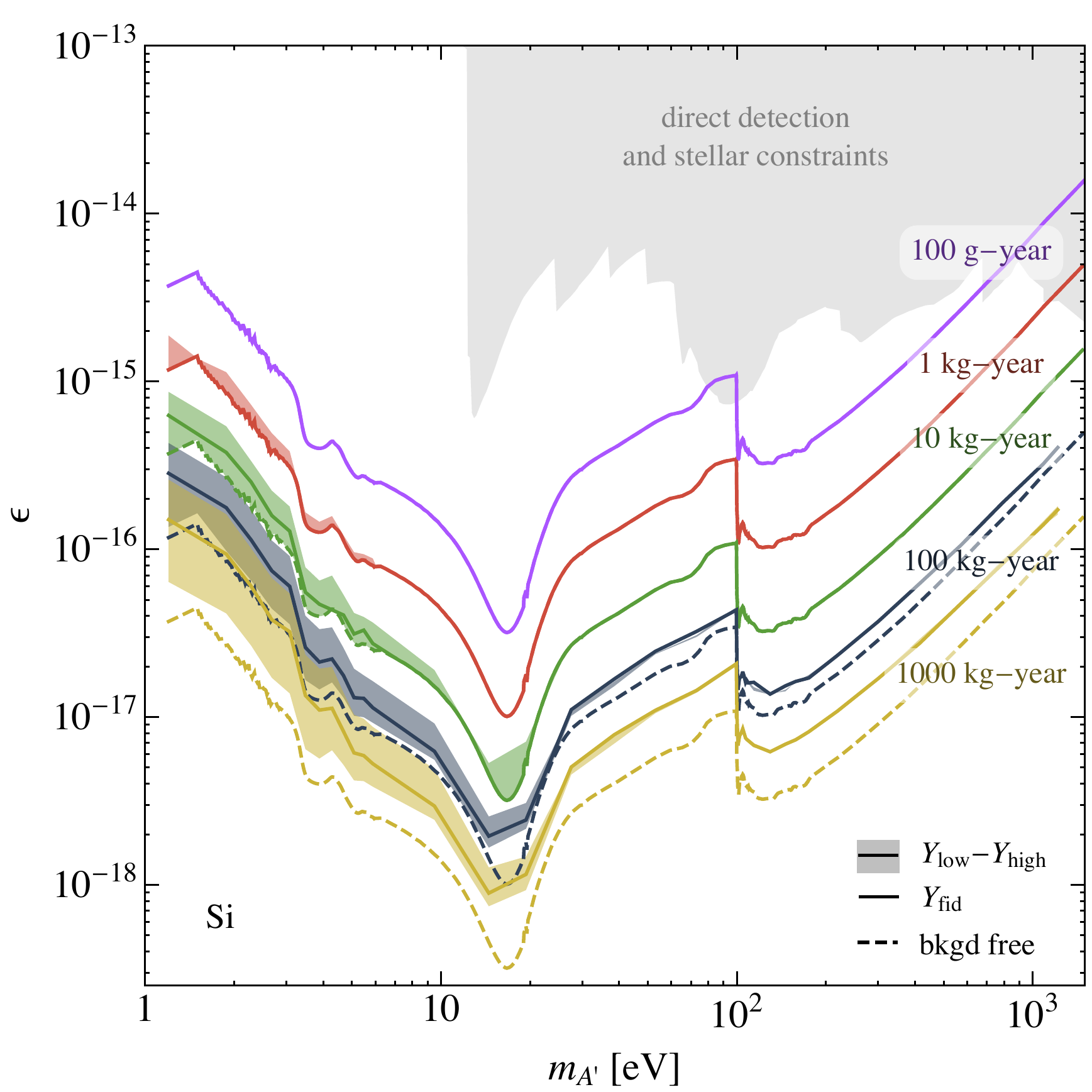}\\
\vspace{-2mm}
\includegraphics[width=0.43\textwidth]{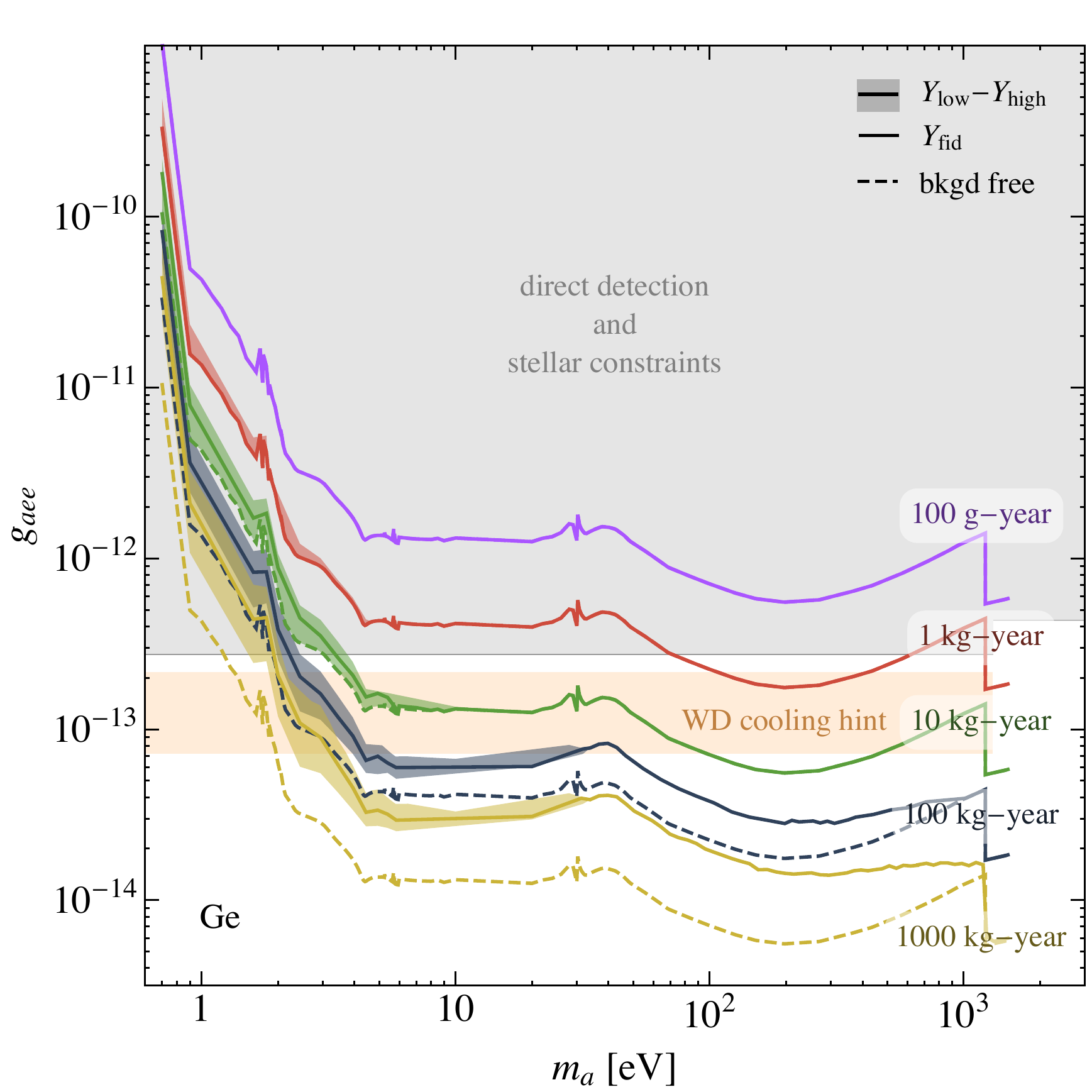} ~~~~~~~
\includegraphics[width=0.43\textwidth]{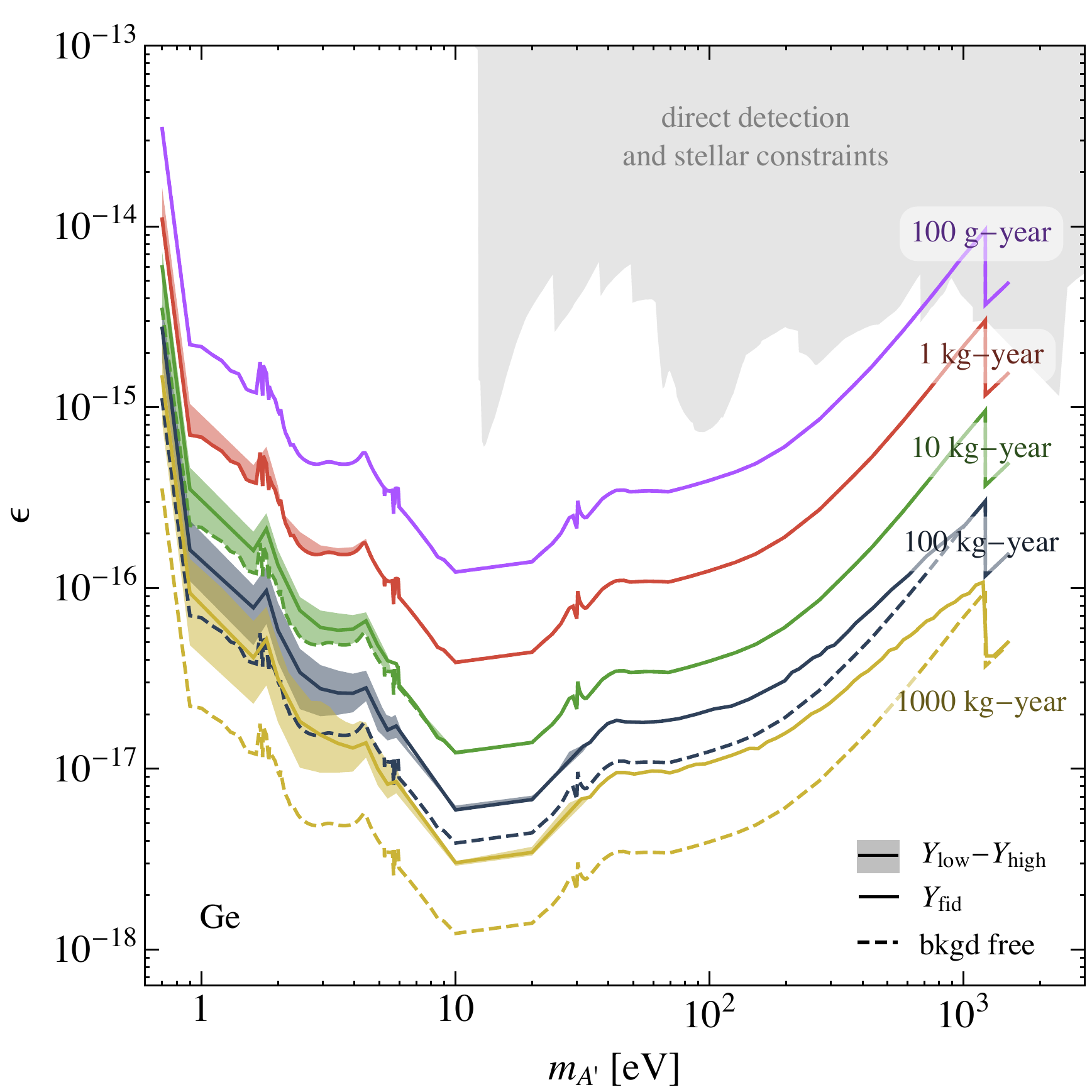}
\vspace{-2mm}
\includegraphics[width=0.43\textwidth]{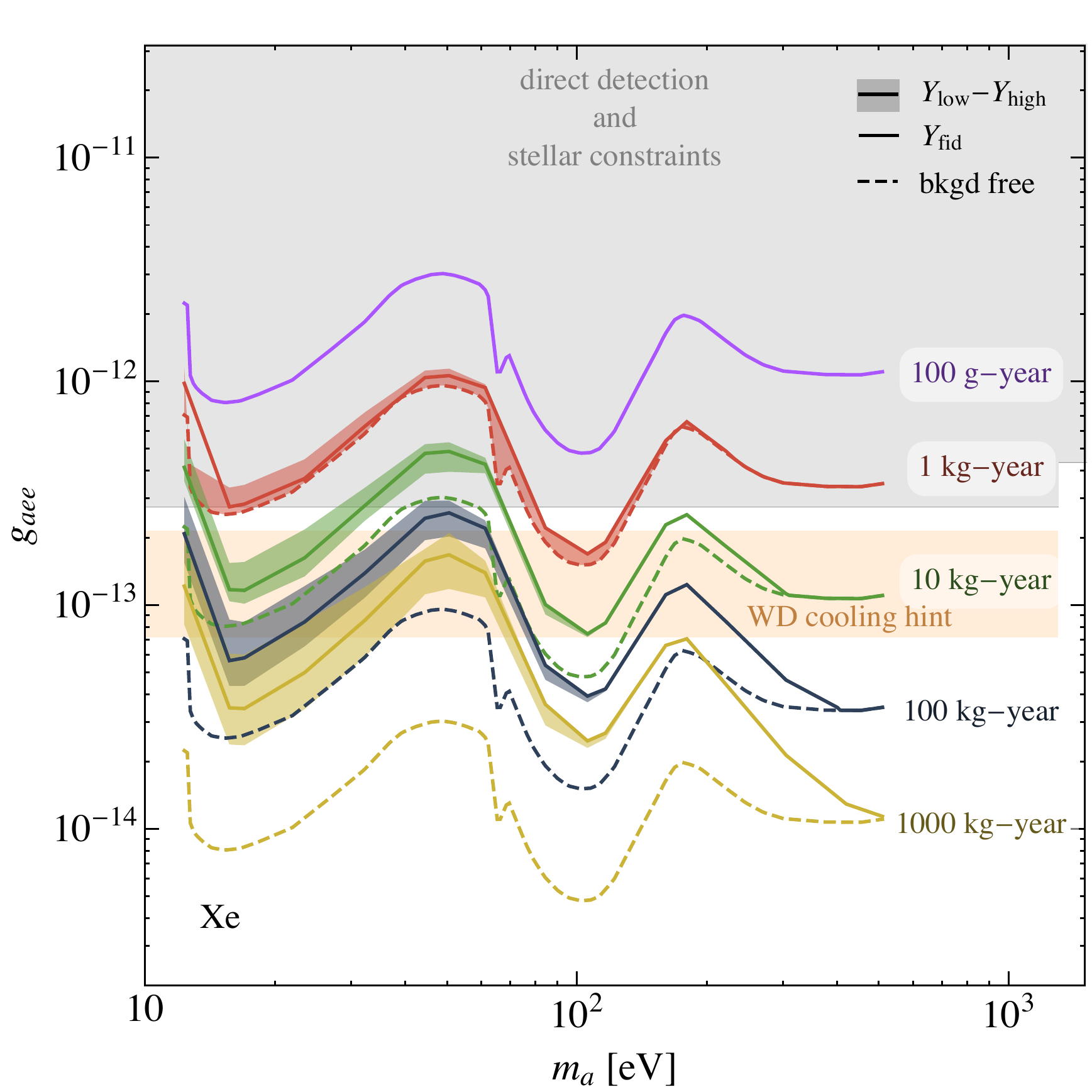} ~~~~~~~
\includegraphics[width=0.43\textwidth]{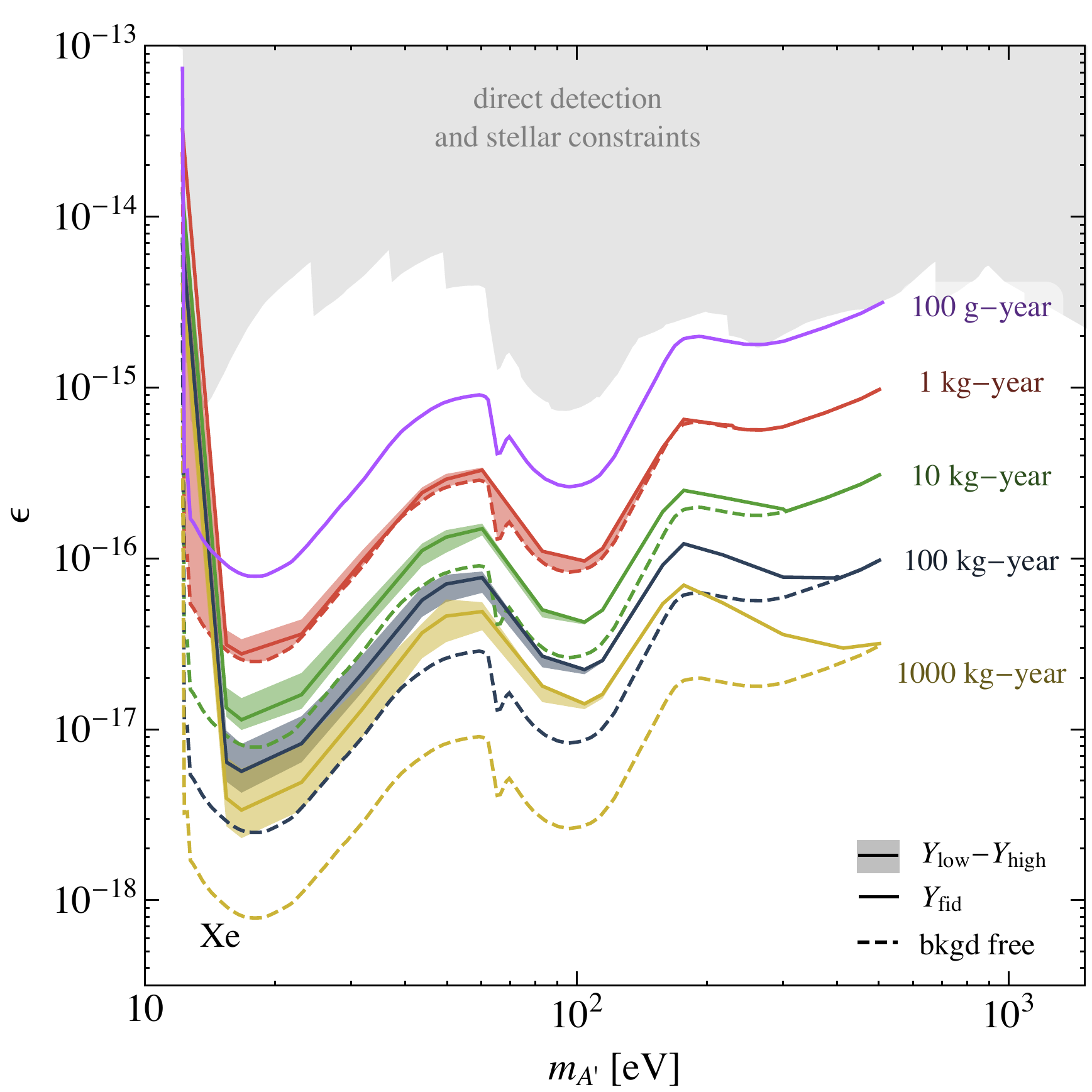}
\caption{
Discovery limits for the absorption of DM ALPs ({\bf left}) and DM $A'$ ({\bf right}) in silicon ({\bf top}), germanium ({\bf middle}), 
and xenon ({\bf bottom}). 
Exposures of 0.1, 1, 10, 100, and 1000~kg-years are shown in {\bf purple}, {\bf red}, {\bf green}, {\bf blue}, and {\bf yellow}, respectively. 
The solid line shows the results assuming the fiducial ionization efficiency, while the shaded bands denote the range between the high and low 
ionization efficiencies.  The dashed lines show the background-free 90\%~C.L.~sensitivities. Note that when the background assuming the fiducial ionization efficiency is negligible, the solid line and the dashed line are indistinguishable, making the dashed line disappear.  
The gray shaded region shows the current direct detection and stellar constraints~\cite{An:2013yfc,Redondo:2013lna,An:2014twa,Bloch:2016sjj}. The shaded orange region in the top panels is consistent with an ALP possibly explaining the white dwarf
luminosity function~\cite{Bertolami:2014wua}.
}
\label{fig:limitsabsorption}
\end{figure*}

In the case of DM-electron scattering, the most relevant neutrino background events are those with $\le 10$~electrons, shown 
in the left and middle panels of Figs.~\ref{fig:neutrinorates} and~\ref{fig:events_xe}.  
We show the 2$\sigma$ discovery limits (defined in Sec.~\ref{sec:likelihood}) 
for silicon, germanium, and xenon targets as a function of DM mass for the exposures of 
 0.1, 1, 10, 100, and 1000~kg-years (indicated by different colors) in Fig.~\ref{fig:limits}.  
 The results have been optimized over all possible thresholds and assume that sensitivity to 1-electron events is possible.  
 For most masses, a 1-electron threshold provides the best sensitivity (see Appendix~\ref{app:2e-threshold} for results 
 that assume the lowest achievable threshold is 2 electrons).  
 The solid lines show the results for the fiducial ionization efficiency, $Y_{\rm fid}$, while the 
edges of the shaded band surrounding each solid lines are defined by the low and high ionization efficiencies ($Y_{\rm low}$ 
and $Y_{\rm high}$).  
We consider two types of DM form factors, in which the DM-electron scattering proceeds through either a 
heavy mediator ($F_{\rm DM}=1$, left plots) or a light mediator ($F_{\rm DM} = \alpha^2m_e^2/q^2$, right plots). 

The discovery limits differ for different ionization efficiencies, as can be seen in Fig.~\ref{fig:limits}. 
$Y_{\rm high}$ ($Y_{\rm low}$) results in a higher (lower) number of solar neutrino background events, and hence 
the discovery cross section is larger (smaller). For comparison, we also include the 90\% C.L.~sensitivity estimates for a background-free experiment (dashed lines), 
corresponding to 2.3~DM events. 
The gray shaded regions show the current limits, derived using XENON10 and XENON100 data, from~\cite{Essig:2012yx,Essig:2017kqs}. 

In semiconductors, we see that for exposures up to 1~kg-year, the neutrino background is small and the cross section at the 
discovery limit corresponds to the cross section for a background-free experiment within $<{\cal{O}}(10\%)$. 
For large exposures the discovery limits can differ significantly from the background-free case, reaching a factor of 
$\sim$10 (7, 30) for $Y_{\rm fid}$ ($Y_{\rm low}$, $Y_{\rm high}$) for 1000 kg-years. 

For xenon, solar neutrinos are already a small background for a 100 gram-year exposure, and affect the sensitivity by 
a factor of $\sim$3 (8) at $m_\chi=1$~GeV for a 1~(10)~kg-year exposure for $F_{\rm DM}=1$ 
(the $F_{\rm DM}\propto1/q^2$ sensitivities are less affected because here the signal is concentrated in the first few bins, in contrast to 
$F_{\rm DM}$=1, for which the signal is spread over a larger number of bins, at least for large enough DM masses).  
This is in sharp contrast to silicon or germanium targets, whose sensitivity is limited by neutrinos only for larger exposures.  
One reason for this is that the CNS rate scales dominantly as the square of the number of neutrons, 
and is thus larger in xenon than in germanium or silicon.  
However, the more important reasons are that the xenon nucleus recoils with lower energy and xenon has a lower ionization efficiency, 
so that neutrinos near the peak of the $^8$B spectrum (which is the main neutrino background component in xenon) 
produce events containing only a few electrons. This coincides with the DM signal spectrum, which predominantly populates the one to 
a few electron bins, see Fig.~\ref{fig:events_xe}; meanwhile, in germanium and silicon targets, 
the $^8$B peak produces a few hundred electrons, which is well above 
where the DM spectrum would dominate, 
while the dominant neutrino components at lower energies ($^7$Be, pep) 
only at most populate the 1-electron bin (depending on the ionization efficiency) and are thus easily distinguished from the DM signal.  

In Fig.~\ref{fig:saturation} we show the discovery limits as a function of exposure.  
The discovery limits scale differently depending on the exposure.  For low exposures, the neutrino backgrounds are negligible, and the 
discovery limits scale as 1/exposure.  For intermediate exposures, neutrinos are a background, but the DM signal can be distinguished 
from the neutrino background via its distinct spectrum; the discovery limits scale as 
$1/\sqrt{\rm exposure}$.  For very large exposures, the systematic uncertainties in the neutrino fluxes dominate, and 
it could become difficult to distinguish the DM signal from the neutrinos, especially if the spectral shapes are similar.  In this case the 
discovery limits would saturate and remain constant as a function of the exposure.  However, in our case, the discovery limits do not 
reach this saturation regime, at least not up to 10,000~kg-years.  

In the case of DM absorption, 
the signal peaks at the DM mass, which we vary from the band gap/binding energy to $\sim 1$~keV.  
We thus need to consider also the neutrino backgrounds that yield a few hundred electron-hole pairs 
(right panels of Figs.~\ref{fig:neutrinorates} and~\ref{fig:events_xe}).  
We show the discovery limits for ALPs and $A'$ in silicon, germanium, and xenon as a function of DM mass for exposures of 
0.1--1000~kg-years, incrementing the exposure in powers of ten in Fig.~\ref{fig:limitsabsorption}.  
The colored bands reflect the uncertainty in the ionization efficiency at low masses; at high masses, the available data constraints 
the ionization efficiency.  
As for the case of DM-electron scattering, the neutrino backgrounds are larger for xenon than the semiconductor targets.  

For DM absorption, we again observe a similar scaling of the discovery limits with exposure as we did for DM scattering, 
but with one notable difference: the scaling is prominently mass-dependent in the absorption case.  As the DM mass increases, 
the neutrino background is dominated by the $^8$B component, which keeps decreasing as a function of energy.  
Searches for large DM masses therefore remain background free for large exposures.

\section{Results: Solar Neutrinos as a Signal in Dark Matter Searches with Electron Recoils} \label{sec:results-coherent}

\setlength{\tabcolsep}{0.5em} 
{\renewcommand{\arraystretch}{1.3} 
\begin{table}[b]
\caption{The ionization threshold above which a search for $^8$B neutrinos is essentially free of backgrounds from other solar neutrino 
components, in silicon, germanium, and xenon for various ionization efficiencies.  
Also shown are the corresponding number of signal events per kg-year.}
\begin{center}
\begin{tabular}{l c c c}
\hline
&Threshold & No. of events (per kg-year)\\ \hline
\multirow{3}{*}{\large{Si}}
\multirow{1}{*}{~~~~~~high}&10 $e^-$  & 0.1305 \\ 
~~~~~~~~~~{fiducial}&~2 $e^{-}$ & 0.1324\\ 
\multirow{1}{*}{~~~~~~~~~~low} &~1 $e^-$ & 0.1182\\
\hline
\multirow{3}{*}{\large{Ge}}
\multirow{1}{*}{~~~~high}&~3 $e^-$  & 0.4528 \\ 
~~~~~~~~~~{fiducial}&~3 $e^{-}$ & 0.4474\\ 
\multirow{1}{*}{~~~~~~~~~~low} &~1 $e^-$ & 0.4396\\
\hline
\multirow{3}{*}{\large{Xe}}
\multirow{1}{*}{~~~~high}&~3 $e^-$  & 1.3661 \\ 
~~~~~~~~~~{fiducial}&~1 $e^{-}$ & 1.4178\\ 
\multirow{1}{*}{~~~~~~~~~~low} &~1 $e^-$ & 0.8392\\
\hline
\end{tabular}
\end{center}
\label{tab:B8}
\vskip -5mm
\end{table}%

\begin{figure}[t!]
\includegraphics[width=0.43\textwidth]{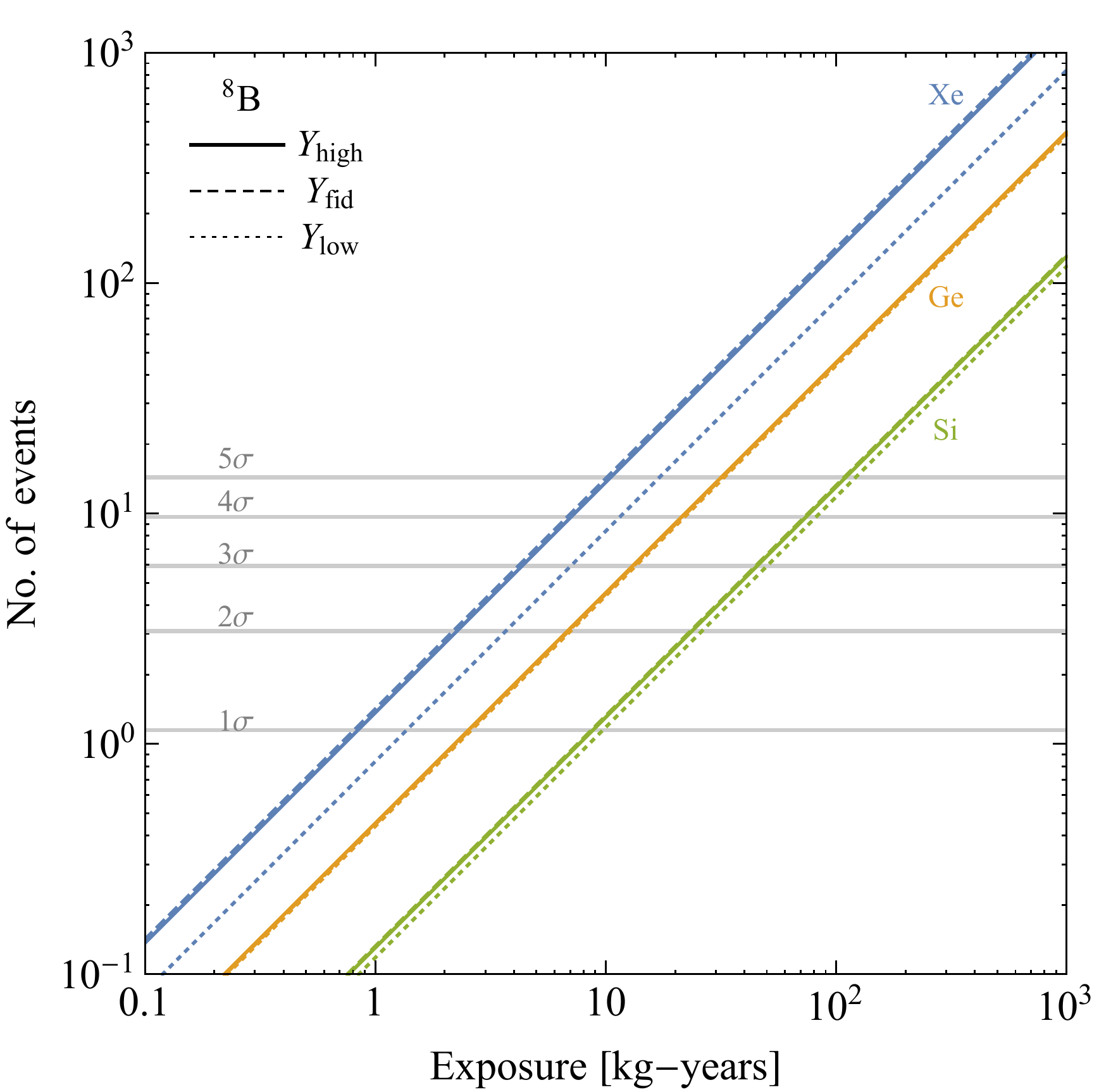}
\caption{The number of events expected for the signal of $^8$B versus the exposure with the thresholds given in Table ~\ref{tab:B8}. The horizontal gray lines show the detection significance.  
}
\label{fig:B8}
\end{figure}

In this section, we treat the neutrinos as the signal and discuss how well future direct-detection experiments that are 
sensitive to electron recoils can detect solar neutrinos via their CNS signal. 

The CNS signal from $^8$B is the easiest to detect and essentially free of backgrounds from other solar neutrino components 
for a sufficiently large threshold, whose value depends 
on the ionization efficiency.  Table~\ref{tab:B8} shows these thresholds for the three different ionization efficiencies, in silicon, germanium, and xenon. 
We also show the respective number of expected events per kg-year. Fig.~\ref{fig:B8} shows the number of expected 
events versus exposure with the thresholds in Table~\ref{tab:B8}.  
We see that $^8$B neutrinos are easiest to detect in xenon.  Assuming an idealized experiment free from all other (non-neutrino) 
backgrounds, a 2$\sigma$ (5$\sigma$) observation requires an exposure of $\sim$ 2 (10)~kg-years.  For example, 
a 10-kg target as envisioned in~\cite{UA'1,Battaglieri:2017aum} could observe the $^8$B component at  
$5\sigma$ after running for 1 year. 
In germanium (silicon), 2$\sigma$ evidence requires an exposure of $\sim$ 7 (24) kg-years.

\begin{figure*}[t!]
\includegraphics[width=0.415\textwidth]{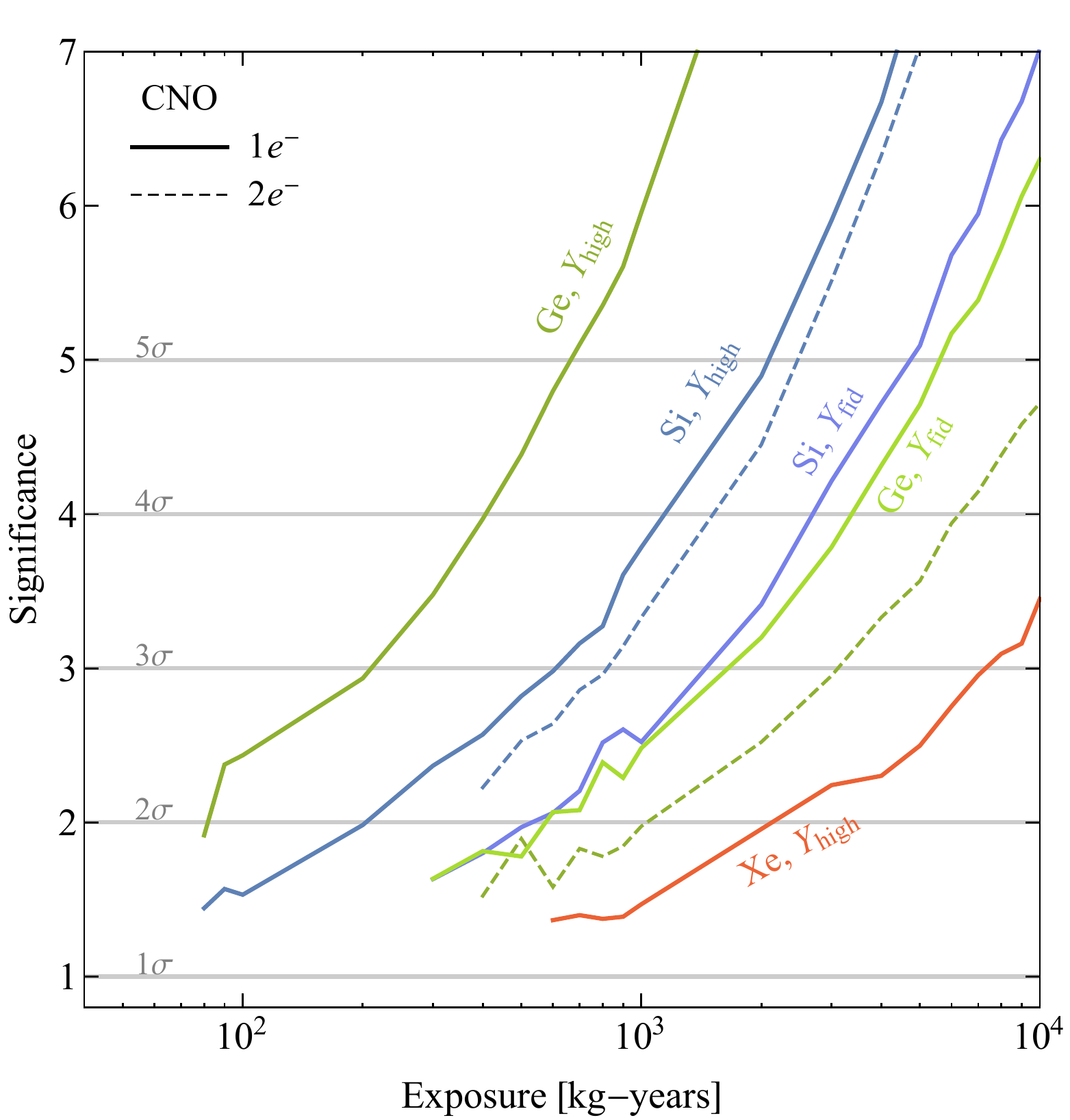} ~~
\includegraphics[width=0.43\textwidth]{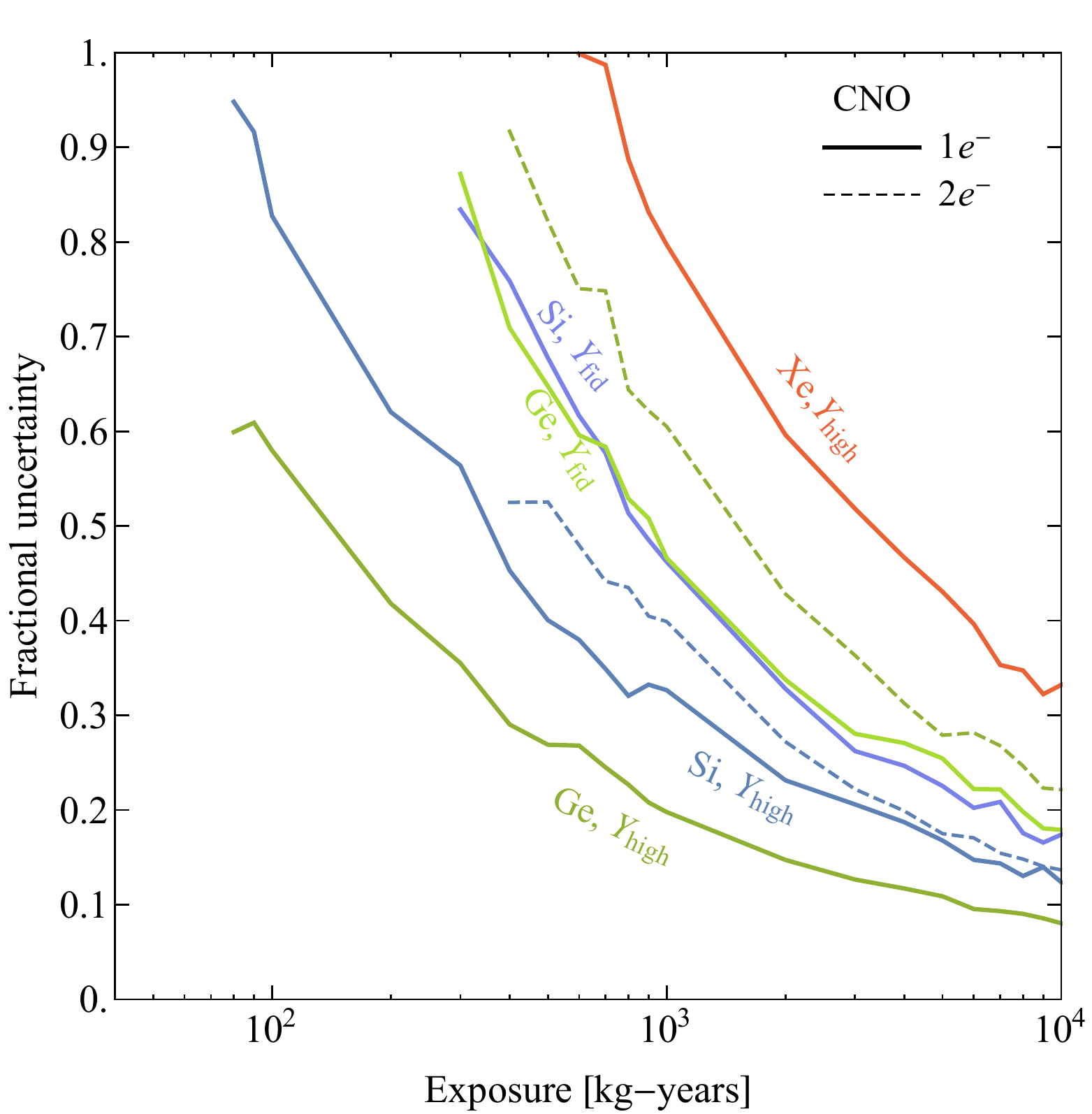}
\caption{Expected significance ({\bf left}) and fractional uncertainty ({\bf right}) for detecting CNO neutrinos. 
Solid (dashed) lines assume a 1-(2-)electron threshold.}
\label{fig:cnoresults}
\end{figure*}

Fig.~\ref{fig:cnoresults} shows the expected significance, $\sigma$, and fractional uncertainty (standard deviation/mean) to detect the CNO flux. 
A detection of the CNO flux requires large exposures, and is only possible if the ionization efficiency is sufficiently large. 
For the high ionization-efficiency model in silicon and germanium, the CNO flux contributes to the 1 and 2-electron bins, and one could detect the CNO flux even with a 2-electron threshold (dashed lines in Fig.~\ref{fig:cnoresults}). In contrast, for the other efficiency models, the CNO flux 
only contributes to the 1-electron bin. 
We also observe that detecting the CNO flux with a xenon target is more difficult than with a semiconductor target, since in xenon 
most of the flux produces no ionization (and does so only for $Y_{\rm high}$) and the $^8$B neutrinos are a larger background.   

\begin{figure*}[t!]
\includegraphics[width=0.43\textwidth]{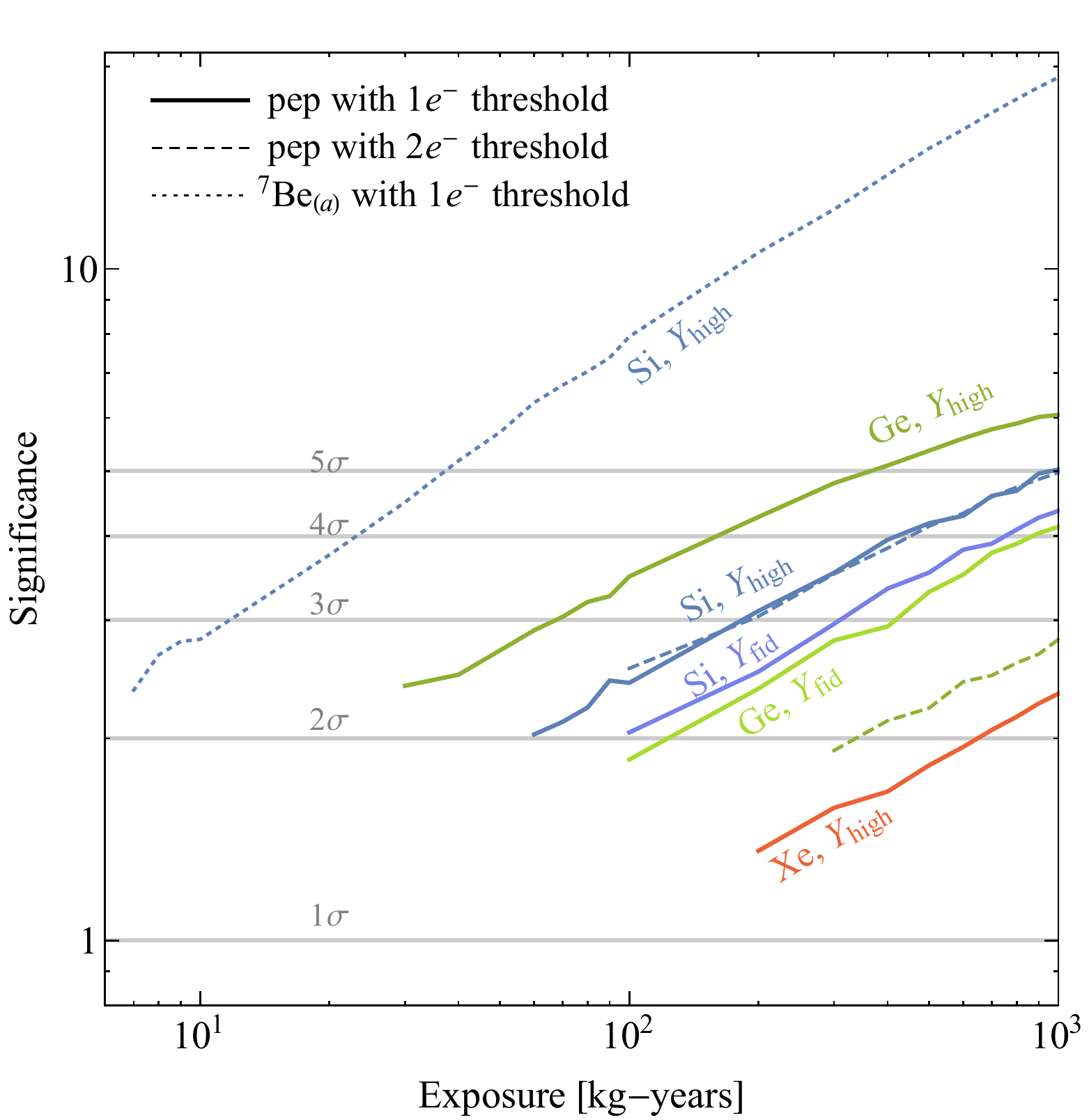} ~~
\includegraphics[width=0.43\textwidth]{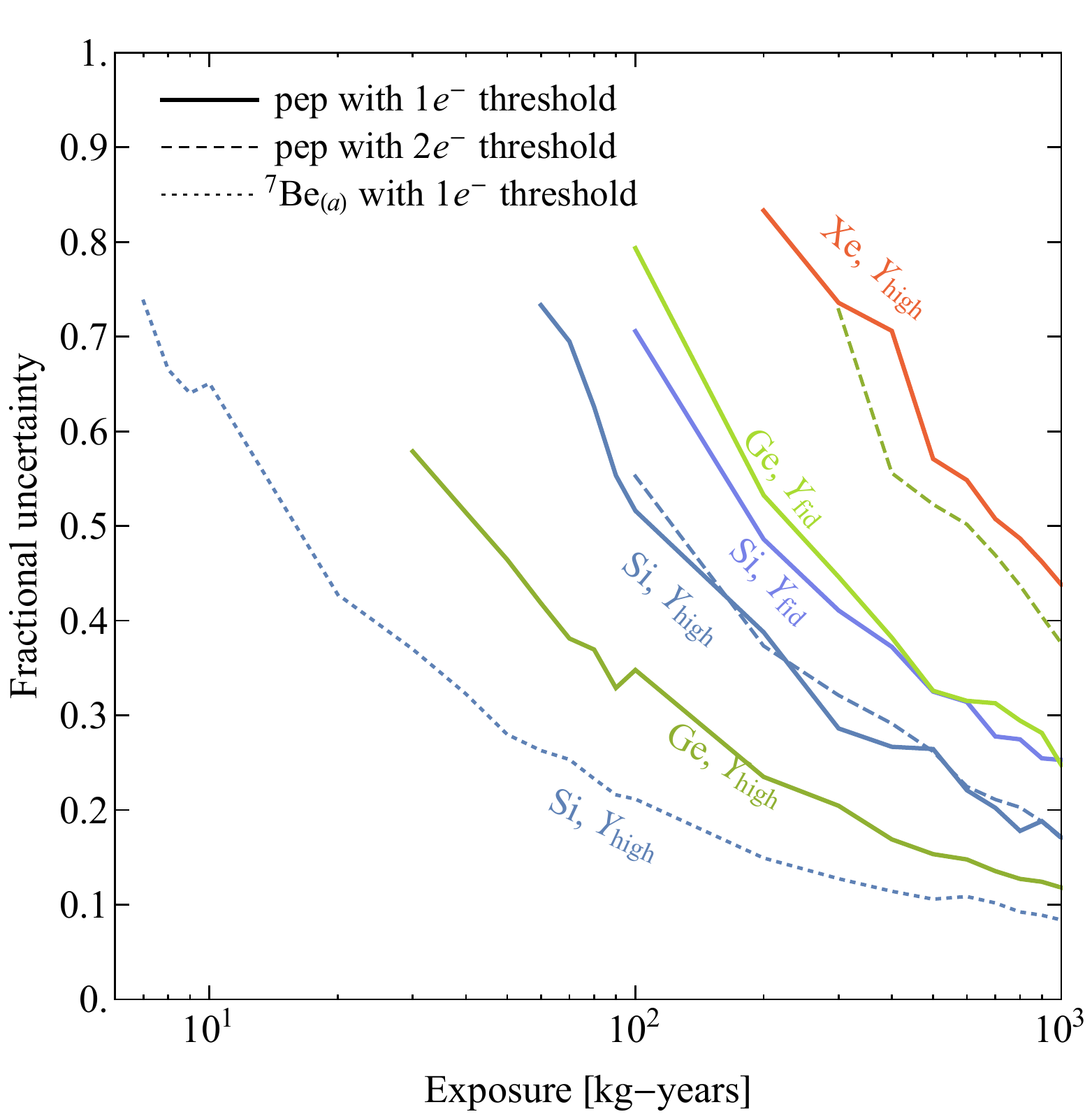}
\caption{Expected significance ({\bf left}) and fractional uncertainty ({\bf right}) for detecting 
the pep and  $^7{\rm Be}_{\rm (a)}$ neutrino components. }
\label{fig:pepberesults}
\end{figure*}

Fig.~\ref{fig:pepberesults} shows the results for the  $^7{\rm Be}_{\rm (a)}$ and pep components. 
The  $^7{\rm Be}_{\rm (a)}$ component can be detected only if the ionization efficiency is high, and then dominates in the 1-electron bin. 
A 5$\sigma$ detection is possible with silicon for an exposure of $\sim$ 40~kg-years.   
The pep components can be detected as well for large enough exposures if the ionization efficiency is sufficiently large.  

\section{Results: Non-Standard Neutrino Interactions}\label{nsi}

Non-standard neutrino interactions (NSI) betweens neutrinos and quarks can modify the CNS cross section. 
An effective Lagrangian describing non-SM interactions of neutrinos with hadrons can be written as~\cite{Barranco:2005yy},
\beqa\label{eq:nsi1}
\mathcal{L}^{NSI}_{\nu{\rm -Had}} & = &-\frac{G_F}{\sqrt{2}}\sum_{\substack{q=u,d \\ \alpha, \beta =e, \mu, \tau}}[\overline{\nu}_{\alpha}\gamma^{\mu}(1-\gamma^{5})\nu_{\beta}] \\
& &\times  (\epsilon^{qL}_{\alpha \beta}[\overline{q}\gamma_{\mu}(1-\gamma^{5})q] + \epsilon^{qR}_{\alpha \beta}[\overline{q}\gamma_{\mu}(1+\gamma^{5})q]),  \nonumber
\eeqa
where NSI are parametrized by $\epsilon^{qP}_{\alpha \beta}$ corresponding to the interaction of neutrinos with flavors $\alpha$ and $\beta$ ($\alpha$, $\beta$ = $e$, $\mu$, $\tau$) with quark $q$ ($q$ = $u$, $d$) of chirality $P$ ($P$ = $L$, $R$). 
For non-universal (flavor-changing) interactions, $\alpha$ = $\beta$ ($\alpha \neq \beta$). 
Assuming only non-universal interactions and 
neglecting the contribution of the axial hadronic current, 
the differential CNS cross section for neutrino flavor $\alpha$ to scatter off a nucleus is modified to 
\begin{eqnarray}\label{nsi2}
\left( \frac{d\sigma}{dE_{\rm{NR}}}\right)_{\nu_{\alpha}A}&=& \frac{G_{F}^{2}}{\pi}m_{N}\left(1-\frac{m_{N}E_{\rm{NR}} }{2E_{\nu_{\alpha}}^{2}}\right)F^{2}(E_{\rm{NR}}) \nonumber\\
& \times& \left\{[Z (g_{V}^{p}+2 \epsilon^{uV}_{\alpha \alpha}+ \epsilon^{dV}_{\alpha \alpha})\right. \\
& +&\left. N (g_{V}^{n}+ \epsilon^{uV}_{\alpha \alpha}+ 2\epsilon^{dV}_{\alpha \alpha})]^2 \right\}, \nonumber
\end{eqnarray}     
where $g_{V}^{p}$ = $(\frac{1}{2} - 2 \sin^{2} \theta_W)$, $g_{V}^{n}$ = -$\frac{1}{2}$ are the SM contributions.  
Assuming $\epsilon^{qV}_{ee}$ = $\epsilon^{qV}_{\mu \mu}$ = $\epsilon^{qV}_{\tau \tau} \equiv \epsilon^{qV}$, 
Eq.~(\ref{nsi2}) can be written in a flavor-independent way as 
 \begin{eqnarray}\label{nsi3}
\Big( \frac{d\sigma}{dE_{\rm{NR}}}\Big)_{\nu A}&=&\frac{G_{F}^{2}}{\pi}m_{N}\left(1-\frac{m_{N}E_{\rm{NR}} }{2E_{\nu}^{2}}\right)F^{2}(E_{\rm{NR}}) \nonumber \\
& & \times \left\{[Z (g_{V}^{p}+2 \epsilon^{uV}+ \epsilon^{dV})  \right.\\
& & \left.+ N (g_{V}^{n}+ \epsilon^{uV}+ 2\epsilon^{dV})]^2 \right\}. \nonumber
\end{eqnarray}      
Convolving this with the total solar neutrino flux gives the modified CNS rates.  
Measurements of the solar neutrino fluxes by direct-detection experiments can constrain NSI parameters. 
As an illustration, Fig.~\ref{fig:nsi} shows the 2$\sigma$ confidence level allowed regions on $\epsilon^{uV}$ and $\epsilon^{dV}$ 
for a xenon target, for two exposures (10 and 100 kg-years), which could be obtained from measuring $^8$B neutrinos.  
Significant improvements over existing constraints from CHARM and COHERENT are possible. 

In Appendix~\ref{app:magnetic_moment}, we will briefly investigate the effect of a non-zero neutrino magnetic moment on solar-neutrino-electron 
scattering.  

\begin{figure}[t!]
\begin{center}
\includegraphics[width=0.43\textwidth]{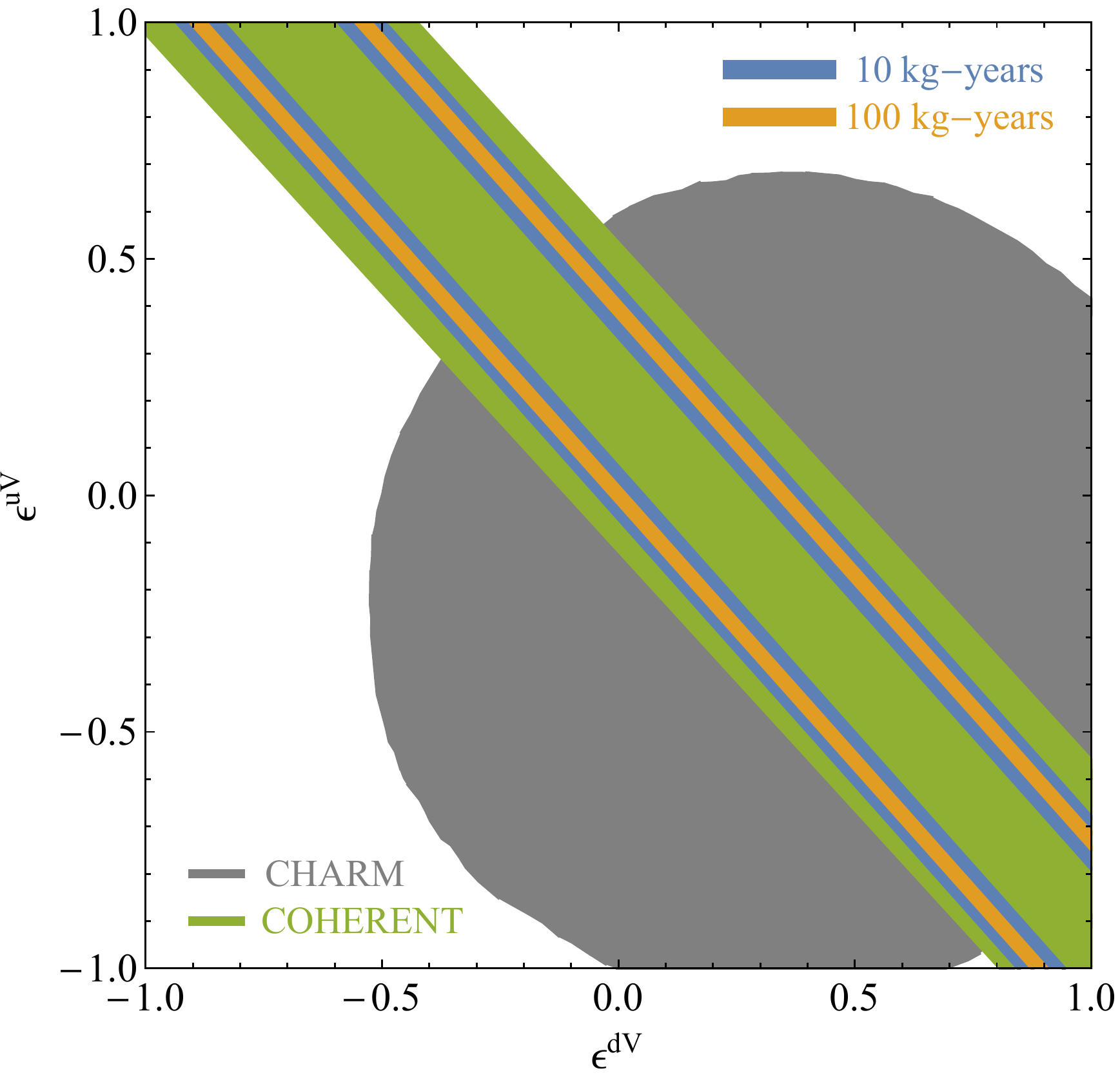}
\caption{Shaded regions show 2$\sigma$-confidence-level allowed regions on NSI parameters $\epsilon^{uV}$ and $\epsilon^{dV}$ for 
a xenon target, assuming a detection of $^8$B neutrinos at the SM predicted value, our fiducial ionization efficiency, and an exposure 
of 10 (100) kg-years for the blue (orange) regions.  Currently allowed regions on $\epsilon_{ee}^{uV}$ and $\epsilon_{ee}^{dV}$ 
 from CHARM~\cite{Davidson:2003ha} (gray) and COHERENT~\cite{Akimov:2017ade} are shown in gray and green, respectively, assuming 
 $\epsilon_{\mu}^{uV}=\epsilon_{\mu}^{dV}=\epsilon_{\tau}^{uV}=\epsilon_{\tau}^{dV}=0$.  
}
\label{fig:nsi}
\end{center}
\end{figure}

\section{Discussion and Conclusions} \label{sec:conclusions}

In this work, we investigated how the ionization produced when solar neutrinos scatter coherently off nuclei limits the future 
sensitivities of (low-threshold) DM direct-detection experiments that search for electron recoils from DM-electron interactions  
(we ignore all other possible backgrounds).  We consider both DM-electron scattering (for different form factors), 
and DM absorption by electrons.  
We also investigated the sensitivity of such experiments to various components of the solar neutrino flux, including the $^8$B, $^7$Be, 
pep, and CNO components.  
We considered silicon, germanium, and xenon as the target material, and for each target we considered three models for the 
ionization efficiency, which is unknown at the low energy range of interest, a ``low'', ``fiducial'', and ``high'' ionization efficiency, see Figs.~\ref{fig:conv} and \ref{fig:convxe}. 

Assuming our fiducial ionization efficiency, neutrinos are expected to generate about 0.076 (0.131) events per kg-year in silicon (germanium), 
see Fig.~\ref{fig:neutrinorates}.  
We can expect at least one neutrino event in 10\% of the experiments for exposures of $\sim$1.4 (0.8) kg-years for silicon (germanium). 
The corresponding exposures for the high and low ionization efficiencies are 0.2 (0.3) kg-years and 9.7 (2) kg-years. In xenon, for the fiducial conversion scheme, neutrinos generate about 1.24 events per kg-year (Fig.~\ref{fig:events_xe}). 
Hence, we can expect at least one neutrino event in 10\% of the experiments for exposures of $\sim$ 0.085 kg-years. 
The corresponding exposures for the high and low ionization efficiencies are 0.05 kg-years and 0.16 kg-years, respectively.     

For larger exposures than the ones listed above, the sensitivity to a DM search is limited by neutrinos, 
as shown in Fig.~\ref{fig:limits} for DM-electron scattering and Fig.~\ref{fig:limitsabsorption} for DM absorption by electrons. 
For very large exposures, it becomes increasingly difficult to probe to lower cross sections, but note that there is no absolute 
neutrino ``floor'' beyond which no improvement is possible (Fig.~\ref{fig:saturation}). 

Treating the neutrinos as a signal, rather than as a background to a DM search, we considered the detection of $^8$B, pep, $^7$Be, and CNO 
neutrinos.  The required exposures to detect the CNO fluxes at a significance greater than $3\sigma$ are very large, 
$\sim$600 kg-years in silicon, $\sim$210 kg-years in germanium, and $\sim$7350 kg-years in xenon (Fig.~\ref{fig:cnoresults}).  
Moreover, the fractional uncertainty at these exposures is quite high ($\sim 0.4$), while a fractional uncertainty of 0.05 is needed to distinguish between low- and high-metallicity models~\cite{Bergstrom:2016cbh}.  

In contrast, $^8$B neutrinos are more easily detected and an accurate measurement of the fluxes can have significant impact on our understanding of neutrinos. A $3\sigma$ observation is possible for exposures greater than $\sim$5 kg-years in xenon, $\sim$15 kg-years in germanium, and $\sim$40 kg-years in silicon (Fig.~\ref{fig:B8}), assuming our fiducial ionization efficiency. 
Detecting the $^8$B at low energies would give a first direct measurement of the electron-neutrino survival probability in the transition between 
the energy region dominated by vacuum oscillations to the region dominated by the MSW-effect.  
Moreover, it provides a window to NSI between neutrinos and quarks, which could increase the coherent scattering 
rates; Fig.~\ref{fig:nsi} shows an example of how measuring the $^8$B neutrinos with a xenon target would constrain NSI parameters.  

Detecting pep and $^7$Be neutrinos at lower energies 
would directly measure the survival probability in the vacuum-oscillation-dominant region.  
This, however, is challenging.  
The pep neutrinos could be detected in silicon and germanium, but a $3\sigma$ observation requires already exposures of 
at least 65 (200) kg-years in germanium (silicon), 
assuming a high ionization efficiency, and even larger exposures in xenon (Fig.~\ref{fig:pepberesults}).  
The $^7$Be$_{(a)}$ neutrinos are detectable, but only in silicon assuming a high ionization efficiency, where a 3$\sigma$ observation 
is possible with in $\sim$10 kg-years (in other cases not even a single electron is produced).  

In summary, direct-detection experiment sensitive to electron recoils will have both an opportunity to detect solar neutrinos via CNS, but 
will also eventually have to contend with them as an important background. 

\section*{Acknowledgments} 
We thank Louis Strigari for useful discussions and for comments on the draft. 
We also thank Rafael Lang, Giacinto Piacquadio, Oren Slone, and Peter Sorensen 
for useful discussions. We also thank Tongyan Lin for pointing out a mistake in Fig.~\ref{fig:limitsabsorption} in the previous version of the paper.   
R.E.~and M.S.~are supported by DoE Grant DE-SC0017938.  
T.-T.Y. thanks the hospitality of the Aspen Center for Physics, which is supported by National Science Foundation grant PHY-1607611. 


\vspace{5mm}
\appendix

\section{Scintillators}\label{app:scintillators}

\begin{figure}[t!]
\begin{center}
\includegraphics[width=0.45\textwidth]{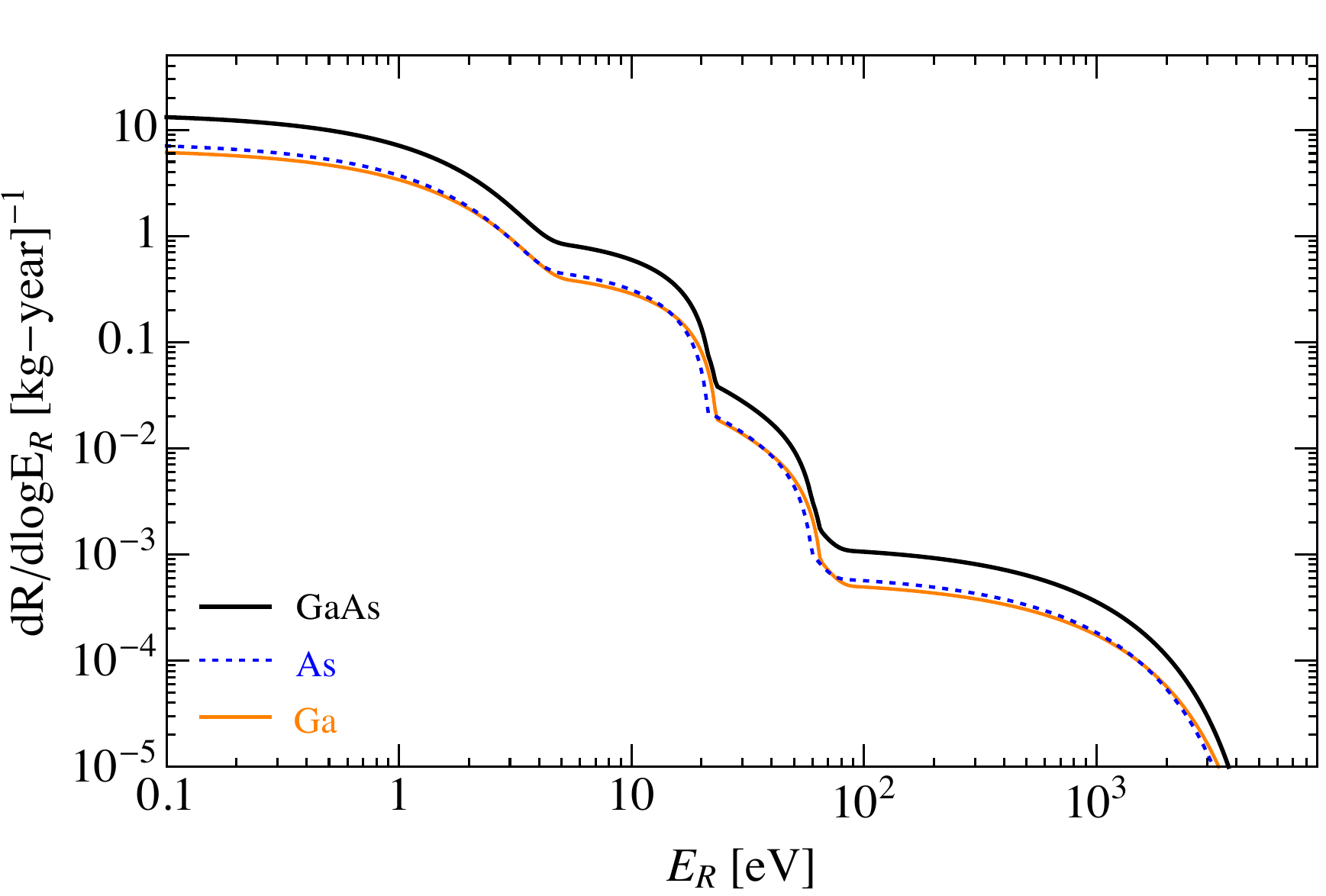}\\
\includegraphics[width=0.45\textwidth]{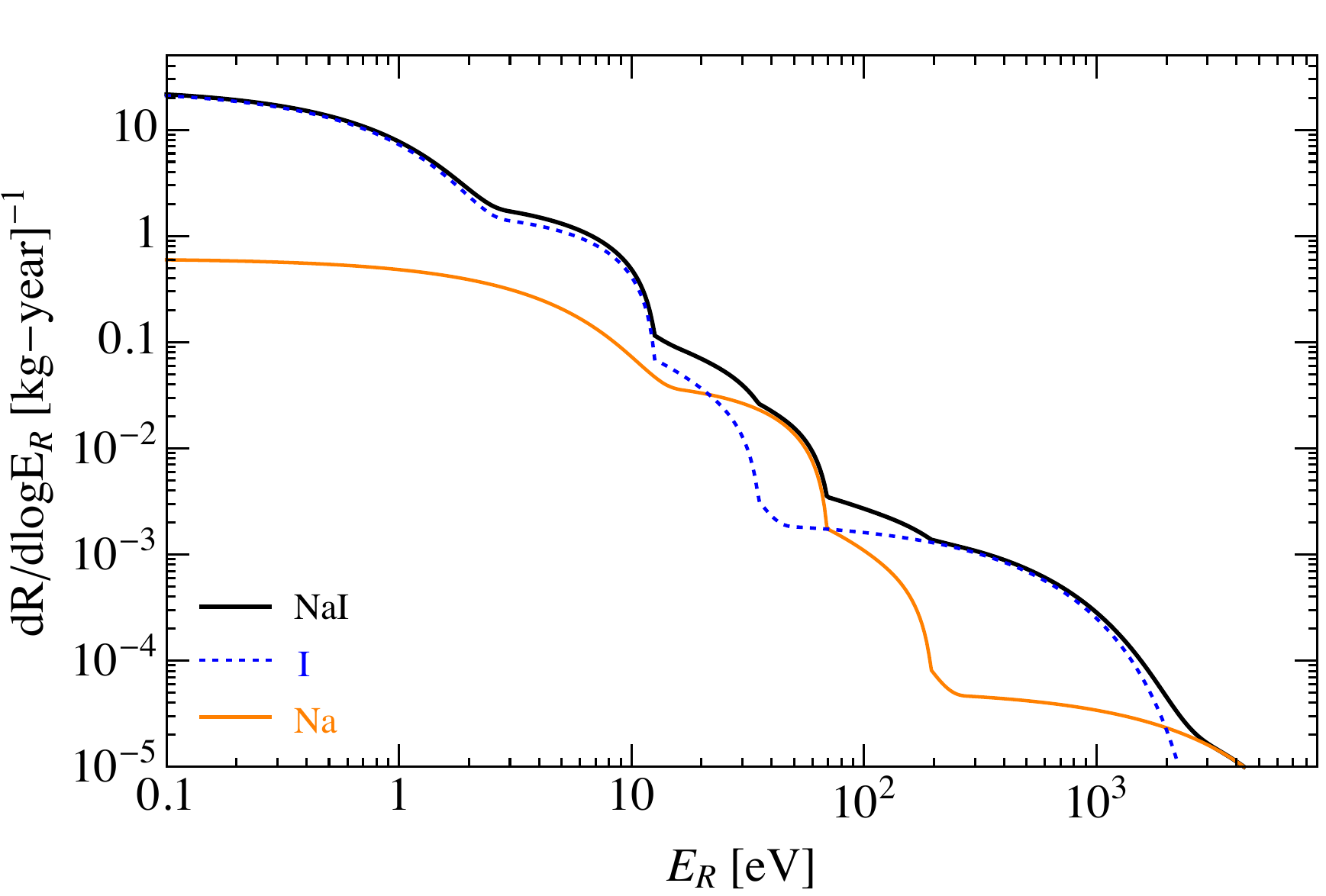}
\caption{Coherent solar neutrino-nucleus scattering event rates per kg-year for GaAs ({\bf top}) and NaI ({\bf bottom}) targets (black lines). 
The contributions from the two individual elements are given in orange and blue. }
\label{fig:scintillatorrate}
\end{center}
\end{figure}

The quenching factor for scintillators can be described through semi-empirical methods~\cite{Tretyak:2009sr}. The light yield suppression of highly ionizing particles was described by Birks~\cite{birks2013theory} as
\beq
L(E)=\int_0^E \frac{S dE}{1+kB\frac{dE}{dr}}\,,
\eeq
where $E$ is the released energy and $S$ is the absolute scintillation factor. $B\frac{dE}{dr}$ is the density of excitations along the track $r$, while $k$ is the quenching factor. The combination of $kB$ is known as the Birks factor and is commonly treated as a single parameter. 

The quenching factor for ions is defined as the ratio of the light yield of ions to that of electrons,
\beq
Q_i(E)=\frac{L_i(E)}{L_e(E)}=\frac{\int_0^E \frac{dE}{1+kB\left(\frac{dE}{dr}\right)_i}}{\int_0^E \frac{dE}{1+kB\left(\frac{dE}{dr}\right)_e}}\,.
\eeq
One sees that the factor of $S$ cancels in the ratio and that the quenching factor depends only on the Birks factor $kB$. The value of the Birks factor for different materials is determined empirically, and depends on the experimental conditions. Importantly, the Birks factor varies 
with energy, especially at low energies. However, given the lack of low-energy data for GaAs and NaI, we will not investigate in detail the neutrino signals in these materials.  Instead, we only show the neutrino-nucleus scattering rates for GaAs and NaI targets in Fig.~\ref{fig:scintillatorrate}. 

\section{Discovery Limits for Two-Electron Thresholds}\label{app:2e-threshold}
We show the discovery limits assuming a threshold of 2-electrons in semiconductors and xenon in Fig.~\ref{fig:limits2ethreshold}. 
We present the results for our three different ionization efficiencies for each material, 
as well as two DM form factors. For comparison, we show the curves for a 1-electron threshold using the fiducial conversion scheme. 
We see that the 2-electron and 1-electron thresholds are similar, except the former of course has a slightly higher mass threshold. 

\begin{figure*}[t!]
\vspace{-4mm}
\includegraphics[width=0.43\textwidth]{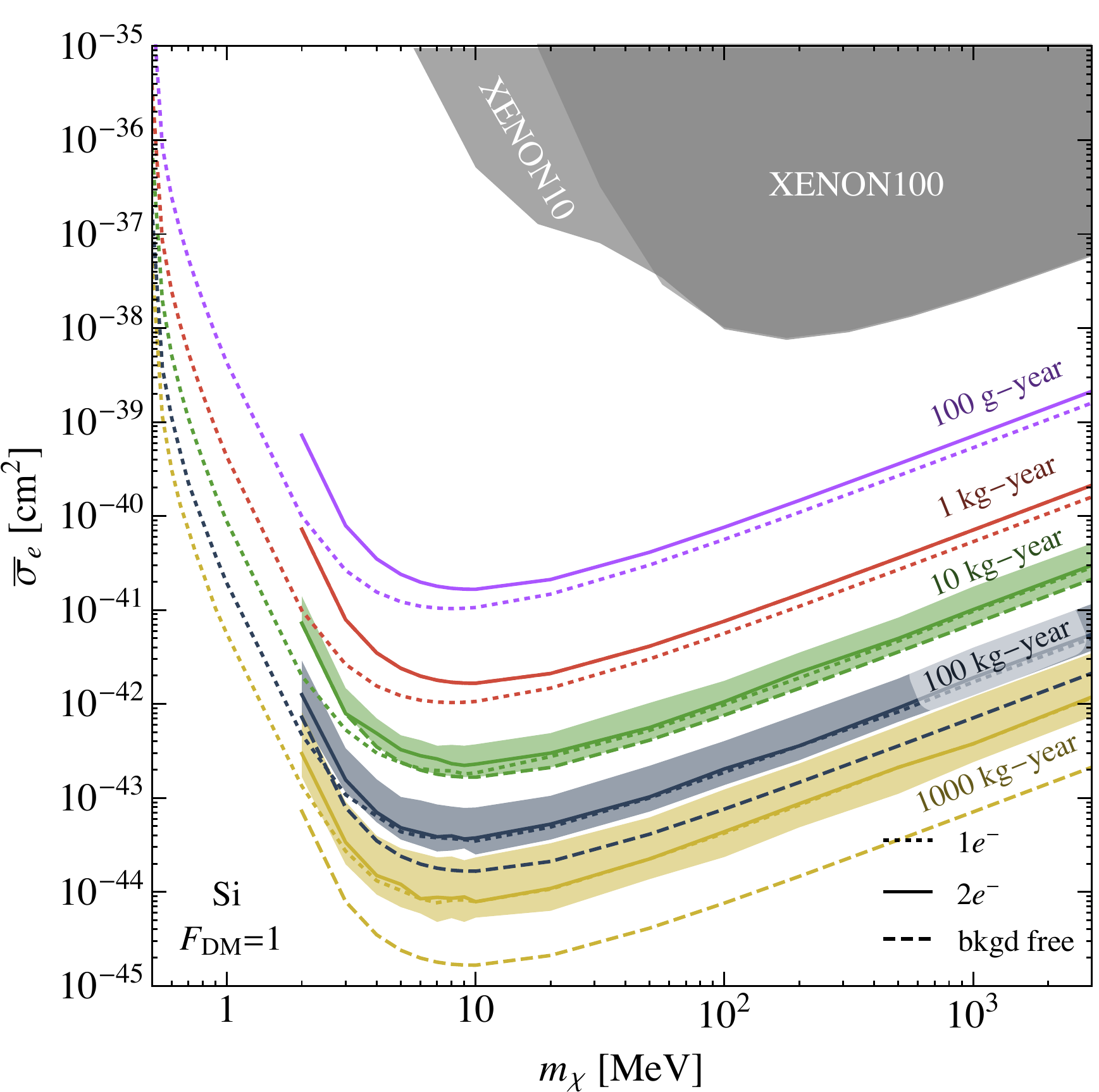} ~~~~~~~
\includegraphics[width=0.43\textwidth]{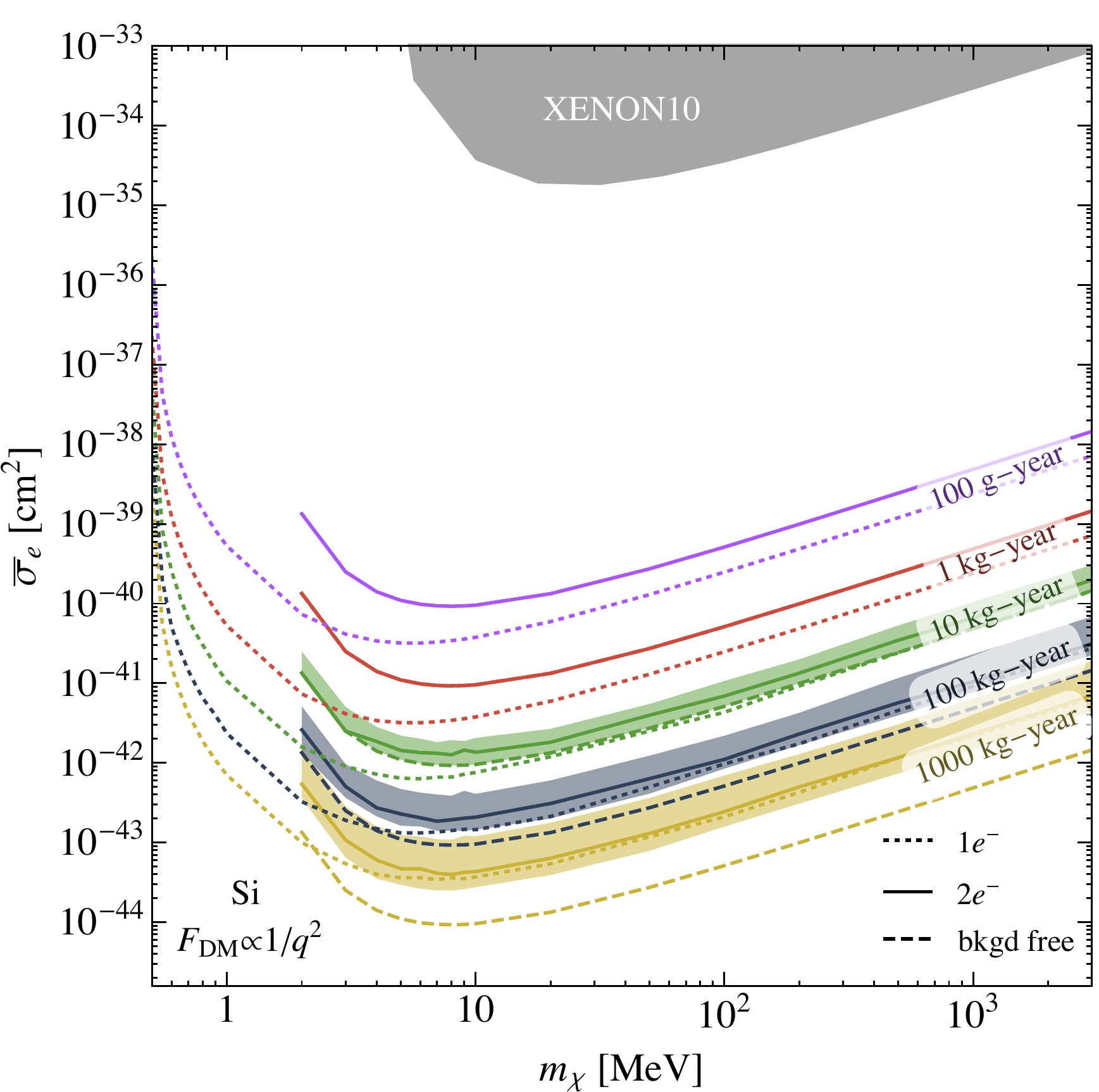} \\
\vspace{-2mm}
\includegraphics[width=0.43\textwidth]{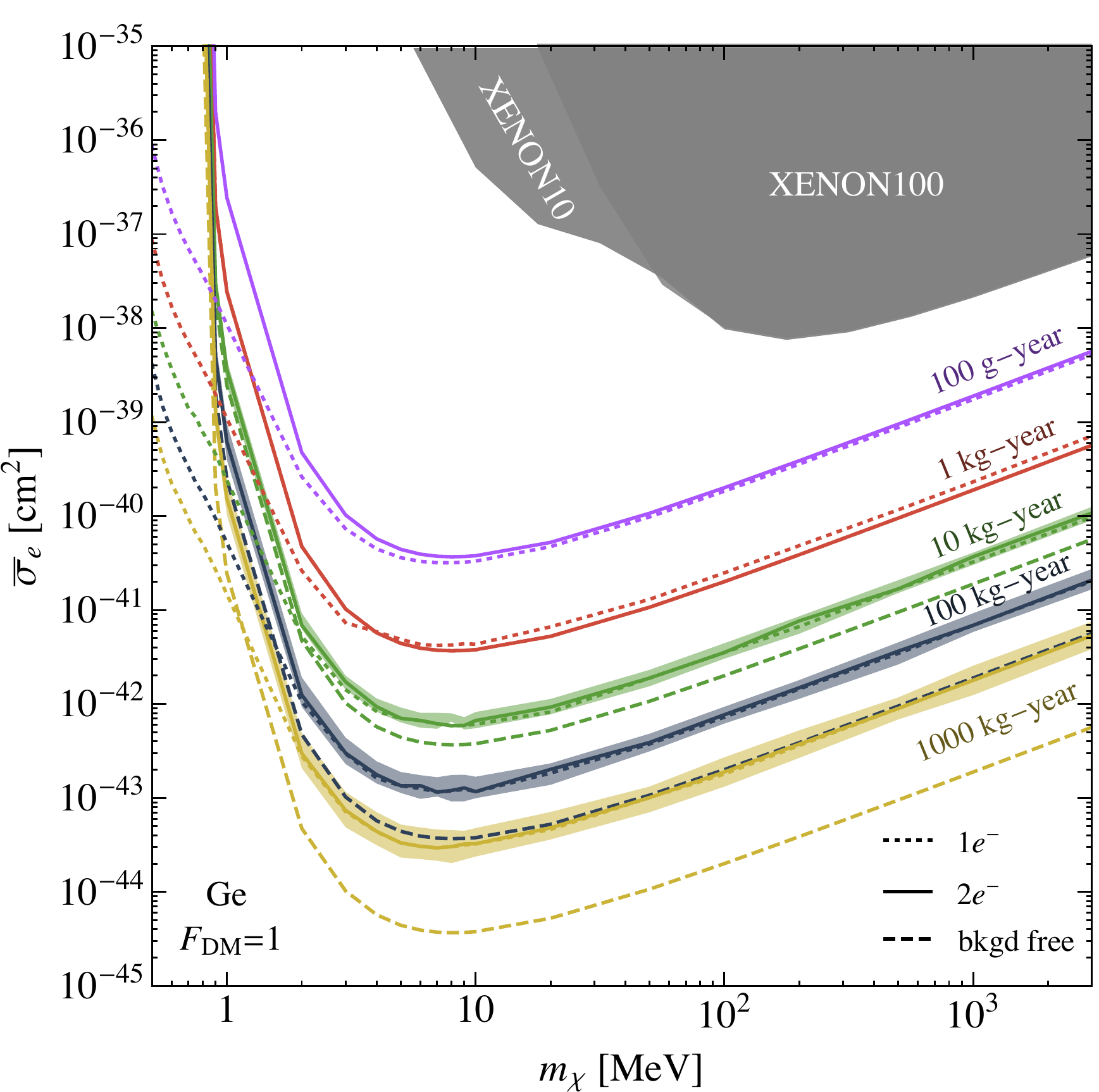}~~~~~~~
\includegraphics[width=0.43\textwidth]{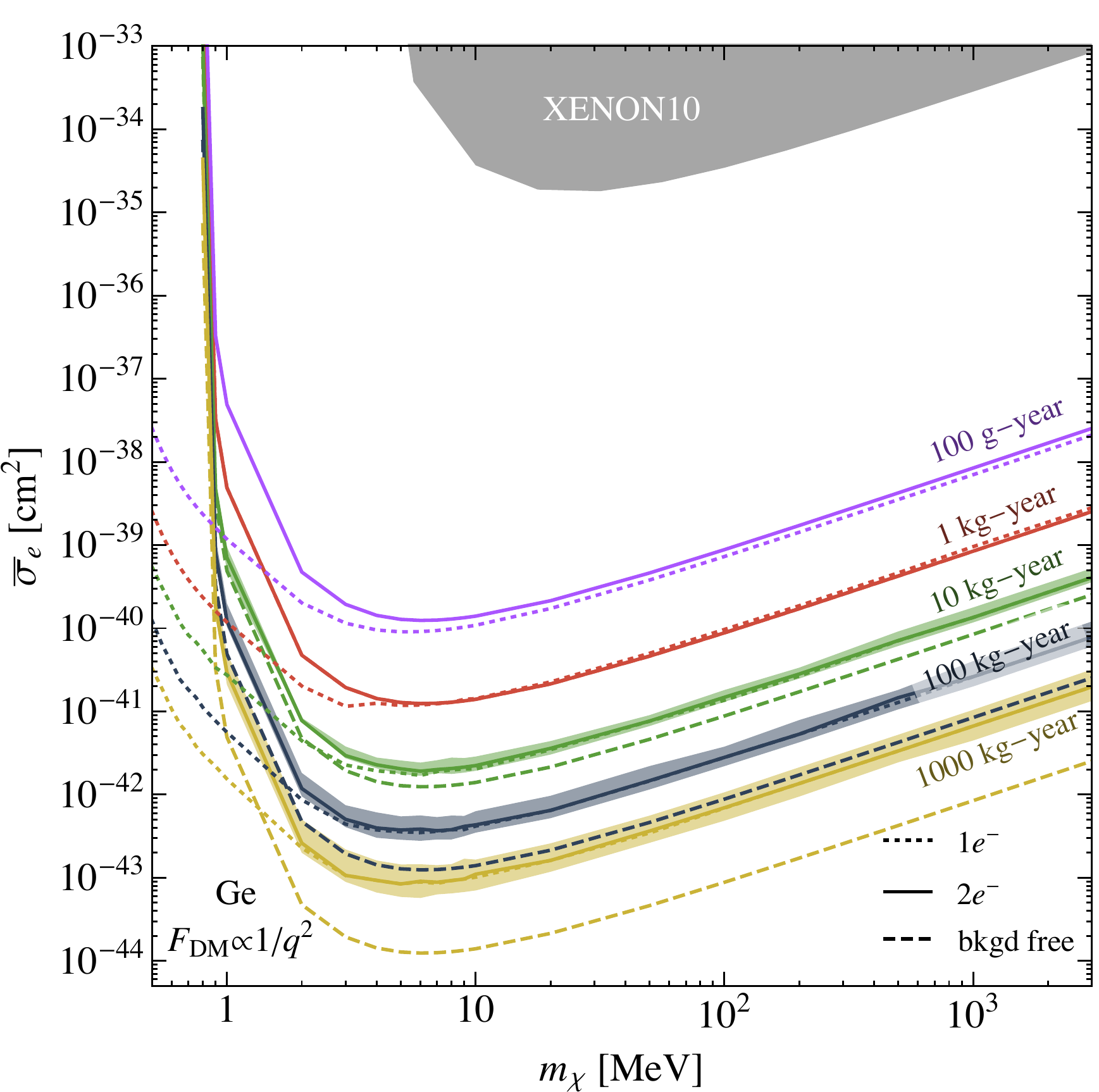}\\
\vspace{-2mm}
\includegraphics[width=0.43\textwidth]{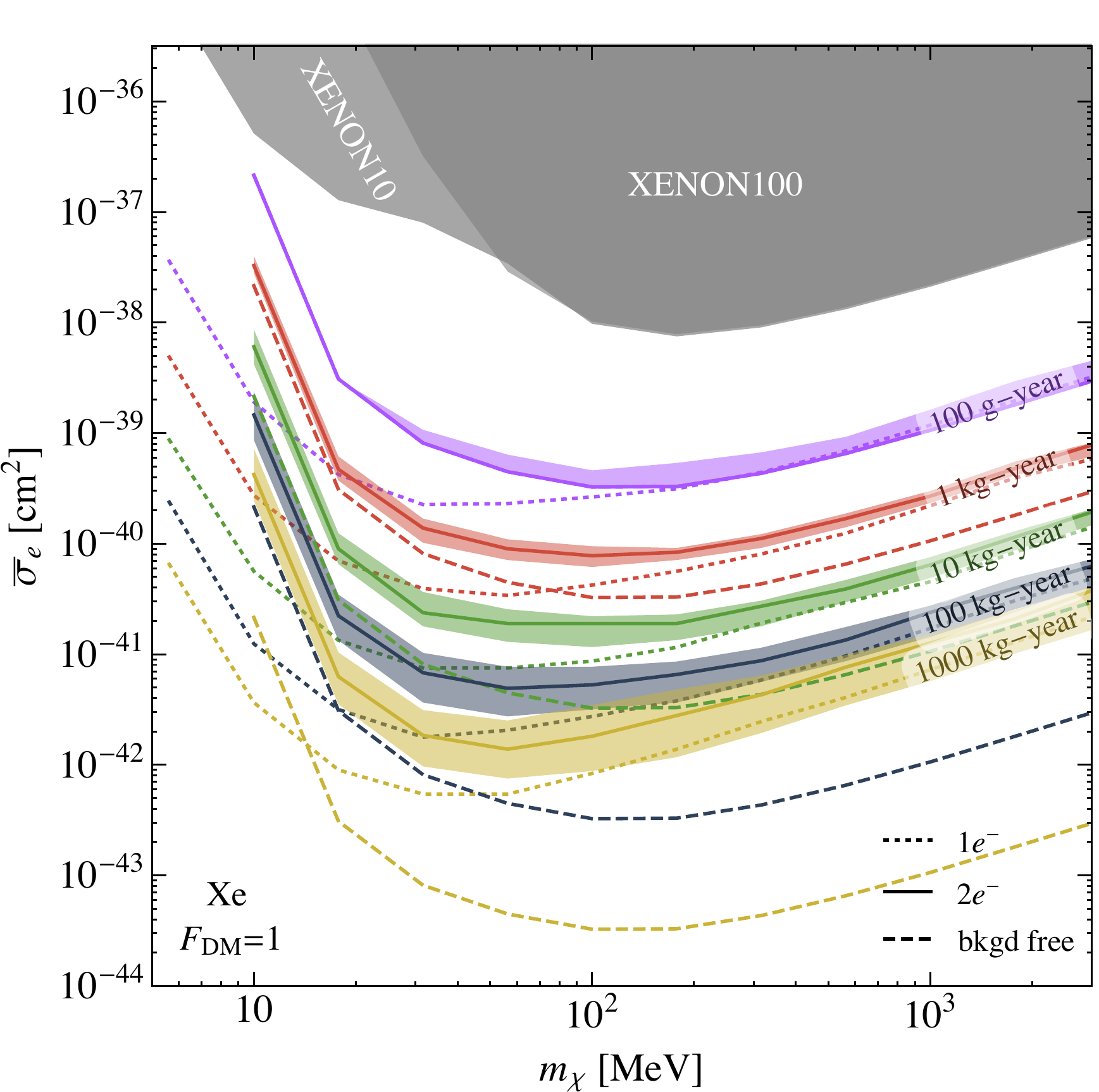} ~~~~~~~
\includegraphics[width=0.43\textwidth]{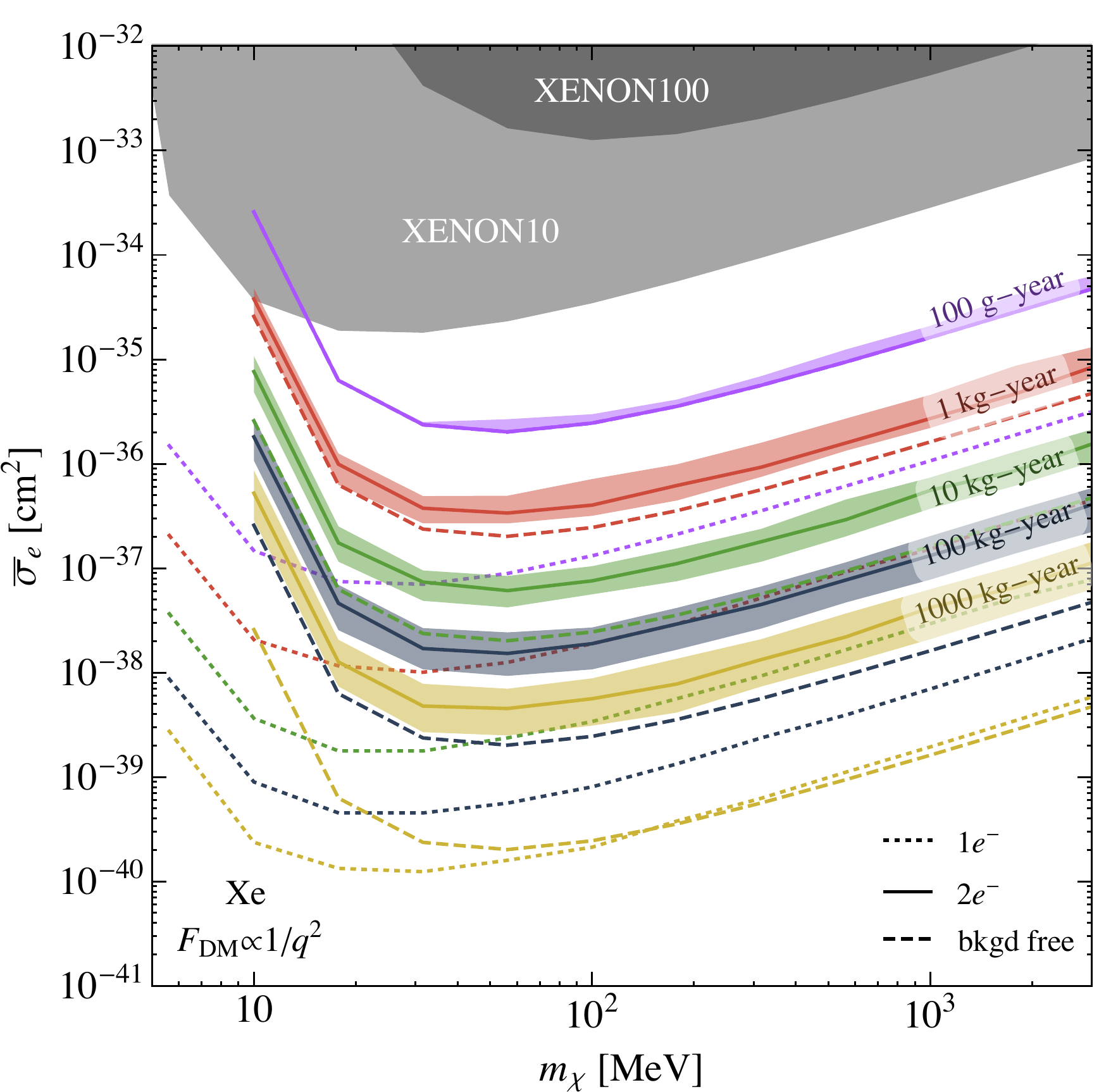}
\caption{Discovery limits for DM-electron scattering in silicon ({\bf top}), germanium ({\bf middle}), and xenon ({\bf bottom}) 
{\bf assuming a 2-electron threshold}. 
The panels on the {\bf left} ({\bf right}) 
assume the scattering is mediated by a heavy (light) particle, i.e.~$F_{\rm DM}=1$ ($F_{\rm DM} = \alpha^2m_e^2/q^2$). 
Exposures of 0.1, 1, 10, 100, and 1000~kg-years are shown in {\bf purple}, {\bf red}, {\bf green}, {\bf blue}, and {\bf yellow}, respectively. 
The solid line shows the results assuming the fiducial ionization efficiency, while the shaded bands denote the range between the high and low 
ionization efficiencies.  The dashed lines show the background-free 90\%~C.L.~sensitivities. Note that when the background assuming the fiducial ionization efficiency is negligible, the solid line and the dashed line are indistinguishable, making the dashed line disappear.  
The gray shaded region shows the current direct-detection limits on DM-electron scattering from~\cite{Essig:2017kqs}. 
}
\vskip 1cm
\label{fig:limits2ethreshold}
\end{figure*}

\section{Constraints on neutrino magnetic moment}\label{app:magnetic_moment} 
In the minimal extensions of the SM in which neutrinos have Dirac masses, $m_{\nu}$, one-loop corrections will induce a neutrino magnetic moment $\mu_{\nu}$, which is given by~\cite{Giunti:2008ve,Marciano:1977wx} 
\beq \label{eq:sm_moment} 
\mu_{\nu} = 3.2 \times 10^{-19} \Big(\frac{m_{\nu}}{\rm{1 eV}}\Big) \mu_{B},
\eeq
where $\mu_{B}$ = $\sqrt{4 \pi \alpha}/2m_{e}$ is the Bohr magneton. 
However, there are other extensions of the SM that predict a significantly higher magnetic moment~\cite{Marciano:1977wx,PhysRevD.14.3000,PhysRevD.17.1395,georgi1990h,PhysRevD.59.013010,PhysRevD.70.057301}.  This would sigificantly enhance both the coherent neutrino-nucleus and neutrino-electron scattering cross sections. 
Currently, the strongest constraint comes from the GEMMA experiment~\cite{beda2012results}, with an upper limit 
of $\mu_\nu<2.9 \times 10^{-11} \mu_{B}$. 

The magnetic moment contribution to the coherent neutrino-nucleus scattering is given by~\cite{vogel1989neutrino}, 
\beq \label{eq:mag_mom_cnns}
\frac{d \sigma_\mu (\nu N \rightarrow \nu N)}{dE_{\rm{NR}}}=\mu_{\nu}^{2} \alpha Z^{2}F^{2}(E_{\rm{NR}})\Big(\frac{1}{E_{\rm{NR}}}-\frac{1}{E_{\nu}}\Big),
\eeq  
where $Z$ is the atomic number and $F(E_{\rm{NR}})$ is the nuclear form factor, which we assume, as before, to be 1. The enhancement in the neutrino-electron cross section is given by,
 \beq \label{eq:mag_mom_nue}
\frac{d \sigma_\mu (\nu e \rightarrow \nu e)}{dE_{e}}=\mu_{\nu}^{2} \alpha \Big(\frac{1}{E_{e}}-\frac{1}{E_{\nu}}\Big).
\eeq  
Both of these enhancements would affect the expected signal from solar neutrinos in terrestrial detectors, including direct-detection experiments, 
especially at low thresholds.  

We investigate the enhancement from a non-zero neutrino magnetic moment, and whether it is visible in 
a future direct-detection experiment.  
We find that even the enhancement from a neutrino magnetic moment with a value given by the current GEMMA bound is too small to be 
detected in searches for coherent solar-neutrino-nucleus scattering, at least in the energy range of interest of upcoming experiments ($E_{\rm NR}\gtrsim 10$~eV).  
The enhancement of the neutrino-electron scattering cross section is more important.  
However, even here a 100 kg-year experiment (without any other ionization backgrounds) would only constrain 
$\mu_\nu$ to be less than than $1.31 \times 10^{-11} \mu_{B}$, $2.28 \times 10^{-11} \mu_{B}$, and  $2.45 \times 10^{-11} \mu_{B}$ at $2\sigma$ confidence level in silicon, germanium and xenon, respectively.  
Direct-detection experiments are thus not very sensitive to a neutrino magnetic moment from measurements of the solar neutrino flux.  

\bibliography{neutrinos}
\end{document}